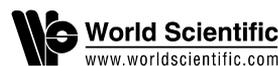

# Collective dynamics of polarized spin-half fermions in relativistic heavy-ion collisions

Rajeev Singh

*Center for Nuclear Theory,*
*Department of Physics and Astronomy,*
*Stony Brook University, Stony Brook, NY 11794, USA*

*Institute of Nuclear Physics, Polish Academy of Sciences,*
*Krakow PL-31-342, Poland*

*Department of Physics, Brookhaven National Laboratory,*
*Upton, NY, 11973-5000, USA*
*rajeev.singh@stonybrook.edu*

Standard relativistic hydrodynamics has been successful in describing the properties of the strongly interacting matter produced in the heavy-ion collision experiments. Recently, there has been a significant theoretical advancement in this field to explain spin polarization of hadrons emitted in these processes. Although current models have successfully explained some of the experimental data based on the coupling between spin polarization and vorticity of the medium, they still lack a clear understanding of the differential measurements. This is commonly interpreted as an indication that the spin needs to be treated as an independent degree of freedom whose dynamics is not entirely bound to flow circulation. In particular, if the spin is a macroscopic property of the system, in equilibrium its dynamics should follow hydrodynamic laws. Here, we develop a framework of relativistic hydrodynamics which includes spin degrees of freedom from the quantum kinetic theory for Dirac fermions and use it for modeling the dynamics of matter. Following experimental observations, we assume that the polarization effects are small and derive conservation laws for the net baryon current, the energy–momentum tensor and the spin tensor based on the de Groot–van Leeuwen–van Weert definitions of these currents. We present various properties of the spin polarization tensor and its components, analyze the propagation properties of the spin polarization components, and derive the spin-wave velocity for arbitrary statistics. We find that only the transverse spin components propagate, analogously to the electromagnetic waves. Finally, using our framework, we study the space–time evolution of the spin polarization for the systems respecting certain space–time symmetries and calculate the mean spin polarization per particle, which can be compared to the experimental data. We find that, for some observables, our spin polarization results agree qualitatively with the experimental findings and other model calculations.

## 1. Introduction

Understanding the origin of our universe, its physical properties, its composition, etc. are among the primary goals of modern science. According to the standard model of cosmology, the theory of the Big Bang proposed by Lemaître[1] in the year 1927,[a] has been considered the most reliable framework to date to study the evolution of the universe. Extrapolated back in time the Big Bang model gives us crucial information about its origin. According to this model, our universe is expanding and it was much hotter and much smaller in the beginning. The experimental validation of its expansion came after galaxy red-shift measurements by Hubble in 1929.[7] His observational evidence along with the cosmic microwave background[8] and Big Bang nucleosynthesis[9] became the key milestones of modern cosmology. To explore the physical properties and the matter content of our universe, physicists have considered different approaches. One of the approaches is the direct astrophysical and astronomical observation, meaning observation at large length scales using telescopes probing different electromagnetic (EM) wavelengths. The other approach is to reproduce the conditions which existed during the very early stages of the universe in terrestrial laboratories producing hot and dense medium in relativistic heavy-ion collisions.[10,11] According to the thermal history of our universe, its temperature and its age, at the very beginning, during the Planck era, were of the order of $10^{19}$ GeV and $10^{-43}$ s, respectively.[12] The universe, being in a very hot and dense early stage, then cooled and diluted due to its expansion. It has been speculated that due to the cooling, our universe also went through several phase transitions, for instance, grand unification,[b] electroweak and quantum chromodynamics (QCD) phase transitions, etc. at different time scales.[13,14] Among these phase transitions, the dynamics of the QCD one is the most important for our current discussion. QCD transition which can also be understood as the transition from the deconfined quark–gluon matter, more commonly known as the quark–gluon plasma (QGP), to confined hadronic matter occurred when the temperature was of the order of $T \sim \Lambda_{\text{QCD}} \approx 200$ MeV and the age of the universe was a few microseconds ($\sim 10^{-6}$ s) after the Big Bang.[14] It turns out that the QCD transition is the only phase transition of the early universe history that can be studied in the relativistic heavy-ion collision experiments. Due to the direct experimental access, we can explore various aspects of the dynamics of QCD phase transition, which is otherwise not possible. We should emphasize that the study of the QCD phase transition dynamics is also crucial for the existence of nuclear matter in the universe. As the universe cooled across the QCD transition scale, bound states of elementary particles such as baryons were formed out of quark–gluon matter. Subsequent evolution of the universe and due to the Big Bang nucleosynthesis, protons and neutrons combined to form certain species of atomic nuclei. Today most of the observed matter in our universe is in the form of atomic nuclei. Therefore, understanding the QCD dynamics in the early universe is of paramount importance. Moreover, large compact objects, such as neutron stars may also help us to understand the properties of matter under extreme conditions. These compact objects can also contain various phases of QCD matter, e.g. QGP, hadronic matter, etc. Since we cannot access the QCD plasma of the early universe or the interior of astrophysical objects directly, relativistic heavy-ion collision experiments across a wide range of collision energies provide us a unique opportunity to explore the QCD dynamics.[15–18]

---

[a]Historically speaking, Friedmann was the first who came up with the idea that the universe can be dynamic[2] after finding the solutions of Einstein's field equations. However, he did not link his findings to any astronomical observations. There were many attempts to obtain the dynamic universe solution from Einstein's equations but none of them suggested that the universe is expanding.[3] It was Lundmark in 1924, who estimated extra-galactic distances and suggested that galaxies are red-shifted.[4] Then in 1927, Lemaître discovered new dynamical solutions to Einstein's equations and obtained the relationship between linear velocity and distance, which he then linked to astronomical observations. This solution was also independently discovered later by Robertson[5] and Walker.[6]

[b]This is a phase transition where the strong force got disentangled from the weak and EM forces.

## 1.1. *Quark–gluon plasma*

The dynamics of strongly interacting matter namely quark–gluon matter and the hadronic matter is governed by the fundamental theory of QCD. Two key aspects of the QCD are *asymptotic freedom* and *color confinement*. The QCD coupling constant becomes very small at the high-energy scale or a large momentum transfer scale in microscopic collisions. This phenomenon is known as the asymptotic freedom[19,20] which can be shown analytically using the methods of perturbative QCD and has been well tested in deep inelastic scattering experiments. Interestingly, the momentum dependence of the QCD coupling also indicates that in the low-energy scale of $\Lambda_{\mathrm{QCD}}$, this coupling is rather large leading to the breakdown of the perturbative QCD techniques. Such nonperturbative nature at the low-energy scale is very interesting and it can be argued to be associated with the nonabelian nature of the QCD gauge fields. Due to this nonabelian nature, gluons can self-interact leading to the phenomenon of color confinement which prevents us to observe directly the "colored" partons, i.e. quarks, antiquarks and gluons. These colored partons are confined within color singlet (color neutral) hadrons[c] which can be observed in the experiments. Fundamental properties of QCD are also manifested in its phase transition dynamics. In the QCD medium, at a very high temperature ($T \gg \Lambda_{\mathrm{QCD}}$) or high baryon number density ($\mathcal{N} \gg \Lambda_{\mathrm{QCD}}^3$),[d] partons are expected to be liberated from individual hadrons and form a deconfined phase of quarks and gluons. In this deconfined medium, partons can move across length scales that are larger than the confinement length scale, i.e.

---

[c] Hadrons can be considered as the low-energy states of QCD, although a framework describing color confinement from a microscopic point of view is still lacking. Nevertheless, various lattice QCD numerical calculations unambiguously predict the existence of various hadronic states starting from the fundamental QCD Lagrangian.

[d] Understanding the extreme properties has always been important in paving the way to discover new states of matter.[21] The first exploration of the properties of matter at high densities was performed by Oppenheimer and Volkoff[22] in the context of star formation, followed by the investigations in other areas of physics.[23,24]

the mean size of hadrons. The transition from the confined hadronic phase to the deconfined QGP phase is known as the confinement–deconfinement phase transition. We should note that, although the confinement–deconfinement transition was predicted as a true phase transition, lattice QCD calculations (with physical quark masses) showed that this is not a phase transition but rather a crossover. Such conclusions can be drawn by considering the variation of the thermal expectation value of the Polyakov loop with various physical scales. The thermal expectation value of the Polyakov loop can be considered as an order parameter of the confinement–deconfinement transition.[25–27] At vanishing baryon chemical potential with physical quark masses, the thermal expectation value of the Polyakov loop changes continuously from the confined phase to the deconfined phase with temperature. On the other hand in the pure gauge sector, QCD phase transition is of first order.[28]

Apart from the $SU(3)_c$ gauge symmetry, QCD Lagrangian can also have other global symmetries. One such symmetry of the massless QCD Lagrangian is the chiral symmetry, denoted as $SU(3)_V \otimes SU(3)_A$ with $V$ and $A$ representing the vector and axial-vector transformation in the fermionic sector of the QCD Lagrangian. In the high-energy scale ($\gg \Lambda_{QCD}$), one can safely neglect masses of light quarks (including strange quark). The chiral symmetry of QCD is exact only in the vanishing quark mass limit, however, in reality, chiral symmetry is only an approximate symmetry. Note that, if we only consider $u$ and $d$ quarks, the physical masses of these quarks can be neglected in comparison to $\Lambda_{QCD}$ scale. Thus, in the nonstrange light quark sector, chiral symmetry is physically motivated. If chiral symmetry is a symmetry of nature then one would expect parity doublets of observed hadrons, which are not observed.[e] This implies that although the chiral symmetry is the symmetry of the QCD Lagrangian, in the QCD vacuum this symmetry is not manifested. Hence, the chiral symmetry is "spontaneously" broken, $SU(3)_V \otimes SU(3)_A \rightarrow SU(3)_V$. For two light-flavor quarks ($u$ and $d$) case, pions are the Goldstone bosons of the spontaneous chiral symmetry breaking.[29,30] Spontaneous breaking of chiral symmetry can be argued in various QCD-inspired effective models due to the presence of nonvanishing quark–antiquark scalar condensate (chiral condensate) which is an order parameter of the chiral phase transition.[31,32] Various QCD effective model calculations indicate that in the high temperature and/or high chemical potential limit the chiral condensate has a vanishing value. This phase is called the chirally symmetric or chirally restored phase. On the other hand, in the low-temperature limit, chiral condensate can have a nonvanishing value. This phase is known as the chiral symmetry broken phase. The quark matter phase is expected to be a chiral symmetry restored phase and in the hadronic phase, the chiral symmetry is spontaneously broken. Spontaneous breaking of chiral symmetry also gives rise to nonperturbative corrections to the masses of quarks and hadrons.[29] One should emphasize that the chiral symmetry group is only defined in the fermionic sector of the QCD Lagrangian. On the other hand, the local $SU(3)_c$ symmetry deals with color gauge fields. Hence, it is not expected that the chiral phase transition and the confinement–deconfinement transitions are connected. However, lattice QCD calculations suggest that the chiral transition temperature and confinement–deconfinement transition temperature coincide indicating that these two transitions can be connected.[33]

---

[e]Pions have specific parity, and pions with opposite parity have not been observed in low-energy experiments.

Based on the concept of asymptotic freedom, Collins and Perry in 1975[34] proposed the idea of superdense matter in the interiors of neutron star and noted that there is a qualitative difference between QCD at high temperature (and/or density) and QCD at low temperature (and/or density). In the same year, Cabibbo and Parisi gave a schematic phase diagram of baryonic matter and suggested that Hagedorn temperature[35] might be the temperature of phase transition between the quark matter and hadronic matter.[36] They also noted that hadronic matter, at high temperature and/or density, should have a phase transition into a new state where quarks are deconfined. Shuryak called this new phase of deconfined matter $QGP$.[37,38] The idea of investigating the characteristics of highly dense and hot matter in laboratory conditions came in 1970's[39–41] after the study of supernovae and neutron stars. This led to the idea that the collisions of heavy ions at very high energy may be able to produce dense matter like the QGP giving birth to a new field of physics, namely relativistic heavy-ion collisions.[42–51] To mimic the conditions existing in the very early universe, the experimental study of the QGP properties was initiated using ultra-relativistic collisions of heavy ions[10,11,52] in facilities like the Conseil Européen pour la Recherche Nucléaire (CERN), founded in 1952 and the Brookhaven National Laboratory (BNL), founded in 1947.

### 1.1.1. *Search for QGP and its possible signatures*

Significant efforts have been put forward through the years to understand the phase transition dynamics of QCD in heavy-ion collision experiments, e.g. at CERN (SPS, LHC) and BNL (AGS, RHIC). Fixed target experiments at AGS and SPS where one collides $Au + Au$ at up to 11 GeV per nucleon beam energy (AGS) and $Pb + Pb$ at up to 160 AGeV (SPS) indicated the possible existence of a new form of hot and dense state of QCD matter. In these low-energy heavy-ion collision experiments the target and the projectile nuclei cannot pass through each other due to the nuclear stopping and strong QCD coupling. Such low-energy collisions provide an opportunity to systematically study the baryon-rich matter or the QGP at high baryon density which is expected to exist in the interior of neutron stars. On the other hand, with the help of the heavy-ion collision program at RHIC (BNL) with the collision energy up to $\sqrt{s_{NN}} = 200$ GeV and at LHC (CERN) with the center-of-mass collision energy per nucleon $\sqrt{s_{NN}} = 2.76$ TeV and higher, we can explore the QCD dynamics at temperature and baryon chemical potential relevant for understanding the QCD phase transition dynamics in the early universe.

Figure 1 shows the possible evolution stages of the matter created in the ultra-relativistic collisions. Contrary to a fixed target experiment, in collider experiments heavy-ion beams collide at a very high-energy scale where QCD coupling is relatively small and, due to the nuclear transparency, the target and projectile nuclei almost pass through each other.[54] Therefore, in the high-energy collider experiments, the interaction region remains almost baryon free. Despite the nuclear transparency at ultra-relativistic energies, when the highly Lorentz contacted nuclei pass through each other, a large amount of energy density can be deposited in the reaction zone forming secondary partons. Due to a large number of re-scatterings, these secondary partons redistribute energy among themselves to produce a (locally) thermally equilibrated plasma which is expected to be the QGP phase, which then expands due to high internal pressure. Because of the expansion, after some time, the fireball cools down and it undergoes the quark-hadron transition to form a hadronic medium at freeze-out.[f] These intermediate stages are not observed in the experiments. Moreover, in detectors, we observe quantities such as momentum spectra and multiplicities of hadrons, but not their actual positions.[56–58]

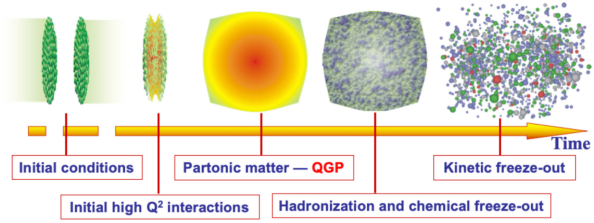

Fig. 1.   Schematic picture of the evolution stages of ultra-relativistic heavy-ion collision. Time in the horizontal axis has the unit of fm/c and $Q^2$ denotes four-momentum transfer squared.[53]

Since QGP is a color-charged medium with a very short lifetime (of the order of femtometers), it cannot be detected in the experiments directly. Thus, one needs to propose physical observables which can provide information about the QGP indirectly. Some signatures of the QGP formation which are commonly studied are as follows.

• **Hard thermal dilepton and photon radiation:**[59,60] Direct photon radiation and dileptons such as $e^+$, $e^-$ and $\mu^+$, $\mu^-$ contain the thermodynamic information of the QGP fireball. Even though they are difficult to interpret, they are probably the best observables to study the relativistic collisions of heavy ions as they penetrate the matter easily and are not affected by the process of hadronization. They are also helpful in providing the initial stage information as their momentum distributions depend on the partons temperature. Since the temperature of the QGP phase is supposed to be higher than the hadronic phase, there should be an increase in higher transverse momenta due to the momentum of direct photons.

---

[f]Freeze-out is the stage where the interaction between particles stops and final hadrons are freely streaming to the detector. There are two types of freeze-outs: *thermal* and *chemical*. *Thermal* or *kinetic* freeze-out is the stage where final hadrons stop interacting completely, whereas, *chemical* freeze-out is the stage where there are no inelastic collisions between the particles.[50] Some experimental data can be explained with the so-called single-freeze-out model where these two types of freeze-outs coincide.[55]

- **Quarkonia suppression:**[61−63] As heavy quarks are produced in the initial hard scattering processes, these quarks witness the entire evolution of the strongly interacting matter produced in the heavy-ion collisions. Heavy quarks and anti-quarks interact to produce bound states which are known as quarkonia, for instance, the bound states of charm and anticharm (charmonia) and bottom and antibottom (bottomonia) quarks.[g] It was proposed in 1986 that suppression of $J/\psi$ (bound state of $c$ and $\bar{c}$) can be used as a probe for QGP. The key concept can be understood as follows: $c$ and $\bar{c}$ are produced at a very early stage of QGP formation. The high gluon density in the deconfined matter, due to the color deconfinement, causes Debye screening of $c$ and $\bar{c}$ effective interaction. This Debye screening can be characterized by the Debye mass or Debye screening length scale which depends on the temperature of the medium. Whenever the Debye screening length becomes smaller than the radius of charmonium, the charm quarks and antiquarks cease to interact with each other giving rise to the decay, and hence, suppression of charmonium states. Moreover, such Debye screened $c$ and $\bar{c}$ particles may form hadrons, like $D$ and $\bar{D}$ mesons, with their light quark partners. Since the Debye screening length is sensitive to the temperature of the medium, different quarkonia states having different binding energies will dissociate in the medium at different temperatures. Hence, quarkonia suppression can also be considered a "thermometer" for the QGP.

- **Strangeness enhancement:**[64,65] Enhancement in the strange quark production relative to the up and down quarks in the heavy-ion collisions has been proposed in the year 1982 as a probe to study QGP properties as more pairs of $s\bar{s}$ are supposed to be produced in the QGP because of high-energy density and fusion of gluons contributing almost 80% to the production of strange quarks. Note that colliding nuclei do not contain strange quarks, therefore all strangeness must be created during the collisions. We should emphasize that the strangeness production mechanism in a hadronic medium can be significantly different from the one in the QGP. Since strangeness is conserved in QCD, strange particles and antiparticles must be produced in pairs. Therefore, the threshold for strange hadron production in a hadronic medium is rather high compared to others, e.g. the threshold energy of strangeness production in a pion nucleon scattering $\pi + N \to \Lambda + K$ is ∼540 MeV. On the other hand, in the QGP medium, due to the chiral symmetry restoration,[66] the *constituent* quark masses can be replaced by the *current* mass. Therefore, above the chiral symmetry restoration threshold, strange–antistrange quark pair production gets reduced, enhancing the strangeness production in the deconfined partonic medium. Assuming that the produced strange particles and antiparticles take part in the hadronization, the strangeness abundance in the partonic phase should be imprinted in the relative yield of the strange and non-strange hadrons. As compared to the $e^+$–$e^-$ collision or proton–proton collision, in the heavy-ion collisions strangeness enhancement has been, indeed, observed.[67]

---

[g]Top quarks form bound state but they have a very short lifetime and thus decay quickly.

- **Jet quenching:**[68,69] Jets originating from the initial hard scattering of partons can be considered as an important probe for the hot and dense QGP matter created in heavy-ion collisions. Jets are composed of highly collimated and correlated partons having large transverse momentum ($p_T$) (transverse to the beam direction) and are mostly produced back to back due to the conservation of energy and momentum (although three and four jet events can also take place). These high $p_T$ particles lose their energy through collinear radiation. Jets produced at the initial stage traverse through the thermalized medium before escaping the QGP. While passing through the medium, they lose their energy much more significantly than the ones traversing hadronic matter, as jet energy loss or parton energy loss can be significantly different in the medium from energy loss in the vacuum. As a result, jets that are produced at the edge of the QGP fireball can escape easily, while jets emitted in the opposite direction traverse a longer length through the thermalized medium and undergo significant medium modification. Such phenomenon is known as "jet quenching" in a medium. Medium effects on jets can also be quantified by the jet nuclear modification factor (jet-$R_{AA}$). Quantitatively, values of jet-$R_{AA}$ less than one imply medium modification. The experimentally observed "jet quenching" indicates the formation of a partonic medium.[70,71]

Evidences provided by CERN,[72] in 2000, and by RHIC at BNL,[54,73] in 2005, confirmed the production of a new state of hot and dense matter. This was possible due to both high-quality experimental data, coming from the experiments, and theoretical models, which gave a precise quantitative description. Opposite to theoretical expectations, it was shown that this new state of matter behaves as a strongly-coupled system whose evolution follows the dynamics of a perfect fluid.[74,75,h] This is due to the very small kinematic shear viscosity ($\eta/s$) obtained from the transverse-momentum spectra of the charged particles; here $\eta$ is the dynamic shear viscosity coefficient and $s$ is the entropy density. In fact, $\eta/s$ of the QGP, which, by the way, is the smallest value observed in nature, is also very close to the so-called Kovtun–Son–Starinet bound, $\eta/s = 1/(4\pi)$,[79,80] derived using gauge–gravity correspondence.[81]

## 1.2. *Spin polarization of Lambda hyperons*

Recently, new directions in heavy-ion collision physics have opened up after the speculation that the noncentral relativistic collisions can be the source of a large magnetic field and angular momentum. The presence of a large magnetic field (of the order of $\Lambda_{QCD}^2$) can give rise to charge–parity-violating effects, e.g. chiral magnetic effect,[82,83] etc. which are not yet confirmed by the experiments. Moreover, noncentral nuclear collisions at high energies can induce, in quark–gluon matter, a large angular momentum ($\sim 10^5 \hbar$)[84] which leads to measurable spin polarization of the emitted hadrons.[85]

---

[h] See Refs. 76–78 for the studies related to the strongly interacting systems in the area of condensed matter physics.

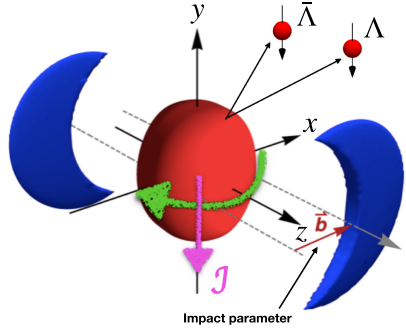

Fig. 2.    Schematic diagram of the initial angular momentum orientation in noncentral heavy-ion collision.[86]

In 2005, it was proposed[87] that the spatial inhomogeneity of the colliding system in noncentral heavy-ion collisions leads to the deposition of large initial orbital angular momentum in the produced matter along the direction orthogonal to the reaction plane ($y$-axis), whose significant fraction may be transferred in partonic collision processes to the spin of the QGP constituents, see Fig. 2. As a result, due to angular momentum conservation, it may lead to polarization of quarks and antiquarks, which can be subsequently transferred to the hadrons resulting in their global spin polarization along $y$-direction. A macroscopic interpretation of this phenomenon was provided in 2008 using the hydrodynamic considerations.[84] According to it, the spatial inhomogeneities, arising from the initial orbital angular momentum, lead to the generation of fluid vorticity. The particles emerging from such a vortical system, through the so-called spin–vorticity coupling, should exhibit a global polarization along the same direction.[i] Among various spin-polarizable hadrons, weakly decaying $\Lambda(\bar{\Lambda})$ hyperons are of special interest for quantitative estimation of the vorticity-driven polarization. In the parity-violating decay of the $\Lambda$ particle the outgoing proton is preferentially emitted along the spin direction of its parent particle, see Fig. 3. Hence, measuring the direction of the proton's momentum in the hyperon's rest frame can provide information about the spin polarization of the hyperon. Such "self-analyzing" weak-decay channel of $\Lambda(\bar{\Lambda})$ hyperons makes them excellent probes for the polarization measurements. The first positive experimental observation of *global spin polarization*[j] of $\Lambda(\bar{\Lambda})$ hyperons[k] published by STAR Collaboration[85,91] provided experimental evidence of vortical structure of the QGP. Using the spin–vorticity coupling, which holds for spin-polarized systems in thermal equilibrium,[92] the experimentally observed value of polarization, see Fig. 4, leads to prediction of the value of vorticity of the order of $10^{21}\,\mathrm{s}^{-1}$ which is much larger than the vorticity of any other physical system. From Fig. 4, one may notice that the magnitude of the average global spin polarization of both $\Lambda$ and $\bar{\Lambda}$ increases with the decrease in the collision energy which naturally makes the low- and mid-energy reactions of special interest. The differences between particle and anti-particle polarization may be explainable by their interaction with the initial EM fields supposedly generated in early stages, however, they are not yet quantitatively described theoretically.

[i]Interestingly, a nonrelativistic analog of such a mechanism has been observed in experiments by Einstein and de Haas[88] and Barnett.[89]
[j]Here, "global" means the average polarization in the event with respect to the direction of the global orbital angular momentum of the colliding system, see Fig. 2.
[k]The discovery of $\Xi$ and $\Omega$ hyperon global polarization was also reported recently.[90]

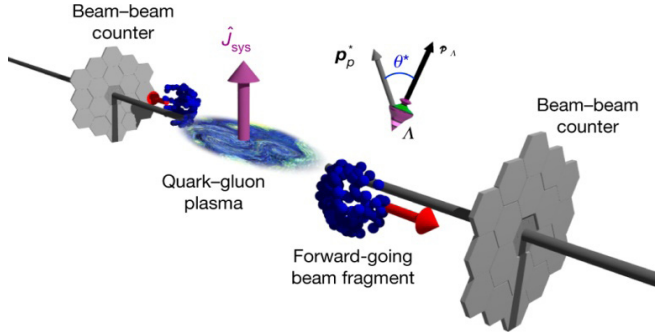

Fig. 3.   Schematic diagram of a Au + Au collision (not to scale) where $\widehat{J}_{\mathrm{sys}}$ is the direction of the angular momentum of the collision, see Ref. 85.

This measurement has opened up new directions in the studies of QGP,[86,93–100] providing opportunities to gain new physics insights such as chiral dynamics of strongly interacting and deconfined matter.[101,102] New phenomena such as chiral vortical effect,[103–113] due to vortical structure of QGP, analogous to the chiral magnetic effect[82,83,114–122] generated huge interests.

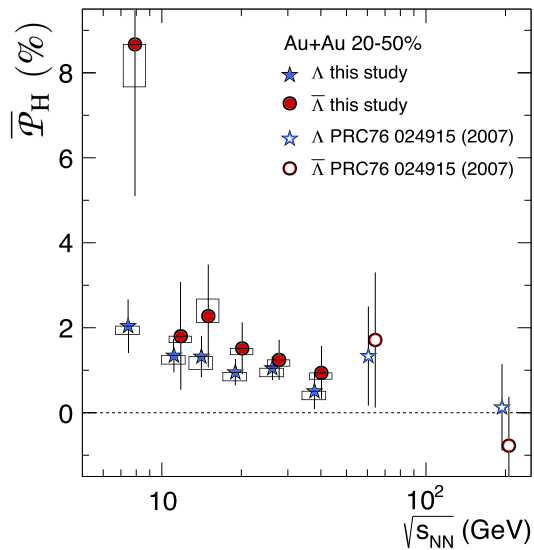

Fig. 4.   Average global spin polarization for $\Lambda(\bar{\Lambda})$ hyperons in 20–50% centrality Au + Au collisions as a function of collision energy.[85]

Apart from global vorticity along the direction of the total angular momentum, namely the $y$-component, the matter produced in the heavy-ion collisions may also exhibit local vortical structures which can lead to nonvanishing values of other components of polarization. Of particular interest is the *longitudinal polarization*, meaning the component of polarization along the beam direction ($z$-axis), as it probes the vorticity of the velocity field generated in the transverse plane, see Fig. 5. Indeed, the measurements of $\Lambda(\bar{\Lambda})$ longitudinal polarization[125] confirm this reasoning. Figure 6 shows the cosine of the polar angle of the daughter particle in the rest frame of its parent particle that is averaged over all $\Lambda(\bar{\Lambda})$ particles. One may notice the quadrupole dependence of the longitudinal polarization which matches the one for longitudinal vorticity resulting from the elliptic flow structure in Fig. 5. While theoretical calculations based on transport models[126] confirm observed signal, hydrodynamic models using spin–vorticity coupling did not predict it correctly.[123,127,128] Note that due to symmetry, the longitudinal component of averaged polarization is supposed to be zero. Spin polarization[l] of the $\Lambda(\bar{\Lambda})$ hyperons has also been observed in the lowest[132] and the highest energies[133,134] which became the subject of intense investigations.[135–139]

### 1.3. *Relativistic hydrodynamics*

If due to initial multi-particle scattering, thermalization can be achieved locally in the early stages of heavy-ion collisions and the interaction among quarks and gluons

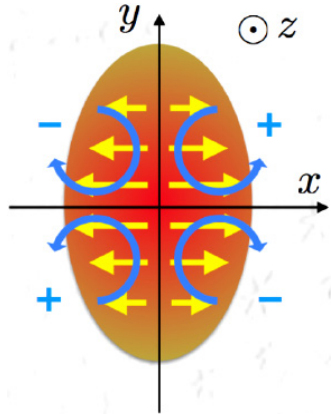

Fig. 5.   Schematic view of the flow structure in the transverse plane which may generate longitudinal polarization.[123,124]

[l]Study of spin physics and vorticity in the context of relativistic heavy-ion collisions may also help in clear understanding of spin–orbit coupling in the field of spintronics.[129–131]

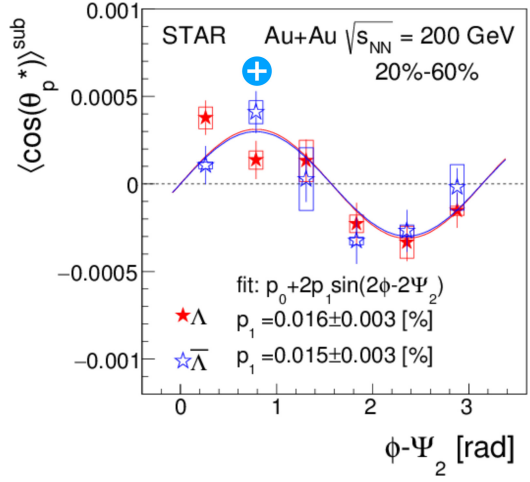

Fig. 6. Cosine of the polar angle of the proton in the rest frame of the parent $\Lambda(\bar{\Lambda})$ that is averaged over all $\Lambda(\bar{\Lambda})$ particles as a function of azimuthal angle $(\phi)$ relative to second-order event plane $(\psi)$.[125] Note the correlation with flow structure in Fig. 5 represented by "+" sign.

is strong enough to maintain local thermodynamic equilibrium (LTE), then the space–time evolution of the strongly interacting matter can be described by hydrodynamic principles.[50,51,140–148]

Contrary to microscopic quantum field theory (QFT), hydrodynamics is an effective theory defined at a length scale larger than the mean free path of microscopic particles, but smaller than the system size. Such scale differences can be argued if the microscopic and the macroscopic length scales associated with the system are sufficiently separated. Generically a hydrodynamic model describes the space–time evolution of macroscopic thermodynamic variables, such as local energy density, pressure, temperature, fluid flow velocity, etc. In fluid or hydrodynamic description the system is expected to be continuous. If we neglect any nonequilibrium effects then every fluid cell (infinitesimal volume element) is assumed to be in LTE throughout its evolution. This implies that each fluid element must be large enough (compared to the microscopic scales) to properly define the thermodynamic variables including the notion of thermal equilibrium, and, at the same time, must be small enough to ensure the continuum limit. Although relativistic hydrodynamics is an effective approach, detailed knowledge of the microscopic dynamics enters it through the equation of state (EoS) relating pressure, energy density and net baryon density. Hydrodynamic models have been used to describe collective dynamics of different physical systems including cosmological and astrophysical plasma,[149] the strongly interacting plasma produced in heavy-ion collision experiments, etc. Relativistic hydrodynamics has been extensively studied to model the bulk evolution of the strongly interacting medium produced in relativistic heavy-ion collisions, and turned out to be successful in describing multiple collective phenomena. Both ideal, as well as dissipative frameworks, have been developed to explain the observed hadron spectra in experiments.

Considering its successes, it is fair to say that the hydrodynamic framework along with the description of the initial state and the prescription for hadronization constitutes the "standard model" of heavy-ion collision physics.

In this paper, we will not discuss dissipative phenomena, explicitly, but a few comments on dissipative hydrodynamics are in order here. Although the relativistic generalization of the ideal hydrodynamic framework is rather straightforward, the generalization of dissipative nonrelativistic hydrodynamics for a relativistic system is a nontrivial task and it is a timely research topic in high-energy physics. A relativistic generalization of Navier–Stokes (NS) theory was proposed by Eckart[150] and Landau[151] independently. Eckart's and Landau's theories are also known as first-order theories of dissipative hydrodynamics as the dissipative current can be expressed as a first-order gradient of the fundamental fluid variables, i.e. temperature, chemical potential (associated with any global conserved charge) and fluid four-velocity. However, the relativistic NS frameworks as given by Eckart and Landau are plagued with acausality and instability problems.[152,153] The reason for the acausality is that in the NS formalism, the dissipative currents are linearly proportional to gradients of temperature, chemical potential and velocity giving rise to parabolic differential equations which do not preserve causality which, in turn, can be the source of instability. The second-order generalization of the NS framework as put forward in various studies solves the problem of acausality by introducing a "time delay" between the gradients of the primary fluid-dynamical variables and resulting dissipative currents. In such a theory, apart from temperature, chemical potential and fluid four-velocity, one also has to consider various dissipative quantities as dynamic variables satisfying evolution equations describing their relaxation towards the NS limit. In this improved framework, the hydrodynamic equations are generically hyperbolic preserving causality. A more detailed discussion on the development of second-order dissipative hydrodynamics can be found in Refs. 145, 154 and 155.

In this work, we focus on the perfect-fluid hydrodynamic framework and its possible generalization to include the space–time evolution of spin degrees of freedom. Therefore, we first discuss briefly the salient features of standard perfect-fluid hydrodynamics without spin for completeness. The details of perfect-fluid hydrodynamics with spin will be discussed in the subsequent sections of this paper.

### 1.3.1. *Kinetic-theory-wise formulation of hydrodynamics*

A covariant hydrodynamic theory can be formulated using the macroscopic conservation laws. On the macroscopic level, a many-particle system can be expressed in terms of currents such as the energy–momentum tensor, number current associated with any conserved quantity, etc. Using the single-particle distribution function $(f(x,p))$ we can write the energy–momentum tensor $(T^{\mu\nu})$, and number current $(N^\mu)$ in the following manner:

$$T^{\mu\nu} = \int \mathrm{dP}\, p^\mu p^\nu f(x,p), \quad N^\mu = \int \mathrm{dP}\, p^\mu f(x,p), \tag{1}$$

where $\mathrm{dP} = d^3p/((2\pi)^3 E_p)$ is the momentum integration measure with $E_p = \sqrt{\boldsymbol{p}^2 + m^2}$ denoting the on-shell particle energy. It can easily be proved that the momentum integration measure dP is a Lorentz-invariant quantity and the distribution function $f(x,p)$ transforms as a Lorentz scalar. Physically, the components $T^{00}$, $T^{0i}$, $T^{i0}$ and $T^{ij}$ can be interpreted as the energy density, energy flow, momentum density and momentum flow, respectively. Similarly, $N^0$ is the particle number density, and $N^i$ is the particle number current. In relativistic kinetic theory, $N^\mu$ and $T^{\mu\nu}$ can be considered as the first and the second moment in momentum space of the distribution function.[156,157] It should be emphasized that to define the energy–momentum tensor (1) one implicitly assumes that the system is dilute such that the interaction energy among different particles can be neglected as compared to their kinetic energy.

Equations (1) can be generalized to a multi-component system in a straightforward way[156]

$$T^{\mu\nu} = \sum_{k=1}^{N} \int \frac{d^3p_k}{(2\pi)^3 p_k^0} p_k^\mu p_k^\nu f_k(x, p_k), \quad N^\mu = \sum_{k=1}^{N} \int \frac{d^3p_k}{(2\pi)^3 p_k^0} p_k^\mu f_k(x, p_k). \quad (2)$$

Here, $k = 1, 2, \ldots, N$ denotes different particle components of the system, e.g. if we consider a hadronic medium then $k$ runs over particle species. *A priori*, the conservation of $T^{\mu\nu}$ and $N^\mu$ is not evident from Eqs. (1). Conservation of these macroscopic currents can be obtained using the evolution equation for the distribution function within the framework of relativistic kinetic theory. Assuming only binary collisions along with the hypothesis of molecular chaos, the kinetic equations determining dynamics of distribution functions $f_k(x, p_k)$ can be expressed as

$$p_k^\mu \partial_\mu f_k(x, p_k) = \frac{1}{2} \sum_{i,j,l=1}^{N} \int \frac{d^3p_l}{(2\pi)^3 p_l^0} \frac{d^3p_i}{(2\pi)^3 p_i^0} \frac{d^3p_j}{(2\pi)^3 p_j^0} (f_i f_j W_{ij \to kl} - f_k f_l W_{kl \to ij}). \quad (3)$$

From the above equation, one can identify the collision term for the process $kl \to ij$

$$\mathcal{C}_{kl} \equiv \frac{1}{2} \sum_{i,j=1}^{N} \int \frac{d^3p_l}{(2\pi)^3 p_l^0} \frac{d^3p_i}{(2\pi)^3 p_i^0} \frac{d^3p_j}{(2\pi)^3 p_j^0} (f_i f_j W_{ij \to kl} - f_k f_l W_{kl \to ij}), \quad (4)$$

where $W_{kl \to ij}$ is the transition rate for the collision process $kl \to ij$ which includes both elastic and inelastic scattering processes. The factor of $1/2$ in the front takes into account the identical particles in the final state. By the virtue of conservation of energy and momentum in microscopic collisions, the collision term $\mathcal{C}_{kl}$ satisfies the following relations[156]:

$$\sum_{k,l=1}^{N} \int \frac{d^3p_k}{(2\pi)^3 p_k^0} \mathcal{C}_{kl}(x, p_k) = 0, \quad \sum_{k,l=1}^{N} \int \frac{d^3p_k}{(2\pi)^3 p_k^0} p_k^\mu \mathcal{C}_{kl}(x, p_k) = 0. \quad (5)$$

Once we identify the properties of the collision terms, the conservation of the number current can be obtained in a straightforward way

$$\partial_\mu N^\mu = \sum_{k=1}^{N} \int \frac{d^3 p_k}{(2\pi)^3 p_k^0} p_k^\mu \partial_\mu f_k(x, p_k) = \sum_{k,l=1}^{N} \int \frac{d^3 p_k}{(2\pi)^3 p_k^0} \mathcal{C}_{kl}(x, p_k) = 0. \quad (6)$$

Similarly, the conservation of the energy–momentum tensor is expressed as

$$\partial_\mu T^{\mu\nu} = \sum_{k=1}^{N} \int \frac{d^3 p_k}{(2\pi)^3 p_k^0} p_k^\mu p_k^\nu \partial_\mu f_k(x, p_k) = \sum_{k,l=1}^{N} \int \frac{d^3 p_k}{(2\pi)^3 p_k^0} p_k^\nu \mathcal{C}_{kl}(x, p_k) = 0.$$

In general, $T^{\mu\nu}$ and $N^\mu$ can be decomposed into ideal and dissipative parts. Since this paper is based particularly on the perfect-fluid hydrodynamics, we discuss the latter without spin as a warm-up. The energy–momentum tensor and the number current for a perfect fluid (single component) can be expressed as

$$T_{\mathrm{eq}}^{\mu\nu} = \int \mathrm{dP} p^\mu p^\nu f_{\mathrm{eq}}(x, p), \quad N_{\mathrm{eq}}^\mu = \int \mathrm{dP} p^\mu f_{\mathrm{eq}}(x, p), \quad (7)$$

where $f_{\mathrm{eq}}(x, p)$ is the distribution function in local equilibrium, hence, the suffix "eq". An ideal fluid is defined assuming that all fluid elements must be locally exactly in thermodynamic equilibrium. Therefore, we can define the local temperature $T(x)$, chemical potential $\mu(x)$ and fluid four-velocity $U^\mu(x)$ (normalized as $U^\mu U_\mu = 1$). Now our task is to write the conserved energy–momentum tensor and number current in terms of $T$, $\mu$ and $U^\mu$. As $T_{\mathrm{eq}}^{\mu\nu}$ is symmetric in its indices, it can be decomposed in terms of $U^\mu$ and $g^{\mu\nu}$ whereas $N_{\mathrm{eq}}^\mu$ can be expressed only in terms of $U^\mu$ as[142,158]

$$T_{\mathrm{eq}}^{\mu\nu} = c_1 U^\mu U^\nu + c_2 g^{\mu\nu}, \quad N_{\mathrm{eq}}^\mu = c_3 U^\mu. \quad (8)$$

The coefficients $c_1, c_2, c_3$ can be uniquely determined by considering the expression of the energy–momentum tensor in the local rest frame (LRF) of the fluid element where $U^\mu = (1, 0, 0, 0)$. In the LRF, there is no flow of energy and the momentum flux is isotropic. Moreover, in this frame, there is no particle three current. Therefore, LRF characterizes a static equilibrium. It can be easily observed that, in LRF, the energy–momentum tensor, and number four-current take the following forms[142,158]:

$$T_{\mathrm{LRF}}^{\mu\nu} = \begin{pmatrix} \mathcal{E} & 0 & 0 & 0 \\ 0 & \mathcal{P} & 0 & 0 \\ 0 & 0 & \mathcal{P} & 0 \\ 0 & 0 & 0 & \mathcal{P} \end{pmatrix}, \quad N_{\mathrm{LRF}}^\mu = \begin{pmatrix} \mathcal{N} \\ 0 \\ 0 \\ 0 \end{pmatrix}, \quad (9)$$

respectively, where $\mathcal{E}$, $\mathcal{P}$ and $\mathcal{N}$ are the energy density, pressure and number density, respectively. Equations (8) and (9) allow us to identify the coefficients $c_1, c_2$ and $c_3$ as

$$c_1 = \mathcal{E} + \mathcal{P}, \quad c_2 = -\mathcal{P}, \quad c_3 = \mathcal{N}. \quad (10)$$

Therefore, the conserved currents of the perfect fluid (8) can be written as

$$T_{\mathrm{eq}}^{\mu\nu} = (\mathcal{E} + \mathcal{P}) U^\mu U^\nu - \mathcal{P} g^{\mu\nu}, \quad N_{\mathrm{eq}}^\mu = \mathcal{N} U^\mu. \quad (11)$$

Using Eqs. (7) and (11), one can obtain the energy density, pressure and number density in terms of the equilibrium distribution function as

$$\mathcal{E} = \int \mathrm{dP}(U \cdot p)^2 f_{\mathrm{eq}}(x, p), \tag{12}$$

$$\mathcal{P} = \frac{1}{3} \int \mathrm{dP}((U \cdot p)^2 - m^2) f_{\mathrm{eq}}(x, p), \tag{13}$$

$$\mathcal{N} = \int \mathrm{dP}(U \cdot p) f_{\mathrm{eq}}(x, p). \tag{14}$$

Note that $\mathcal{E}$, $\mathcal{P}$ and $\mathcal{N}$ are local functions of $T$, $\mu$ and $U^\mu$. The space–time evolution of these quantities is governed by the conservation laws

$$\partial_\mu T_{\mathrm{eq}}^{\mu\nu} = 0, \quad \partial_\mu N_{\mathrm{eq}}^\mu = 0. \tag{15}$$

It is a common practice to split the conservation equation of the energy–momentum tensor into two parts, one along the direction of $U^\mu$ and another orthogonal to $U^\mu$. Therefore, taking the projection of conservation law of energy–momentum tensor parallel and orthogonal to the fluid four-velocity together with the conservation of particle four-current, we find the equations of motion of the perfect-fluid hydrodynamics

$$U_\nu \partial_\mu T_{\mathrm{eq}}^{\mu\nu} \equiv U^\mu \partial_\mu \mathcal{E} + (\mathcal{E} + \mathcal{P}) \partial_\mu U^\mu = 0, \tag{16}$$

$$\Delta_\nu^\alpha \partial_\mu T_{\mathrm{eq}}^{\mu\nu} \equiv (\mathcal{E} + \mathcal{P}) U^\mu \partial_\mu U^\alpha - \Delta^{\alpha\beta} \partial_\beta \mathcal{P} = 0, \tag{17}$$

$$\partial_\mu N_{\mathrm{eq}}^\mu \equiv U^\mu \partial_\mu \mathcal{N} + \mathcal{N} \partial_\mu U^\mu = 0, \tag{18}$$

where $\Delta^{\mu\nu} = g^{\mu\nu} - (U^\mu U^\nu)/U \cdot U$ is the spatial projection operator orthogonal to $U$. It is important to emphasize that the conservation of the energy–momentum tensor and number four-current provide five evolution equations for six unknowns $\mathcal{E}$, $\mathcal{P}$, $\mathcal{N}$ and three independent components of $U^\mu$ (note normalization constraint of $U$). Therefore, to close the system of equations we need an EoS relating different state variables, i.e. $\mathcal{P} = \mathcal{P}(\mathcal{E}, \mathcal{N})$. Once the EoS is properly defined, ideal hydrodynamic equations can be used to obtain the space–time evolution of energy density, pressure, number density and fluid four-velocity.

## 1.4. *Theoretical efforts to explain spin polarization*

Spin polarization is one of the most interesting properties of the QGP measured recently, allowing to probe its quantum properties directly. With the arrival of spin polarization data, discussed in Subsec. 1.2, it became necessary to develop theoretical models for the qualitative and quantitative interpretation of the data. This in turn gives a hope for a better understanding of the vorticity as well as spin dynamics in the QGP. This paper focuses on this particular problem. The first formal understanding of the spin polarization phenomena came from the "spin-thermal models". These models assume that a large orbital angular momentum created in the noncentral collisions induces vorticity in the QGP, which, if the spin is thermalized together with other degrees of freedom, gives rise to the spin polarization of the particles due to the so-called *spin–vorticity coupling*, with the vorticity quantified by the so-called *thermal vorticity*, $\varpi_{\mu\nu} = -(1/2) \times (\partial_\mu \beta_\nu - \partial_\nu \beta_\mu)$[92] where $\beta_\mu = U_\mu/T$.

Based on these assumptions, several hydrodynamic and transport models have been used to study the global polarization and the azimuthal angle dependence of the longitudinal spin polarization in heavy-ion collisions.[92,127,159–171] These models are successful in describing the center-of-mass energy dependence of the global polarization measurements, see Fig. 4. Within a hydrodynamic framework, addressing the spin–vorticity coupling is straightforward, as gradients of the flow can be obtained naturally. On the other hand, transport models which describe microscopic particle dynamics can also be used to obtain flow gradients by considering suitable coarse-graining procedures. The "spin-thermal" models, which explain the global polarization data within the systematic uncertainty range, fail to explain the azimuthal angle dependence of longitudinal polarization shown in Fig. 6. In fact, theoretical model prediction is opposite (having opposite sign) to the data, see Refs. 123, 126–128 and 172. This disagreement may indicate that the assumptions of LTE for spin degrees of freedom may not be fully satisfied under those extreme conditions or spin polarization may not be solely determined by the thermal vorticity in equilibrium. A possible explanation of this puzzle was proposed recently by including the *thermal shear* component, $\xi_{\mu\nu} = (1/2)(\partial_\mu \beta_\nu + \partial_\nu \beta_\mu)$, along with the thermal vorticity component,[173–178] however, enforcing strong assumptions such as neglecting temperature gradients at the freeze-out which may only be true for the high-energy collisions or correcting the masses of the particles of interest. The lack of theoretical understanding of the longitudinal spin polarization motivates us to consider new hydrodynamic and kinetic theory approaches where spin polarization is an independent dynamical variable, not necessarily respecting the spin–vorticity coupling. The development of such approaches is currently under intense investigation, see Subsec. 1.6.

In this paper, we will discuss one such approach where the spin degrees of freedom are incorporated into standard relativistic perfect-fluid hydrodynamics treating spin as a dynamical quantity like other standard hydrodynamic variables such as temperature and baryon chemical potential and use this formalism to study various physical systems.

## 1.5. *Pseudogauge transformation*

Before diving into the details of hydrodynamics with spin, we would first like to convey that treating spin as a dynamical quantity requires extra care, meaning that, in addition to the conservation of energy–momentum tensor, we also need to include the evolution of the spin through the conservation of total angular momentum.

More specifically, from the knowledge of QFT,[179,180] it is known that, for a system with spin, total angular momentum tensor ($\hat{J}^{\lambda,\mu\nu}$) consists of both the orbital ($\hat{L}^{\lambda,\mu\nu}$) and spin ($\hat{S}^{\lambda,\mu\nu}$) contributions, namely

$$\hat{J}^{\lambda,\mu\nu} = \hat{L}^{\lambda,\mu\nu} + \hat{S}^{\lambda,\mu\nu} = x^\mu \hat{T}^{\lambda\nu} - x^\nu \hat{T}^{\lambda\mu} + \hat{S}^{\lambda,\mu\nu}, \tag{19}$$

which, with the help of total angular momentum conservation[156,180]

$$\partial_\lambda \hat{J}^{\lambda,\mu\nu} = \partial_\lambda \hat{L}^{\lambda,\mu\nu} + \partial_\lambda \hat{S}^{\lambda,\mu\nu} = \hat{T}^{\mu\nu} - \hat{T}^{\nu\mu} + \partial_\lambda \hat{S}^{\lambda,\mu\nu} = 0, \tag{20}$$

leads to

$$\partial_\lambda \hat{S}^{\lambda,\mu\nu} = \hat{T}^{\nu\mu} - \hat{T}^{\mu\nu}. \tag{21}$$

From the above relation, one may conclude that, although the total angular momentum is conserved, neither the spin tensor nor the orbital angular momentum tensor is separately conserved. Physically, it means that orbital angular momentum can be transformed to spin angular momentum and vice versa (spin–orbit coupling). However, the forms of the energy–momentum tensor $T^{\mu\nu}$ and spin tensor $S^{\lambda\mu\nu}$ in Eq. (21) are not uniquely defined by Noether's theorem. One can, in principle, construct various forms of the energy–momentum tensor and the spin tensor through the so-called *pseudogauge transformation*[95,181–183] such that the total energy, momentum and angular momentum are preserved.

In particular, Eq. (21) holds for the system of massive spin-half fermions where we use the Dirac Lagrangian for the free fields[m]

$$\mathcal{L}_D(x) = \frac{i}{2}\bar{\psi}(x)\overleftrightarrow{\partial\!\!\!/}\psi(x) - m\bar{\psi}(x)\psi(x), \tag{22}$$

with $\psi$ and $\bar{\psi} \equiv \psi^\dagger \gamma^0$ being the Dirac field operator and its adjoint, respectively, and Noether theorem to arrive at[179]

$$\partial_\mu \hat{T}_{\text{Can}}^{\mu\nu} = 0, \quad \partial_\lambda \hat{J}_{\text{Can}}^{\lambda,\mu\nu} = 0, \tag{23}$$

$$\partial_\lambda \hat{S}_{\text{Can}}^{\lambda,\mu\nu} = \hat{T}_{\text{Can}}^{\nu\mu} - \hat{T}_{\text{Can}}^{\mu\nu}. \tag{24}$$

Here, by the label "Can" we denote canonical procedure of obtaining the currents. One can check that the resulting energy–momentum tensor

$$\hat{T}_{\text{Can}}^{\mu\nu} = \frac{i}{2}\bar{\psi}\gamma^\mu \overleftrightarrow{\partial}^\nu \psi - g^{\mu\nu}\mathcal{L}_D, \tag{25}$$

is asymmetric, whereas the spin tensor $\hat{S}_{\text{Can}}^{\lambda,\mu\nu}$ is defined as[n]

$$\hat{S}_{\text{Can}}^{\lambda,\mu\nu} = \frac{i}{8}\bar{\psi}\{\gamma^\lambda, [\gamma^\mu, \gamma^\nu]\}\psi = -\frac{1}{2}\epsilon^{\lambda\mu\nu\alpha}\bar{\psi}\gamma_\alpha\gamma_5\psi, \tag{26}$$

where we adopt the convention $\epsilon^{0123} = 1$ and $\gamma_5$ matrix is defined in terms of the other Dirac gamma matrices, $\gamma_5 \equiv \gamma^5 = i\gamma^0\gamma^1\gamma^2\gamma^3$.

---

[m] $\partial\!\!\!/$ represents the Feynman not notation with $\partial\!\!\!/ = \gamma^\mu\partial_\mu$ and $\overleftrightarrow{\partial\!\!\!/} \equiv \overrightarrow{\partial\!\!\!/} - \overleftarrow{\partial\!\!\!/}$.
[n] Here, $\{A, B\} = AB + BA$ and $[A, B] = AB - BA$.

One may construct a new pair of "improved" energy–momentum and spin tensors in such a way that the improved energy–momentum tensor is symmetric, and, as a result, the improved spin tensor is separately conserved. The new pair of tensors may be obtained from the canonical one by means of the pseudogauge transformation[86,95,156,182]

$$\hat{T}^{\mu\nu} = \hat{T}^{\mu\nu}_{\text{Can}} + \frac{1}{2}\partial_\lambda(\hat{\Pi}^{\lambda,\mu\nu} + \hat{\Pi}^{\nu,\mu\lambda} + \hat{\Pi}^{\mu,\nu\lambda}), \quad \hat{S}^{\lambda,\mu\nu}$$

$$= \hat{S}^{\lambda,\mu\nu}_{\text{Can}} - \hat{\Pi}^{\lambda,\mu\nu} + \partial_\rho \hat{\Upsilon}^{\mu\nu,\lambda\rho}, \tag{27}$$

where the superpotentials $\hat{\Pi}^{\lambda,\mu\nu}$ and $\hat{\Upsilon}^{\mu\nu,\lambda\rho}$ satisfy

$$\hat{\Pi}^{\lambda,\mu\nu} = -\hat{\Pi}^{\lambda,\nu\mu}, \quad \hat{\Upsilon}^{\mu\nu,\lambda\rho} = -\hat{\Upsilon}^{\nu\mu,\lambda\rho} = -\hat{\Upsilon}^{\mu\nu,\rho\lambda}. \tag{28}$$

In principle, one may have several choices of $\hat{\Pi}$ and $\hat{\Upsilon}$, nevertheless, the improved tensors do not change the total four-momentum and total angular momentum and Eqs. (23) are still valid with these improved tensors.

We mention some well-explored pseudogauge choices relevant for theoretical modeling of the medium produced in heavy-ion collisions:

- The first trivial choice is $\hat{\Pi}^{\lambda,\mu\nu} = \hat{\Upsilon}^{\mu\nu,\lambda\rho} = 0$. This gives back the canonical currents, Eqs. (25) and (26).
- Another choice is the well-known Belinfante–Rosenfeld (BR) pseudogauge[181,184,185] where $\hat{\Pi}^{\lambda,\mu\nu} = \hat{S}^{\lambda,\mu\nu}_{\text{Can}}$ and $\hat{\Upsilon}^{\mu\nu,\lambda\rho} = 0$. In this case, energy–momentum tensor and spin tensor become[183]

$$\hat{T}^{\mu\nu}_{\text{BR}} = \frac{i}{2}\bar{\psi}\gamma^\mu \overleftrightarrow{\partial^\nu}\psi - \frac{i}{16}\partial_\lambda(\bar{\psi}\{\gamma^\lambda,[\gamma^\mu,\gamma^\nu]\}\psi), \quad \hat{S}^{\lambda,\mu\nu}_{\text{BR}} = 0, \tag{29}$$

respectively. The above tensors are particularly important from the point of view of General Relativity (GR). Since in GR, the symmetric energy–momentum tensor is considered to be the source of the gravitational field, its canonical form cannot be used.[182] Symmetric energy–momentum tensor is also natural in GR as it can be obtained using the variation of the Einstein-Hilbert action with respect to the space–time metric. However, the formalism of GR based on the Einstein–Hilbert action is not suitable to capture the spin dynamics.

- The third choice is Hilgevoord–Wouthuysen (HW) form of the pseudogauge[95,186,187] where $\hat{\Pi}^{\lambda,\mu\nu} = M^{[\mu\nu]\lambda} - g^{\lambda[\mu}M^{\nu]\rho}_\rho$ with $M^{\lambda\mu\nu} \equiv \frac{i}{4m}\bar{\psi}\sigma^{\mu\nu}\overleftrightarrow{\partial^\lambda}\psi$ and $\hat{\Upsilon}^{\mu\nu,\lambda\rho} = -\frac{1}{8m}\bar{\psi}\times(\sigma^{\mu\nu}\sigma^{\lambda\rho} + \sigma^{\lambda\rho}\sigma^{\mu\nu})\psi$, which yields

$$\hat{T}^{\mu\nu}_{\text{HW}} = \hat{T}^{\mu\nu}_{\text{Can}} + \frac{i}{2m}(\partial^\nu\bar{\psi}\sigma^{\mu\beta}\partial_\beta\psi + \partial_\alpha\bar{\psi}\sigma^{\alpha\mu}\partial^\nu\psi) - \frac{i}{4m}g^{\mu\nu}\partial_\lambda(\bar{\psi}\sigma^{\lambda\alpha}\overleftrightarrow{\partial_\alpha}\psi),$$

$$\hat{S}^{\lambda,\mu\nu}_{\text{HW}} = \hat{S}^{\lambda,\mu\nu}_{\text{Can}} - \frac{1}{4m}(\bar{\psi}\sigma^{\mu\nu}\sigma^{\lambda\rho}\partial_\rho\psi + \partial_\rho\bar{\psi}\sigma^{\lambda\rho}\sigma^{\mu\nu}\psi). \tag{30}$$

- Finally, we come to de Groot–van Leeuwen–van Weert (GLW) pseudogauge[86,156] where $\hat{\Pi}^{\lambda,\mu\nu} = \frac{i}{4m}\,\bar{\psi}(\sigma^{\lambda\mu}\overleftrightarrow{\partial}^\nu - \sigma^{\lambda\nu}\overleftrightarrow{\partial}^\mu)\psi$ and $\hat{\Upsilon}^{\mu\nu,\lambda\rho} = 0$, which give

$$\hat{T}^{\mu\nu}_{\mathrm{GLW}} = -\frac{1}{4m}\,\bar{\psi}\,\overleftrightarrow{\partial}^\mu\,\overleftrightarrow{\partial}^\nu\,\psi,$$

$$\hat{S}^{\lambda,\mu\nu}_{\mathrm{GLW}} = \bar{\psi}\left[\frac{\sigma^{\mu\nu}}{4} - \frac{1}{8m}(\gamma^\mu\overleftrightarrow{\partial}^\nu - \gamma^\nu\overleftrightarrow{\partial}^\mu)\right]\gamma^\lambda\psi + \mathrm{h.c.},$$

(31)

where $\sigma^{\mu\nu} = (i/2)[\gamma^\mu, \gamma^\nu]$ is the Dirac spin operator. HW and GLW pseudogauges provide symmetric energy–momentum tensors, and therefore, one may obtain conserved spin tensors.[o]

## 1.6. *Objective of the paper*

Different pseudogauge choices can give rise to different hydrodynamic frameworks for spin-polarized media. Although these frameworks can be formulated from equivalent (through pseudogauge) sets of the energy–momentum and the spin tensors, the resulting dynamical evolution of spin may not be equivalent, particularly in the local thermal equilibrium, as, in particular, densities of the hydrodynamic quantities change. This can be observed in the following manner. In the case of BR pseudogauge, the energy–momentum tensor is symmetric and there is no explicit presence of spin, hence, there is no independent evolution of the spin degree of freedom. Nevertheless, in this case, one can still completely determine the polarization as the latter is bound to the evolution of velocity gradients. This reasoning has been used in "spin-thermal" framework of Refs. 92 and 174, where spin polarization arises through spin–vorticity coupling.

On the other hand, the canonical energy–momentum tensor is asymmetric, hence, the spin tensor is not separately conserved. Finally, in the case of GLW and HW, the energy–momentum tensor is symmetric by construction, making the spin tensor conserved independently. These pseudogauges can be used to formulate hydrodynamics with spin. The constitutive relations and the hydrodynamic equations can be different in these frameworks, therefore, *a priori*, it is not evident whether these frameworks can be considered equivalent and well defined as a boundary value problem, particularly as a set of partial differential equations with some initial conditions. Note that, in the context of hyperon polarization observed in the heavy-ion collision experiments, physical implications of the pseudogauge transformation of the energy–momentum tensor and the spin tensor have been extensively discussed.[86,95,182,188,190,191]

As relativistic hydrodynamics is a classical theory, the natural starting point for a proper formulation of hydrodynamics with spin is to define the energy–momentum and the spin currents as ensemble averages of their respective normal-ordered QFT operators, namely

$$T^{\mu\nu} = \langle : \hat{T}^{\mu\nu} : \rangle, \quad S^{\lambda,\mu\nu} = \langle : \hat{S}^{\lambda,\mu\nu} : \rangle.$$

(32)

---

[o]Some more forms of pseudogauge and its details can be found in Refs. 95, 182, 188 and 189.

In the framework of hydrodynamics with spin, in addition to the conservation of energy–momentum, we also need to incorporate conservation of total angular momentum

$$\partial_\mu T^{\mu\nu} = 0, \quad \partial_\lambda S^{\lambda,\mu\nu} = T^{\nu\mu} - T^{\mu\nu}. \tag{33}$$

The evolution of the spin tensor will introduce six additional dynamical equations in addition to the standard hydrodynamic ones.

Having stated different intricacies and conceptual difficulties arising when developing hydrodynamics with spin, in the following we consider the GLW definition of the energy–momentum tensor and the spin tensor. One of the important features of this pseudogauge, as mentioned earlier, is that the spin tensor is separately conserved because the energy–momentum tensor is symmetric, see Eq. (33). Therefore, various components of the energy–momentum tensor and the spin tensor evolve separately, making this framework much simpler to handle mathematically as well as numerically.[p]

The first proposition of *relativistic hydrodynamics with spin* as a dynamical quantity[q] was made in 2017 by Florkowski *et al.* in Ref. 225 using the kinetic theory approach. This formalism[225–228] was further extended to derive the constitutive relations for the net baryon density, the energy–momentum tensor and the spin tensor[229,230] using the Wigner function formalism[156,192,231,232] and semi-classical expansion method[233–237] based on the GLW pseudogauge. Numerical modeling of the spin polarization was investigated in Refs. 238–243 using different hydrodynamic backgrounds. Further extensions to include dissipative effects using relaxation time approximation were performed in Refs. 244–247, for a recent review, see Ref. 86. Keeping the experimental results[85,91] in mind, the spin polarization effects in Refs. 225, 226 and 229 were considered small, leading to decoupling of background dynamics (energy, momentum and baryon number) from the spin dynamics. Moreover, in this framework, the spin polarization is assumed to appear in zeroth order in the semi-classical expansion of the Wigner function satisfying quantum kinetic equations. The crucial aspect of such a kinetic theory is the presence of a second-rank antisymmetric spin chemical potential in the equilibrium distribution function which plays a role of an extra Lagrange multiplier (more precisely six Lagrange multipliers) responsible for angular momentum conservation in the relativistic regime. This is the striking difference between this framework and the "spin-thermal models", as generically, the spin chemical potential is not necessarily related to the thermal vorticity appearing in the spin-thermal models. Recently, propagation properties of the components of spin polarization tensor have also been investigated[248] through the method of linear mode analysis.[218,249–253]

---

[p] We note that use of HW pseudogauge is also possible, and it leads to equivalent hydrodynamic framework as long as we consider only local collisions between the particles. Otherwise, these two frameworks are different, see Refs. 192 and 193 for the details of the framework based on HW pseudogauge considering nonlocal collisions.

[q] Many alternative approaches incorporating spin into hydrodynamics were also developed. They used effective theory,[191,194–197] entropy-current analysis,[198–203] statistical operator method,[204] nonlocal collisions,[192,193,205–210] chiral kinetic theory,[205,211–214] methods of holography,[191,215–219] anomalous (with triangle anomaly) hydrodynamics,[103,220] the Lagrangian method[221,222] and perturbative scattering techniques.[223,224]

This paper focuses on the developments of relativistic perfect-fluid hydrodynamics with spin based on the GLW pseudogauge and modeling of the dynamics of spin-polarized matter using this framework.[238–243,248,254–265]

## 1.7.  *Overview of the paper*

- The quantum kinetic theory formulation of relativistic perfect-fluid hydrodynamics with spin from the collisionless transport equation for the Wigner function for spin-half fermions using GLW pseudogauge (31) and assuming a small spin polarization limit was developed in Ref. 229. As this pseudogauge makes the energy–momentum tensor symmetric, the spin tensor is conserved independently, leading to the decoupling of the orbital angular momentum from spin angular momentum.

  However, as shown in Refs. 192 and 193, the nonlocal collisions at the microscopic scale, which were neglected in Ref. 229, may give rise to coupling between the spin and orbital parts of the total angular momentum through antisymmetric parts of the energy–momentum tensor, see Eq. (21). Hence, to address vorticity sourcing of polarization, it is necessary to include such collisional effects in the formalism. This extension of the framework of Ref. 229 is presented in Sec. 2. It provides a pedagogical introduction to the Wigner function formalism for spin-half massive particles and the semi-classical expansion method, followed by the derivation of the Boltzmann-like spin-kinetic equations using the equations of motion for the components of the Wigner function. For this derivation, we do not assume any specific constraints on the collision terms and consider that the spin polarization effects can appear at both the zeroth ($\hbar^0$) and first ($\hbar^1$) order in $\hbar$. The physical meaning of this assumption is that the spin polarization effects can have both classical and quantum origins, respectively.

- Section 3 briefly reviews the formalism of perfect-fluid relativistic hydrodynamics with spin, presented in Ref. 229. The assumption that the system is in the local equilibrium state, makes the collision terms vanish, which corresponds to the zeroth order of the semi-classical expansion. Such a system can be described by the equilibrium Wigner function, expressed in terms of the Dirac spinors. Due to the spinor representation of the Wigner function, using the Clifford algebra, the Wigner function can be decomposed into irreducible components which transform in specific ways under the Lorentz transformation. Interestingly, these irreducible components can be used to derive the relations constituting the perfect-fluid hydrodynamics with spin.

  In this section, we explicitly calculate these components, using an ansatz for the spin-dependent phase-space distribution function, which are subsequently used to derive the conservation laws for the net baryon current, energy–momentum tensor and spin tensor based on the GLW pseudogauge. These conservation laws are used in the following sections to study the dynamics of spin polarization of $\Lambda(\bar{\Lambda})$ hyperons.

An alternative approach to formulating our framework is based on the classical description of spin.[86] It is compelling to observe that the constitutive relations for the net baryon current, energy–momentum tensor and spin tensor obtained this way in the small spin polarization limit match the relations derived using the above-mentioned Wigner function approach. The results mentioned in this section form the basis for the later parts of the paper.

- In Sec. 3, the spin tensor was derived, for simplicity, using classical Boltzmann statistics. However, to study the polarization of particles obeying quantum statistics, the spin tensor needs to be generalized. This is done in Sec. 4. Using wave propagation analysis, this extended spin tensor is used to study the behavior of the spin polarization components after they are perturbed in the longitudinal direction and to derive the dispersion relation of spin-wave velocity. Then, we obtain spin-wave velocity in two special cases of distribution: Maxwell–Jüttner (MJ) and Fermi–Dirac (FD). We observe that, for the FD gas, the spin-wave velocity is related to sound velocity in the degenerate limit.

- Using the spin hydrodynamic formalism developed in Sec. 3, one may determine the space–time evolution of the thermodynamic and hydrodynamic parameters. However, these are not suitable to be compared with the experimental spin polarization measurements. To make such a comparison feasible, we have to calculate the spin polarization of particles at the freeze-out surface. For this purpose, in Sec. 5, we provide the details required to compute the momentum-dependent and momentum-averaged mean spin polarization per particle using Pauli–Lubański (PL) four-vector. These expressions are then used in Sec. 6 to numerically model the spin polarization of $\Lambda(\bar{\Lambda})$ hyperons.

- Section 6 starts with the derivation of the equations of motion for the conservation laws of net baryon current, energy–momentum tensor and spin tensor presented in Sec. 3 which are then used to study the spin polarization dynamics of the systems respecting certain space–time symmetries.

  We begin our study with the Bjorken model, which is commonly employed in the phenomenology of heavy-ion collisions. We determine the time evolution of temperature, baryon chemical potential and spin components and calculate mean spin polarization at the freeze-out. In the following sections, we relax the symmetry of boost-invariance, keeping homogeneity in the transverse plane in order to capture some nontrivial dynamics of realistic systems produced in the low-and mid-energy heavy-ion collisions. Our spin polarization results for the nonboost-invariant case qualitatively agree with other model calculations and experimental data.

- As EM fields may be present in the early stages of relativistic collisions of heavy ions, they may have some effects on the physical observables. In particular, it is suggested that the splitting of $\Lambda$ and $\bar{\Lambda}$ spin polarization seen in experiments may have a source in the interaction of the produced polarized matter with EM fields.

To find how these fields affect the evolution of spin polarization within our framework, we incorporate them into our formalism. Section 7 studies the background and spin dynamics in the presence of an external electric field for Bjorken-expanding matter and finds that its presence may play a significant role in spin polarization evolution.

- Section 8 discusses the dynamics of a longitudinally boost-invariant system with inhomogeneous expansion in the transverse direction which respects the so-called Gubser conformal symmetry, assuming that the system is cylindrically symmetric with respect to the beam direction. Using Gubser's prescription, we find transformation rules that the conservation laws, derived in Sec. 3, must respect to be conformally invariant. We find that, while the energy–momentum tensor and net baryon current preserve conformal symmetry, the spin tensor breaks it explicitly. However, this breaking does not spoil the conformal invariance of the background as the background and spin dynamics are decoupled due to the assumption of small spin polarization. This allows us to find approximate novel solutions for the spin.

- Section 9 closes the paper with a brief summary.

## 1.8. *Notations*

In this work, the following notations/conventions have been adopted:

- "Mostly minus" signature for the space–time metric, unless specified otherwise

$$
g_{\alpha\beta} = \begin{pmatrix} 1 & 0 & 0 & 0 \\ 0 & -1 & 0 & 0 \\ 0 & 0 & -1 & 0 \\ 0 & 0 & 0 & -1 \end{pmatrix},
\tag{34}
$$

where $g^{\mu\nu}g_{\mu\nu} = 4$ and $x^\mu = (t, x, y, z)$ in the Cartesian coordinate.

- For the scalar (dot) product of four-vectors: $a \cdot b = a^\alpha b_\alpha = g_{\alpha\beta} a^\alpha b^\beta = a^0 b^0 - \boldsymbol{a} \cdot \boldsymbol{b}$, where bold font denotes three-vectors.

- For Levi-Civita symbol $\epsilon^{\alpha\beta\gamma\delta}$: $\epsilon^{0123} = 1 = -\epsilon_{0123}$.

- We denote symmetrization by $A_{(\mu\nu)} = \frac{1}{2}(A_{\mu\nu} + A_{\nu\mu})$ and antisymmetrization by $A_{[\mu\nu]} = \frac{1}{2}(A_{\mu\nu} - A_{\nu\mu})$.

- We represent commutator and anticommutator between $A$ and $B$ as $[A, B] = AB - BA$ and $\{A, B\} = AB + BA$, respectively.

- The dual of any tensor is denoted by a star and obtained by contracting it with the Levi-Civita symbol as

$$
A^\star{}_{\alpha\beta} = \frac{1}{2} \epsilon^{\alpha\beta\gamma\delta} A_{\gamma\delta}.
\tag{35}
$$

- $\mathrm{A} = \gamma^\mu A_\mu$ represents Feynman not notation.

- $\langle : A : \rangle$ denotes ensemble (statistical) average of the normal-ordered quantity $A$.

- Directional derivatives along basis vectors are denoted as $U^\alpha \partial_\alpha \equiv (\bullet)$, $X^\alpha \partial_\alpha \equiv (\blacksquare)$, $Y^\alpha \partial_\alpha \equiv (\square)$, $Z^\alpha \partial_\alpha \equiv (\circ)$, and divergence of a four-vector $A$ is written as $\partial_\alpha A^\alpha \equiv \theta_A$.
- Einstein summation convention is assumed unless mentioned otherwise.

Throughout the work, we assume natural units, i.e. $c = \hbar = k_B = 1$, unless mentioned otherwise. Here, $c$, $\hbar$ and $k_B$ is speed of light, reduced Planck constant and Boltzmann constant, respectively.

## 2. Kinetic Theory for Dirac Fermions

Hydrodynamics, being a macroscopic, long-wavelength limit of the theory describing systems near equilibrium, follows directly from the continuity equations of conserved quantities, such as energy, momentum and charge, supplemented by the second law of thermodynamics and the EoS. However, the space–time evolution of the dissipative quantities rests upon the specifics of how the system approaches thermal distribution and requires a correct and explicit description of the processes within the underlying microscopic theory, such as kinetic theory.[156,266,267]

Properties of a many-body system depend, in particular, on the details of particle interactions (collisions) and external forces (constraints). To describe the system dynamics near equilibrium, we need to express these quantities in the language of macroscopic state variables, such as temperature, charge density and fluid velocity. Within the classical relativistic kinetic theory, this description can be formulated using the single-particle distribution function $f(x, k)$ that describes the mean number of particles with four-momentum $k$ at space–time position $x$. The specific form of $f(x, k)$ follows from the kinetic (transport) equation, which describes its phase-space evolution. Conserved quantities can then be evaluated through the moments of $f(x, k)$ in momentum space, see Subsec. 1.3 for more details on the introductory level.

However, for the description of the quantum mechanical systems the classical distribution function, due to Heisenberg's uncertainty principle, is not properly defined.[268] Instead, we use the Wigner function, which represents the quantum analog of the classical distribution function. Wigner function is a quasi-probability distribution function that, unlike the classical distribution function, may also give rise to negative probabilities that disappear in the classical limit. However, at the leading order of spatial gradients, the Wigner function can be related to the classical distribution.[269] Densities of macroscopic quantities can be then obtained from the Wigner function after integrating over the momentum variable $k$.[234]

With the motivation of explaining spin polarization of $\Lambda(\bar\Lambda)$ hyperons in mind,[85,91] the authors of Ref. 229 formulated a framework of relativistic hydrodynamics with spin using the collisionless transport equation for the Wigner function with the assumption of small spin polarization. This assumption derives its motivation from the actual magnitude of the hyperon spin polarization observed in the experiments.[85,91] The developed formalism is based on the specific form of the energy–momentum tensor and the spin tensor resulting from GLW pseudogauge,[156] see Subsec. 1.5. The spin tensor in this framework is conserved independently due to the symmetric form of the energy–momentum tensor. Thus, the evolution of spin polarization is governed only by the conservation of spin tensor.

However, in general, antisymmetric parts of the energy–momentum tensor, due to nonlocal collisions of particles, may not vanish which allows the coupling between the spin angular momentum and orbital angular momentum (even though energy–momentum tensor and total angular momentum are conserved separately). A general framework of relativistic hydrodynamics with spin starting from the microscopic quantum kinetic theory and considering both local and nonlocal collisions may allow having such an interaction.[192,205–208] This spin–orbit interaction can give rise to the dissipative phenomena[198,r] which were not taken into account in Ref. 229.

In this section, we study such a case by extending the analysis done in Ref. 229 to include local and nonlocal collisions between particles in a way proposed in Refs. 192 and 193. After discussing the details of the Wigner function approach and semiclassical expansion, we derive equations of motion for the components of the Wigner function. Our work generalizes the results of Refs. 192 and 193[s] and considers that spin effects may have their origin at both the classical and quantum levels (zeroth and first order in $\hbar$). This means that both the zeroth-and the first-order axial-vector components of the Wigner function, in our approach, are nonvanishing. We then derive the general form of a Boltzmann-like spin-kinetic equation that may serve as a starting point in formulating a general formalism of relativistic hydrodynamics with spin. Details of this section are based on Ref. 265.

## 2.1. *Covariant Wigner function and its transport equation*

Let us start by introducing the definition of the Wigner function[t] for spin-half particles having mass $m$ (Dirac fermions) as[156,231,234,u]

$$W_{\alpha\beta}(x,k) = \int \frac{d^4 y}{(2\pi\hbar)^4} e^{-\frac{i}{\hbar}k \cdot y} \langle : \bar{\psi}_\beta(x_+)\psi_\alpha(x_-) : \rangle. \tag{36}$$

Here, $\psi$ and $\bar{\psi} \equiv \psi^\dagger \gamma^0$ are the Dirac field operator and its adjoint, respectively, whereas $x_+ = x + y/2$ and $x_- = x - y/2$ denote two space–time points with $x$ as the center position and $y$ as the relative position. In Eq. (36), $\langle : Q_W : \rangle$ means ensemble (statistical) average of the normal-ordered quantity $Q_W$.

---

[r]In the context of nonrelativistic spin hydrodynamic formalism, see Refs. 130, 270 and 271.

[s]The authors of Refs. 192 and 193 assumed that spin effects arise at the level of $\hbar$, hence, they only consider the first-order axial-vector component nonvanishing making spin a dissipative effect.

[t]This section deals with the semi-classical expansion of the Wigner function that can also be interpreted as an expansion in $\hbar$. Thus, we put $\hbar$ explicitly in the definition of the Wigner function. Although, in this work, we use Wigner function formalism for massive spin-half particles, this method can also be used to develop respective kinetic theory for chiral fermions.[106,214,272–275]

[u]Note that $\alpha$ and $\beta$ in Eq. (36) represent the spinor indices.

The Dirac equation for the system of spin-half particles with interactions is defined as[156,v]

$$(i\hbar\slashed{\partial} - m)\psi(x) = \hbar\rho(x) = -\frac{\partial\mathcal{L}_I}{\partial\bar{\psi}},\tag{37}$$

where $\mathcal{L}_I(x)$ is the interaction Lagrangian density.[w] Using the total Lagrangian density

$$\mathcal{L}(x) = \mathcal{L}_D(x) + \mathcal{L}_I(x),$$

with

$$\mathcal{L}_D(x) = \frac{i\hbar}{2}\bar{\psi}(x)\overleftrightarrow{\slashed{\partial}}\psi(x) - m\bar{\psi}(x)\psi(x),\tag{38}$$

being the Lagrangian density for the free field and $\overleftrightarrow{\slashed{\partial}} \equiv \overrightarrow{\slashed{\partial}} - \overleftarrow{\slashed{\partial}}$, the following transport equation can be derived[156,231]

$$\left(i\hbar\frac{\slashed{\partial}}{2} + \slashed{k} - m\right)W(x,k) = \hbar\mathcal{C}[W(x,k)],\tag{39}$$

with

$$\mathcal{C}_{\alpha\beta}[W(x,k)] \equiv \int\frac{d^4y}{(2\pi\hbar)^4}e^{-\frac{i}{\hbar}k\cdot y}\langle : \rho_\alpha(x_-)\bar{\psi}_\beta(x_+) :\rangle,\tag{40}$$

being the collision kernel which vanishes in the global equilibrium.[x]

As the Wigner function is a $4\times4$ complex matrix, it is difficult to provide some physical insights about the dynamics of the Wigner function and its components by working directly with Eq. (36). Instead, it is more convenient to perform the decomposition of Eq. (36) in terms of 16 independent generators of the Clifford algebra as follows[179,y]:

$$W(x,k) = \frac{1}{4}[\mathbf{1}F(x,k) + i\gamma^5 P(x,k) + \gamma^\mu V_\mu(x,k) + \gamma^5\gamma^\mu A_\mu(x,k)$$
$$+ \Sigma^{\mu\nu}S_{\mu\nu}(x,k)],\tag{41}$$

where $\Sigma^{\mu\nu} \equiv (1/2)\sigma^{\mu\nu} \equiv (i/4)[\gamma^\mu, \gamma^\nu]$ is the Dirac spin operator and $F(x,k)$, $P(x,k)$, $V_\mu(x,k)$, $A_\mu(x,k)$ and $S_{\mu\nu}(x,k)$ are the 16 independent components of the Wigner function. To obtain the latter we first need to multiply the Wigner function $W(x,k)$ by the matrices: $\Gamma_X \in \{\mathbf{1}, -i\gamma_5, \gamma^\mu, \gamma^\mu\gamma_5, 2\Sigma^{\mu\nu}\}$ where $X \in \{F, P, V, A, S\}$, respectively,

---

[v] $\slashed{\partial}$ represents the Feynman not notation with $\slashed{\partial} = \gamma^\mu\partial_\mu$. Note that $\hbar$ is always present with the gradient, hence, $\hbar$ expansion is effectively a gradient expansion.

[w] We assume that $\mathcal{L}_I(x)$ does not contain derivatives of the fields.

[x] In this paper, the collision term describes the nonequilibrium system giving rise to the quantum corrections to the zeroth-order Wigner function. These corrections appear beyond zeroth order in $\hbar$ ($\hbar^0$).

[y] The Clifford (geometric) algebra decomposition is a widely used expansion method, for instance, to derive the transport equations for abelian plasmas,[234,236] the QGP,[233,276] chiral models[237] and spin polarization.[86,192,193,229,275,277–281]

and then calculate its trace. Using the decomposition (41) along with the conjugation relation

$$W(x,k) = \gamma_0 W(x,k)^\dagger \gamma_0, \tag{42}$$

(which is also followed by the Clifford algebra generators), it can be observed that the Wigner function components are real.

Under Lorentz transformations the components $F$, $P$, $V_\mu$, $A_\mu$ and $S_{\mu\nu}$ transform as a scalar, pseudo-scalar, vector, axial-vector and an antisymmetric tensor, respectively.[234] Moreover, $F$ and $P$ can be interpreted as the mass and pseudo-scalar condensate, respectively, whereas, fermion number current density and the polarization density can be represented by $V_\mu$ and $A_\mu$, respectively. Finally, we can interpret the six independent components of $S_{\mu\nu}$ as electric ($S_{0i}$) and magnetic ($S_{ij}$) dipole moments.

Using Eq. (41) in Eq. (39) gives a set of coupled equations of motion for the Wigner function coefficients $F$, $P$, $V^\mu$, $A^\mu$ and $S^{\mu\nu}$, where the real parts of these equations are

$$k^\mu V_\mu - mF = \hbar \mathcal{D}_F, \tag{43}$$

$$-\frac{\hbar}{2}\partial^\mu A_\mu - mP = \hbar \mathcal{D}_P, \tag{44}$$

$$k_\mu F - \frac{\hbar}{2}\partial^\nu S_{\nu\mu} - mV_\mu = \hbar \mathcal{D}_{V,\mu}, \tag{45}$$

$$\frac{\hbar}{2}\partial_\mu P - k^\beta S_{\mu\beta}^\star - mA_\mu = \hbar \mathcal{D}_{A,\mu}, \tag{46}$$

$$\hbar\partial_{[\mu}V_{\nu]} - \epsilon_{\mu\nu\alpha\beta}k^\alpha A^\beta - mS_{\mu\nu} = \hbar \mathcal{D}_{S,\mu\nu}, \tag{47}$$

whereas the imaginary parts read[z]

$$\frac{\hbar}{2}\partial^\mu V_\mu = \hbar \mathcal{C}_F, \tag{48}$$

$$k^\mu A_\mu = \hbar \mathcal{C}_P, \tag{49}$$

$$\frac{\hbar}{2}\partial_\mu F + k^\nu S_{\nu\mu} = \hbar \mathcal{C}_{V,\mu}, \tag{50}$$

$$-k_\mu P - \frac{\hbar}{2}\partial^\beta S_{\mu\beta}^\star = \hbar \mathcal{C}_{A,\mu}, \tag{51}$$

$$-2k_{[\mu}V_{\nu]} - \frac{\hbar}{2}\epsilon_{\mu\nu\alpha\beta}\partial^\alpha A^\beta = \hbar \mathcal{C}_{S,\mu\nu}. \tag{52}$$

The quantities $\mathcal{D}_X$ and $\mathcal{C}_X$ denote the collision terms for the real parts and the imaginary parts of the kinetic equations, respectively, that can be evaluated using

$$\mathcal{D}_X = \Re\mathrm{Tr}[\Gamma_X \mathcal{C}[W(x,k)]] \quad \text{and} \quad \mathcal{C}_X = \Im\mathrm{Tr}[\Gamma_X \mathcal{C}[W(x,k)]].$$

---

[z] One should note here that Eq. (48) has $\hbar$ on both sides of the equation. Thus, one may also consider this equation at the zeroth order by removing $\hbar$ from both sides. However, we consider it to be at the first order keeping $\hbar$.[234]

## 2.2. *Semi-classical expansion*

As one can observe, Eqs. (43)–(52) are coupled which makes them rather difficult to interpret physically.

However, this complexity can be decreased after using the method of semi-classical expansion in $\hbar$ where the terms with zeroth power in $\hbar$ match with the classical terms and higher-order terms in $\hbar$ can be interpreted as the quantum corrections.[268] In the $\hbar$ expansion method, Eqs. (43)–(52) give rise to a set of coupled hierarchical equations at different orders of $\hbar$. This can be achieved by expanding components of the Wigner function and the collision terms in powers of $\hbar$ as $X = \sum_n \hbar^n X^{(n)}$, $\mathcal{C}_X = \sum_n \hbar^n \mathcal{C}_X^{(n)}$ and $\mathcal{D}_X = \sum_n \hbar^n \mathcal{D}_X^{(n)}$. In the following, we expand Eqs. (43)–(52) up to second order in $\hbar$.

### 2.2.1. *Zeroth order*

Below are the real parts of the equations of motion (43)–(52) in the zeroth order in $\hbar$[234]

$$k^\mu V_\mu^{(0)} - m F^{(0)} = 0, \tag{53}$$

$$m P^{(0)} = 0, \tag{54}$$

$$k_\mu F^{(0)} - m V_\mu^{(0)} = 0, \tag{55}$$

$$k^\beta S_{\mu\beta}^{\star (0)} + m A_\mu^{(0)} = 0, \tag{56}$$

$$\epsilon_{\mu\nu\alpha\beta} k^\alpha A^{\beta(0)} + m S_{\mu\nu}^{(0)} = 0. \tag{57}$$

Equation (54) tells that zeroth-order pseudo-scalar component $P^{(0)}$ always vanishes for massive spin-half particles.[229,234,282]

The imaginary parts in the zeroth order of $\hbar$ are

$$k^\mu A_\mu^{(0)} = 0, \tag{58}$$

$$k^\nu S_{\nu\mu}^{(0)} = 0, \tag{59}$$

$$k_\mu P^{(0)} = 0, \tag{60}$$

$$k_{[\mu} V_{\nu]}^{(0)} = 0. \tag{61}$$

If one looks closely on Eqs. (53)–(61), it can be observed that all the Wigner function components can be written in terms of $F^{(0)}$ and $A_\mu^{(0)}$. Thus, we can safely assume these components as the basic independent ones,[aa] provided the orthogonality condition (58) of $A_\mu^{(0)}$ is fulfilled.[234]

---

[aa] This assumption is only true for the kinetic theory of massive spin-half particles and is not valid for the case of massless particles.[214,275]

### 2.2.2. *First order*

Reals parts of Eqs. (43)–(52) in the first order in $\hbar$ are

$$k^\mu V_\mu^{(1)} - mF^{(1)} = \mathcal{D}_F^{(0)}, \tag{62}$$

$$-\frac{1}{2}\partial^\mu A_\mu^{(0)} - mP^{(1)} = \mathcal{D}_P^{(0)}, \tag{63}$$

$$k_\mu F^{(1)} - \frac{1}{2}\partial^\nu S_{\nu\mu}^{(0)} - mV_\mu^{(1)} = \mathcal{D}_{V,\mu}^{(0)}, \tag{64}$$

$$\frac{1}{2}\partial_\mu P^{(0)} - k^\beta S_{\mu\beta}^\star (1) - mA_\mu^{(1)} = \mathcal{D}_{A,\mu}^{(0)}, \tag{65}$$

$$\partial_{[\mu}V_{\nu]}^{(0)} - \epsilon_{\mu\nu\alpha\beta}k^\alpha A^{\beta(1)} - mS_{\mu\nu}^{(1)} = \mathcal{D}_{S,\mu\nu}^{(0)}, \tag{66}$$

whereas the imaginary parts give

$$\frac{1}{2}\partial^\mu V_\mu^{(0)} = \mathcal{C}_F^{(0)}, \tag{67}$$

$$k^\mu A_\mu^{(1)} = \mathcal{C}_P^{(0)}, \tag{68}$$

$$\frac{1}{2}\partial_\mu F^{(0)} + k^\nu S_{\nu\mu}^{(1)} = \mathcal{C}_{V,\mu}^{(0)}, \tag{69}$$

$$-k_\mu P^{(1)} - \frac{1}{2}\partial^\beta S_{\mu\beta}^\star (0) = \mathcal{C}_{A,\mu}^{(0)}, \tag{70}$$

$$-2k_{[\mu}V_{\nu]}^{(1)} - \frac{1}{2}\epsilon_{\mu\nu\alpha\beta}\partial^\alpha A^{\beta(0)} = \mathcal{C}_{S,\mu\nu}^{(0)}. \tag{71}$$

While going from the zeroth order to the first order, we observe from Eq. (68) that the presence of collisions prevents the first-order axial-vector coefficient $A^{(1)}$ to be orthogonal to $k$, cf. Eq. (58).

### 2.2.3. *Second order*

It is necessary to go to the second-order equations of motion for the derivation of the kinetic equations for the first-order Wigner function components. In this case, the real parts yield

$$k^\mu V_\mu^{(2)} - mF^{(2)} = \mathcal{D}_F^{(1)}, \tag{72}$$

$$-\frac{1}{2}\partial^\mu A_\mu^{(1)} - mP^{(2)} = \mathcal{D}_P^{(1)}, \tag{73}$$

$$k_\mu F^{(2)} - \frac{1}{2}\partial^\nu S_{\nu\mu}^{(1)} - mV_\mu^{(2)} = \mathcal{D}_{V,\mu}^{(1)}, \tag{74}$$

$$\frac{1}{2}\partial_\mu P^{(1)} - k^\beta S_{\mu\beta}^\star (2) - mA_\mu^{(2)} = \mathcal{D}_{A,\mu}^{(1)}, \tag{75}$$

$$\partial_{[\mu}V_{\nu]}^{(1)} - \epsilon_{\mu\nu\alpha\beta}k^\alpha A^{\beta(2)} - mS_{\mu\nu}^{(2)} = \mathcal{D}_{S,\mu\nu}^{(1)}, \tag{76}$$

while the imaginary parts give

$$\frac{1}{2}\partial^\mu V^{(1)}_\mu = \mathcal{C}^{(1)}_F, \tag{77}$$

$$k^\mu A^{(2)}_\mu = \mathcal{C}^{(1)}_P, \tag{78}$$

$$\frac{1}{2}\partial_\mu F^{(1)} + k^\nu S^{(2)}_{\nu\mu} = \mathcal{C}^{(1)}_{V,\mu}, \tag{79}$$

$$-k_\mu P^{(2)} - \frac{1}{2}\partial^\beta S^\star_{\mu\beta}{}^{(1)} = \mathcal{C}^{(1)}_{A,\mu}, \tag{80}$$

$$-2k_{[\mu}V^{(2)}_{\nu]} - \frac{1}{2}\epsilon_{\mu\nu\alpha\beta}\partial^\alpha A^{\beta(1)} = \mathcal{C}^{(1)}_{S,\mu\nu}. \tag{81}$$

## 2.3. *Mass-shell conditions*

In order to describe the physical system, the components of the Wigner function must satisfy the mass-shell conditions. We shall derive below the conditions to be satisfied by the zeroth-and first-order components of the Wigner function.

### 2.3.1. *Zeroth order*

Equation (54), as mentioned before, depicts that the zeroth-order pseudo-scalar component always satisfies

$$P^{(0)} = 0,$$

whereas Eq. (55) gives an important relation between the zeroth-order vector coefficient $V^{(0)}_\mu$ and the zeroth-order scalar coefficient $F^{(0)}$[229,234,282]

$$V^{(0)}_\mu = \frac{k_\mu}{m}F^{(0)}. \tag{82}$$

Multiplying Eq. (61) with $k^\mu$ and then using Eqs. (53) and (55), we arrive at the constraint equation for the zeroth-order vector coefficient

$$k^2 V^{(0)}_\mu = m^2 V^{(0)}_\mu. \tag{83}$$

Similarly, plugging Eq. (82) in Eq. (53) we obtain the constraint equation for the zeroth-order scalar coefficient[282]

$$k^2 F^{(0)} = m^2 F^{(0)}. \tag{84}$$

Within the framework of quantum kinetic theory,[283] the axial-vector coefficient $A^\mu$[192,280] holds an important place as spin polarization effects appear through $A^\mu$. In this section, for generality, we assume that the spin polarization effects can appear at, both, the zeroth and the first order in $\hbar$.[bb] Thus, nonzero $A^{\mu}_{(0)}$ implies nonvanishing zeroth-order tensor coefficient $S^{(0)}_{\mu\nu}$ through the relation[229,234]

$$S^{(0)}_{\mu\nu} = -\frac{1}{m}\epsilon_{\mu\nu\alpha\beta}k^{\alpha}A^{\beta}_{(0)}, \tag{85}$$

(see Eq. (57)), whereas the dual form of $S^{(0)}_{\mu\nu}$ is

$$\overset{\star}{S}^{\mu\nu}_{(0)} = \frac{1}{m}(k^{\mu}A^{\nu}_{(0)} - k^{\nu}A^{\mu}_{(0)}). \tag{86}$$

The above relation is crucial to obtain the constraint equation for $A^{(0)}_{\mu}$. We first put Eq. (86) in Eq. (56) and then employ Eq. (58) to get[282]

$$k^2 A^{(0)}_{\mu} = m^2 A^{(0)}_{\mu}. \tag{87}$$

In a similar way, substituting Eq. (85) in Eq. (57), and then plugging Eq. (59), we arrive at the constraint equation for $S^{(0)}_{\mu\nu}$

$$k^2 S^{(0)}_{\mu\nu} = m^2 S^{(0)}_{\mu\nu}. \tag{88}$$

From the discussion above, we observe that all the zeroth-order coefficients of the Wigner function need to fulfill the on-shell condition $(k^2 = m^2)$ for a nontrivial solution with $k$ being the kinetic momentum. We find that Eqs. (82), (85) and (86) not only satisfy Eqs. (53)–(61) provided Eq. (58) is fulfilled, but also can be expressed in terms of independent components $F^{(0)}$ and $A^{(0)}_{\nu}$[229,234] whose on-shell conditions, (84) and (87), give rise to[282]

$$F^{(0)} = \delta(k^2 - m^2)\mathcal{F}^{(0)}, \quad A^{(0)}_{\mu} = \delta(k^2 - m^2)\mathcal{A}^{(0)}_{\mu}. \tag{89}$$

$\mathcal{F}^{(0)}$ and $\mathcal{A}^{(0)}_{\mu}$ are scalar and axial-vector functions, respectively, which are non-singular at $k^2 = m^2$.

### 2.3.2. *First order*

The first-order forms of the pseudo-scalar, vector and tensor coefficients can be obtained from Eqs. (63), (64) and (66), respectively, as

$$P^{(1)} = -\frac{1}{2m}[\partial^{\mu}A^{(0)}_{\mu} + 2\mathcal{D}^{(0)}_{P}], \tag{90}$$

$$V^{(1)}_{\mu} = \frac{1}{m}\left[k_{\mu}F^{(1)} - \frac{1}{2}\partial^{\nu}S^{(0)}_{\nu\mu} - \mathcal{D}^{(0)}_{V,\mu}\right], \tag{91}$$

$$S^{(1)}_{\mu\nu} = \frac{1}{m}[\partial_{[\mu}V^{(0)}_{\nu]} - \epsilon_{\mu\nu\alpha\beta}k^{\alpha}A^{\beta}_{(1)} - \mathcal{D}^{(0)}_{S,\mu\nu}], \tag{92}$$

---

[bb] Our assumption is more general in contrast to Refs. 192 and 193. References 192 and 193 consider that spin is a dissipative effect coming from first order in $\hbar$, hence, they assume $A^{\mu}_{(0)} = 0$, whereas we consider that spin polarization can have both classical and quantum counterparts.

with the dual of $S^{(1)}_{\mu\nu}$ written as

$$S^{\star}_{\mu\beta}(1) = \frac{1}{m}\left[\frac{1}{2}\epsilon_{\mu\beta\sigma\rho}\partial^{[\sigma}V^{\rho]}_{(0)} + 2k_{[\mu}A^{(1)}_{\beta]} - \frac{1}{2}\epsilon_{\mu\beta\sigma\rho}\mathcal{D}^{\sigma\rho}_{S(0)}\right].$$ (93)

To obtain the constraint condition for first-order axial-vector coefficient, we first put Eqs. (54) and (93) in Eq. (65) and then use Eqs. (68) and (82) to arrive at

$$(k^2 - m^2)A^{(1)}_\mu = k_\mu \mathcal{C}^{(0)}_P - \frac{1}{2}\epsilon_{\mu\beta\sigma\rho}k^\beta \mathcal{D}^{\sigma\rho}_{S(0)} + m\mathcal{D}^{(0)}_{A,\mu}.$$ (94)

Multiplying Eq. (64) with $k^\mu$ and then using Eqs. (62) and (85) give the constraint equation of the first-order scalar coefficient

$$(k^2 - m^2)F^{(1)} = k^\mu \mathcal{D}^{(0)}_{V,\mu} + m\mathcal{D}^{(0)}_F.$$ (95)

Contracting Eq. (71) with $k_\mu$ and then using Eqs. (57), (62) and (64) we arrive at the constraint equation for first-order vector coefficient

$$(k^2 - m^2)V^\rho_{(1)} = m\mathcal{D}^{(0)}_{V(0)} + k^\rho \mathcal{D}_{F(0)} - k_\lambda \mathcal{C}^{\lambda\rho}_{S(0)}.$$ (96)

To obtain the constraint equation for first-order pseudo-scalar coefficient, we proceed as follows: we first multiply Eq. (70) by $k_\rho$ and Eq. (63) by $m$, and then subtract the resulting equations. Subsequently, we use this equation in Eq. (56) to get

$$(k^2 - m^2)P^{(1)} = -k_\rho \mathcal{C}^\rho_{A(0)} + m\mathcal{D}^{(0)}_P.$$ (97)

Combining Eqs. (54), (65) and (66) and performing some algebraic manipulations, we obtain

$$2m\partial^{[\rho}V^{\lambda]}_{(0)} - \epsilon^{\alpha\rho\lambda\sigma}\epsilon_{\alpha\gamma\delta\beta}k_\sigma k^\beta S^{\gamma\delta}_{(1)} + 2\epsilon^{\rho\lambda\sigma\alpha}k_\sigma \mathcal{D}^{(0)}_{A,\alpha} - 2m^2 S^{\rho\lambda}_{(1)} = 2m\mathcal{D}^{\rho\lambda}_{S(0)}.$$ (98)

Contracting the Levi-Civita tensors[cc] in Eq. (98) and then using Eqs. (69) and (55) one obtains the constraint equation for first-order tensor coefficient $S^{\rho\lambda}_{(1)}$

$$(k^2 - m^2)S^{\rho\lambda}_{(1)} = 2k^{[\rho}\mathcal{C}^{\lambda]}_{V(0)} + m\mathcal{D}^{\rho\lambda}_{S(0)} - \epsilon^{\rho\lambda\sigma\alpha}k_\sigma \mathcal{D}^{(0)}_{A,\alpha}.$$ (99)

From Eqs. (94)–(97) and (99), one can observe that, in the absence of collisions, all the first-order coefficients remain on-shell.

## 2.4. *General kinetic equation*

In this section, we will obtain the general Boltzmann-like kinetic equation which can be used for the formulation of a general spin hydrodynamic formalism. For that purpose we start with formulating the kinetic equations for the Wigner function components at the zeroth and first order in $\hbar$.

---

[cc] $\epsilon^{\mu\lambda\gamma\delta}\epsilon_{\nu\alpha\beta\delta} = [-g^\mu_\nu g^\lambda_\alpha g^\gamma_\beta + g^\mu_\nu g^\lambda_\beta g^\gamma_\alpha + g^\mu_\alpha g^\lambda_\nu g^\gamma_\beta - g^\mu_\alpha g^\lambda_\beta g^\gamma_\nu - g^\mu_\beta g^\lambda_\nu g^\gamma_\alpha + g^\mu_\beta g^\lambda_\alpha g^\gamma_\nu].$

Using Eqs. (67) and (82), the kinetic equation for the zeroth-order scalar coefficient is expressed as

$$k^\mu \partial_\mu F^{(0)} = 2m \mathcal{C}_F^{(0)}, \tag{100}$$

whereas combining Eqs. (77) and (91) we arrive at the kinetic equation for the first-order scalar coefficient

$$k^\mu \partial_\mu F^{(1)} = 2m \mathcal{C}_F^{(1)} + \partial^\mu \mathcal{D}_{V,\mu}^{(0)}. \tag{101}$$

Similarly, the kinetic equation for the zeroth-order axial-vector coefficient can be obtained by combining Eqs. (63), (70) and (86)

$$k^\beta \partial_\beta A_\mu^{(0)} = 2m \mathcal{C}_{A,\mu}^{(0)} - 2k_\mu \mathcal{D}_P^{(0)}, \tag{102}$$

while using Eqs. (73) and (93) in Eq. (80), we obtain the kinetic equation for the first-order axial-vector coefficient

$$k^\beta \partial_\beta A_\mu^{(1)} = 2m \mathcal{C}_{A,\mu}^{(1)} - 2k_\mu \mathcal{D}_P^{(1)} - \frac{1}{2} \epsilon_{\mu\beta\gamma\delta} \partial^\beta \partial^\gamma \mathcal{D}_{S(0)}^{\gamma\delta}. \tag{103}$$

At this point, it is important to highlight again the difference of the assumptions and considerations taken in Refs. 192 and 193 and this work, where we consider the most general structure of the equations without any assumptions on the collision terms:

- In this work, we have assumed that both zeroth-($A_\mu^{(0)}$) and first-order ($A_\mu^{(1)}$) axial-vector components are nonzero, which means that spin polarization effects can appear at both the classical ($\hbar^0$) and quantum level ($\hbar^1$), which is fundamentally different from the considerations in Refs. 192 and 193 where they consider that spin effects arise at order $\hbar$, thus considering only first-order axial-vector component nonvanishing.
- Moreover, Refs. 192 and 193 assumed zeroth-order collision terms corresponding to pseudo-scalar, axial-vector and tensor components vanishing, i.e. $\mathcal{C}_P^{(0)} = 0$, $\mathcal{C}_A^{\mu(0)} = 0$, $\mathcal{C}_S^{\mu\nu(0)} = 0$, $\mathcal{D}_P^{(0)} = 0$, $\mathcal{D}_A^{\mu(0)} = 0$ and $\mathcal{D}_S^{\mu\nu(0)} = 0$. These assumptions may also have some effect on the mass-shell conditions of the Wigner function components.

  However, we put no constraints on the collision terms, hence, they are, in general, all nonvanishing, leading to $P^{(1)} \neq 0$, $(k^2 - m^2) A_\mu^{(1)} \neq 0$ and $k^\mu A_\mu = \mathcal{O}(\hbar)$.

We use Eqs. (100) and (101) to obtain the kinetic equation for the scalar coefficient

$$k^\mu \partial_\mu \tilde{F} = 2m \tilde{\mathcal{C}}_F, \tag{104}$$

where we introduced the following notation:

$$\tilde{F} = F^{(0)} + \hbar F^{(1)}, \quad \tilde{\mathcal{C}}_F = \mathcal{C}_F^{(0)} + \hbar \left( \mathcal{C}_F^{(1)} + \frac{1}{2m} \partial^\mu \mathcal{D}_{V,\mu}^{(0)} \right). \tag{105}$$

Using Eqs. (102) and (103), we arrive at the analogous kinetic equation for the axial-vector component

$$k^\beta \partial_\beta \tilde{A}_\mu = 2m\tilde{\mathcal{C}}_{A,\mu}, \tag{106}$$

where we used

$$
\begin{aligned}
\tilde{A}_\mu &= A_\mu^{(0)} + \hbar A_\mu^{(1)}, \\
\tilde{\mathcal{C}}_{A,\mu} &= \mathcal{C}_{A,\mu}^{(0)} + \hbar\mathcal{C}_{A,\mu}^{(1)} - \frac{k_\mu}{m}\left(\mathcal{D}_P^{(0)} + \hbar\mathcal{D}_P^{(1)}\right) - \frac{\hbar}{4m}\epsilon_{\mu\beta\gamma\delta}\partial^\beta\mathcal{D}_{S(0)}^{\gamma\delta}.
\end{aligned} \tag{107}
$$

We now define single-particle distribution function extended to spin phase-space as[86,192,284–286]

$$\mathfrak{f}(x, k, s) = \frac{1}{2}\left(\tilde{F}(x, k) - s \cdot \tilde{A}(x, k)\right), \tag{108}$$

where $s^\alpha$ denotes the spin four-vector. The above relation will allow us to connect the quantum description of spin to a classical description which, then, can be used for developing hydrodynamical equations. It also incorporates the dynamics of the kinetic equations (104) and (106) into one Boltzmann-like equation, see Eq. (116), which may lead to a more general interpretation of the conservation laws.

Performing the proper averages in the spin space, it is possible to invert Eq. (108) to get the scalar and axial vector components of the Wigner function. Namely, we can write

$$\tilde{F}(x, k) = \int dS(k)\mathfrak{f}(x, k, s), \quad \tilde{A}^\mu(x, k) = \int dS(k)s^\mu\mathfrak{f}(x, k, s), \tag{109}$$

with the covariant spin measure defined as[192,193]

$$\int dS(k) \equiv \frac{1}{\pi}\sqrt{\frac{k^2}{3}}\int d^4s\,\delta(s \cdot s + 3)\delta(k \cdot s). \tag{110}$$

Note that the spin measure consists of two delta functions which are responsible for the normalization of $s$ and orthogonal condition between $k$ and $s$. The factor outside the integral takes care of the following normalization

$$\int dS(k) = 2, \tag{111}$$

which takes into account the spin degeneracy for the spin-half particles. Using Eq. (111) and other identities for the spin measure[192,193]

$$\int dS(k)s^\mu s^\nu = -2\left(g^{\mu\nu} - \frac{k^\mu k^\nu}{k^2}\right), \quad \int dS(k)s^\mu = 0, \tag{112}$$

we can verify the relation between $\tilde{F}(x, k)$ and $\mathfrak{f}(x, k, s)$ in Eq. (109). To establish the relation between $\tilde{A}^\mu(x, k)$ and $\mathfrak{f}(x, k, s)$ we proceed as follows: using Eq. (108) in Eq. (109), we have

$$\int dS(k)s^\mu\mathfrak{f}(x, k, s) = \tilde{A}^\mu(x, k) - \frac{k^\mu}{k^2}(k \cdot \tilde{A}), \tag{113}$$

where, using Eqs. (58) and (68), we find that $k \cdot \tilde{A} = \hbar \mathcal{C}_P^{(0)}$. On the other hand, using Eqs. (46) and (49), one gets

$$k^\mu \partial_\mu P = 2m\mathcal{C}_P + 2k^\mu \mathcal{D}_{A,\mu}, \tag{114}$$

which, in the zeroth order, gives

$$m\mathcal{C}_P^{(0)} + k^\mu \mathcal{D}_{A,\mu}^{(0)} = 0. \tag{115}$$

Above we have used the knowledge that $P^{(0)} = 0$. Since $\mathcal{D}_{A,\mu}^{(0)}$ is an axial vector under the Lorentz transformation, it can be written as $\mathcal{D}_{A,\mu}^{(0)} = s_\mu \delta A$, with $\delta A$ being a scalar function. Using the orthogonality condition $k \cdot s = 0$ and Eqs. (110) and (115), we obtain $\mathcal{C}_P^{(0)} = 0$.[dd] To summarize, Eqs. (109) give the correct relations[ee] between the Wigner function components and the distribution function $\mathfrak{f}(x, k, s)$.

Combining Eqs. (104), (106) and (108), we get the general Boltzmann equation as follows:

$$k^\mu \partial_\mu \mathfrak{f}(x, k, s) = m\mathfrak{C}(\mathfrak{f}) = m(\tilde{\mathcal{C}}_F - s \cdot \tilde{\mathcal{C}}_A), \tag{116}$$

where $\mathfrak{C}(\mathfrak{f})$ is the collision term. Quasi-particle approximation allows us to write $\mathfrak{f}(x, k, s)$ as $\mathfrak{f}(x, k, s) = m\delta(k^2 - M^2)f(x, k, s)$, where $\delta(k^2 - M^2)$ represents the on-shell singularity for the quasi-particle having mass $M$[ff] with the nonsingular function $f(x, k, s)$.[156,192]

We would like to emphasize here that Eq. (116) is a general kinetic equation considering spin effects both at the zeroth and first order in $\hbar$ that may serve as a starting point in formulating a general framework for relativistic hydrodynamics with spin.

## 3. Formulation of Perfect-Fluid Hydrodynamics with Spin

In the previous section, we derived equations of motion for the Wigner function components in the presence of collisions which, using the semi-classical expansion, led us to the general kinetic equation for the distribution function in the phase-space extended to spin. Using such a quantum kinetic theory approach, one can, in principle, obtain a general dissipative spin hydrodynamic equations.

In this section, we develop perfect-fluid hydrodynamics for a spin-polarized system of Dirac fermions in equilibrium using the GLW framework and assuming that the collisional kernels are vanishing completely, which is a natural expectation for equilibrium. This can be achieved by explicitly calculating the equilibrium Wigner function[gg] using

---

[dd] Another possibility to obtain a relation between $\tilde{A}^\mu(x, k)$ and $\mathfrak{f}(x, k, s)$ in Eq. (109) is to define Eq. (108) with only the components of $\tilde{A}^\mu$ which are orthogonal to $k^\mu$, i.e. $(g^{\mu\nu} - k^\mu k^\nu/k^2)\tilde{A}_\nu$. This actually does not change Eq. (108), since $s^\mu$ is orthogonal to $k^\mu$, so the parallel components vanish, and we obtain the second relation in Eq. (109). (Thanks to Nora Weickgenannt for pointing this out.)

[ee] These relations are important as macroscopic currents such as energy–momentum and spin tensors are expressed in terms of the Wigner function components.[86,95]

[ff] $M$ includes quantum corrections to the particle mass $m$, however, these corrections do not contribute to the Boltzmann equation.[192,193]

[gg] Equilibrium means the leading (or zeroth) order in $\hbar$.

an ansatz for the spin-dependent phase-space distribution functions for spin-half particles. Various hydrodynamic currents, e.g. the energy–momentum tensor, net baryon current, spin-tensor, etc. can be obtained by expressing them in terms of different components of the Wigner function in the Clifford algebra basis. Conservation laws of these currents lead us to the formulation of hydrodynamic equations of interest.[229]

Subsequently, we show that such spin-hydrodynamic equations can also be obtained starting from the classical transport theory in the phase-space extended to spin. In Ref. [86] such a framework has been obtained using the spin-dependent equilibrium distribution function, where one inherently considers the classical description of spin. Interestingly, the constitutive relations for the net baryon current, energy–momentum tensor and spin tensor match in these two cases.[229] The same results from the two completely independent approaches put the GLW framework on a firm mathematical footing.

## 3.1. *Wigner function approach*

In this section, employing an ansatz for a phase-space distribution function in the spin representation, we construct the equilibrium Wigner function which we subsequently use to derive the conservation laws for net baryon current, energy–momentum tensor and spin tensor, constituting perfect-fluid hydrodynamics with spin.

### 3.1.1. *Equilibrium Wigner function*

Let us start with the transport equation for the Wigner function (39), and assume that the system reaches equilibrium state. In such a situation, the collision kernel on the right-hand side of Eq. (39) is expected to vanish, meaning that the rest energy of the particles dominates over the mean interaction energy. In this case, the respective transport equation takes the form[hh]

$$\left( i\hbar \frac{\slashed{\partial}}{2} + \slashed{k} - m \right) W_{\text{eq}}(x,k) = 0, \tag{117}$$

where we introduced the label "eq" to indicate the equilibrium form of the respective Wigner function. Moreover, in the case of global equilibrium one may expect the microscopic nonuniformities in the system to vanish, allowing the gradient term in Eq. (117) to be neglected. From the discussion, in Subsec. 2.2, we know that this situation corresponds to the zeroth order of the semi-classical expansion, where the momenta of the particles satisfy the mass-shell condition.

As shown by GLW,[156] the restrictions discussed above are satisfied by the Wigner function of the form

$$W_{\text{eq}}(x,k) = W_{\text{eq}}^+(x,k) + W_{\text{eq}}^-(x,k),$$

---

[hh] From the transport equation for the Wigner function, one may notice that the gradient expansion is effectively a $\hbar$ expansion.

with the particle and antiparticle contributions given by

$$W_{\text{eq}}^{+}(x, k) = \frac{1}{2} \sum_{r,s} \int dP \delta^{(4)}(k - p) \mathcal{U}^r(p) \bar{\mathcal{U}}^s(p) f_{rs}^{+}(x, p), \tag{118}$$

$$W_{\text{eq}}^{-}(x, k) = -\frac{1}{2} \sum_{r,s} \int dP \delta^{(4)}(k + p) \mathcal{V}^s(p) \bar{\mathcal{V}}^r(p) f_{rs}^{-}(x, p), \tag{119}$$

respectively.[ii] Here, $dP = d^3 p / ((2\pi)^3 E_p)$ is the invariant momentum integration measure with $E_p = \sqrt{\boldsymbol{p}^2 + m^2}$ denoting the on-shell particle energy, $\mathcal{U}_r(p)$ and $\mathcal{V}_r(p)$ denote the Dirac bispinors with the normalizations $\bar{\mathcal{U}}_r(p)\mathcal{U}_s(p) = 2m\delta_{rs}$ and $\bar{\mathcal{V}}_r(p) \mathcal{V}_s(p) = -2m\delta_{rs}$ while indices $r$ and $s$ represent the spin states.

For the equilibrium Wigner function $W_{\text{eq}}(x, k)$ to satisfy the conjugation relation (42), the phase-space distribution functions $f_{rs}^{\pm}(x, p)$ have to be Hermitian matrices. They can be written as[159]

$$\begin{aligned}
[f^{+}(x, p)]_{rs} &\equiv f_{rs}^{+}(x, p) = \frac{1}{2m} \bar{\mathcal{U}}_r(p) X^{+} \mathcal{U}_s(p), \\
[f^{-}(x, p)]_{rs} &\equiv f_{rs}^{-}(x, p) = -\frac{1}{2m} \bar{\mathcal{V}}_s(p) X^{-} \mathcal{V}_r(p).
\end{aligned} \tag{120}$$

In Eq. (120), $X^{\pm}$ are $4 \times 4$ matrices defined as the products of the MJ distributions[287,jj] and matrices $M^{\pm}$[225]

$$M^{\pm} = \exp\left[\pm\frac{1}{2}\omega_{\mu\nu}(x)\Sigma^{\mu\nu}\right], \tag{121}$$

namely

$$X^{\pm} = \exp\left[\pm\xi(x) - \beta_{\mu}(x)p^{\mu}\right] M^{\pm}. \tag{122}$$

The quantity $\beta^{\mu}(x)$ is the ratio of the fluid four-velocity $U^{\mu}(x)$[kk] and the local temperature $T(x)$, $\beta^{\mu} = U^{\mu}/T$ and $\xi(x)$ is the ratio of the baryon chemical potential $\mu_B(x)$ and the temperature, $\xi = \mu_B/T$.[ll] The rank-two antisymmetric tensor $\omega^{\mu\nu}(x)$, also known as the spin polarization tensor,[mm] introduces into the theory, apart from the standard ones, six extra Lagrange multipliers, which, given $\omega^{\mu\nu}(x)$ being to conjugated to the Dirac spin operator $\Sigma^{\mu\nu} = (i/4)[\gamma^{\mu}, \gamma^{\nu}]$ (generators of Lorentz transformation of spinors), are responsible for angular momentum conservation in the system. One should stress here that the quantity $\omega^{\mu\nu}(x)$ in Eq. (121) is, assumed

---

[ii]Note the Dirac delta functions in the equilibrium Wigner function definitions, suggesting that these definitions describe classical motion, i.e. particle energy is always on the mass shell.

[jj]Note that for simplicity, instead of FD statistics used in Ref. 159, herein we consider the classical Boltzmann statistics.

[kk]Due to normalization condition, $U \cdot U = 1$, the four-velocity vector has only three independent components. Hence, one may write $U^{\mu}(x) = \gamma(1, \boldsymbol{v})$ where $\boldsymbol{v}$ is the three-velocity and $\gamma = 1/\sqrt{1 - v^2}$ is the Lorentz factor.

[ll]Note that for $\Lambda(\bar{\Lambda})$ hyperon, $\mu_B = 3\mu_Q$, with $\mu_Q$ being the quark chemical potential.[50]

[mm]Note that in literature instead of $\omega_{\mu\nu}$ one often uses the tensor $\Omega_{\mu\nu} = T\omega_{\mu\nu}$, which, by analogy to the baryon chemical potential, $\mu_B = T\xi$, is called the spin chemical potential.

to be, in general, different from the thermal vorticity, $\varpi_{\mu\nu} = -(1/2)(\partial_\mu \beta_\nu - \partial_\nu \beta_\mu)$ originally used in Ref. 159, allowing us to construct the closed[nn] perfect-fluid hydrodynamic framework based on the conservation equations.

As shown in Ref. 225, Eq. (121) can be put into the form

$$M^\pm = \frac{1}{2}\left[2\cosh(\zeta) \pm \frac{\sinh(\zeta)}{\zeta}\omega_{\mu\nu}\Sigma^{\mu\nu}\right], \qquad (123)$$

where the quantity

$$\zeta = \frac{1}{2\sqrt{2}}\sqrt{\omega_{\mu\nu}\omega^{\mu\nu}}, \qquad (124)$$

is assumed to be real.[oo] In this paper, following experimental observations of small values of polarization, we consider the limit $\omega_{\mu\nu} \ll 1$ (in other words, $\zeta \to 0$), where Eq. (123) reduces to

$$M^\pm = 1 \pm \frac{1}{2}\omega_{\mu\nu}\Sigma^{\mu\nu}.$$

However, in order to keep our discussion general, in what follows, we use Eq. (123), considering the small polarization limit only in the final formulas.

Putting Eqs. (120) in Eqs. (118) and (119), we find

$$W_{\mathrm{eq}}^\pm(x,k) = \frac{1}{4m}\int \mathrm{dP}\delta^{(4)}(k \mp p)(\not{p} \pm m)X^\pm(\not{p} \pm m), \qquad (126)$$

respectively. With the help of Eq. (123), we can further rewrite the above equation as

$$W_{\mathrm{eq}}^\pm(x,k) = \frac{1}{4m}\int \mathrm{dP}e^{-\beta\cdot p \pm \xi}\delta^{(4)}(k \mp p)$$
$$\times \left[2m(m \pm \not{p})\cosh(\zeta) \pm \frac{\sinh(\zeta)}{2\zeta}\omega_{\mu\nu}(\not{p} \pm m)\Sigma^{\mu\nu}(\not{p} \pm m)\right]. \qquad (127)$$

It is instructive to decompose the Wigner function Eq. (127) using the Clifford algebra expansion in an analogous way as in Eq. (41) to obtain equilibrium Wigner function components. In this case, one obtains[229]

$$F_{\mathrm{eq}}^\pm(x,k) = \mathrm{tr}[W_{\mathrm{eq}}^\pm(x,k)] = 2m\cosh(\zeta)\int \mathrm{dP}e^{-\beta\cdot p \pm \xi}\delta^{(4)}(k \mp p), \qquad (128)$$

$$P_{\mathrm{eq}}^\pm(x,k) = -i\mathrm{tr}[\gamma^5 W_{\mathrm{eq}}^\pm(x,k)] = 0, \qquad (129)$$

---

[nn]By closed, we have in mind that the number of conservation equations matches the number of Lagrange multipliers.

[oo]In Refs. 225 and 226, it was shown that the thermodynamic consistency requires the spin polarization tensor to satisfy the following constraints:

$$\omega^{\mu\nu}\omega_{\mu\nu} \geq 0, \qquad \overset{\star}{\omega}{}^{\mu\nu}\omega_{\mu\nu} = 0. \qquad (125)$$

Violating the above conditions may result in imaginary $\zeta$ which leads to the negative values of the thermodynamic quantities. However, one should notice that such constraints are not necessary in the small polarization limit $\omega_{\mu\nu} \ll 1$ in which case one can directly Taylor expand Eq. (121).

$$V_{\text{eq},\mu}^{\pm}(x,k) = \text{tr}[\gamma_{\mu} W_{\text{eq}}^{\pm}(x,k)] = \pm 2\cosh(\zeta) \int \text{d}P e^{-\beta \cdot p \pm \xi} \delta^{(4)}(k \mp p) p_{\mu}, \qquad (130)$$

$$A_{\text{eq},\mu}^{\pm}(x,k) = \text{tr}[\gamma_{\mu}\gamma^5 W_{\text{eq}}^{\pm}(x,k)] = -\frac{\sinh(\zeta)}{\zeta} \int \text{d}P e^{-\beta \cdot p \pm \xi} \delta^{(4)}(k \mp p) \omega_{\mu\nu}^{\star} p^{\nu}, \quad (131)$$

$$S_{\text{eq},\mu\nu}^{\pm}(x,k) = 2\text{tr}[\Sigma_{\mu\nu} W_{\text{eq}}^{\pm}(x,k)]$$
$$= \pm \frac{\sinh(\zeta)}{m\zeta} \int \text{d}P e^{-\beta \cdot p \pm \xi} \delta^{(4)}(k \mp p)[(p_{\mu}\omega_{\nu\alpha} - p_{\nu}\omega_{\mu\alpha})p^{\alpha} + m^2\omega_{\mu\nu}]. \quad (132)$$

One can easily check that for arbitrary form of the fields $\xi$, $\beta$ and $\omega_{\mu\nu}$, the coefficient functions defined in Eqs. (128)–(132) satisfy the following constraints:

$$k^{\mu} V_{\text{eq},\mu}^{\pm}(x,k) = m F_{\text{eq}}^{\pm}(x,k), \quad k_{\mu} F_{\text{eq}}^{\pm}(x,k) = m V_{\text{eq},\mu}^{\pm}(x,k), \qquad (133)$$

$$P_{\text{eq}}^{\pm}(x,k) = 0, \quad k^{\mu} A_{\text{eq},\mu}^{\pm}(x,k) = 0, \quad k^{\mu} S_{\text{eq},\mu\nu}^{\pm}(x,k) = 0, \qquad (134)$$

$$k^{\beta} \overset{\star}{S}_{\text{eq},\mu\beta}^{\pm}(x,k) + m A_{\text{eq},\mu}^{\pm}(x,k) = 0, \quad \epsilon_{\mu\nu\alpha\beta} k^{\alpha} A_{\text{eq}}^{\pm\beta}(x,k) + m S_{\text{eq},\mu\nu}^{\pm}(x,k) = 0. \quad (135)$$

From Eqs. (133)–(135), we can notice that the equilibrium coefficient functions follow constraints of the same form as the ones satisfied by the zeroth-order components of the Wigner function, cf. Eqs. (53)–(61). It suggests that the equilibrium Wigner functions Eqs. (127) are the good candidates for the zeroth-order Wigner function components, allowing us to write[229]

$$F^{(0)} = F_{\text{eq}}, \quad P^{(0)} = 0, \quad V_{\mu}^{(0)} = V_{\text{eq},\mu}, \quad A_{\mu}^{(0)} = A_{\text{eq},\mu}, \quad S_{\mu\nu}^{(0)} = S_{\text{eq},\mu\nu}. \qquad (136)$$

In the following sections, these components will serve as a starting point to construct the perfect-fluid hydrodynamics with spin formalism for spin-polarized particles.

### 3.1.2. *Conservation laws*

In the absence of collisions, the zeroth-order scalar and axial-vector coefficients of the Wigner function defined by Eqs. (136) satisfy kinetic equations (100) and (102) found at the first order of the semi-classical expansion,[pp] with the vanishing collision kernels, $\mathcal{C}_{\mathcal{X}} = \mathcal{D}_{\mathcal{X}} = 0$, supplemented with the constraint (58). Hence, we have

$$k^{\mu}\partial_{\mu} F_{\text{eq}}(x,k) = 0, \quad k^{\mu}\partial_{\mu} A_{\text{eq}}^{\nu}(x,k) = 0, \quad k_{\nu} A_{\text{eq}}^{\nu}(x,k) = 0, \qquad (137)$$

where kinetic momentum $k$ is supposed to be on the mass shell. First-order contributions to other coefficients of Wigner function can be written as

$$P^{(1)} = -\frac{\partial^{\mu} A_{\text{eq},\mu}}{2m}, \quad V_{\mu}^{(1)} = -\frac{\partial^{\nu} S_{\text{eq},\nu\mu}}{2m}, \quad S_{\mu\nu}^{(1)} = \frac{1}{2m}\left(\partial_{\mu} V_{\text{eq},\nu} - \partial_{\nu} V_{\text{eq},\mu}\right). \qquad (138)$$

---

[pp]Following Eqs. (68), (101) and (103), we observe that the zeroth-order and first-order coefficients decouple in the absence of particle interactions, hence, in the following, we can safely assume $F^{(1)}(x,k) = A_{\mu}^{(1)}(x,k) = 0$.

Similarly to the treatment of spinless particles,[229,267,288] left-hand sides of collisionless Boltzmann-like kinetic equations (137) are satisfied exactly in the global equilibrium. In this case, using (128) and (131), one can show that Eqs. (137) are satisfied if $\beta^\alpha$ is a Killing field fulfilling $\partial_\alpha\beta_\beta + \partial_\beta\beta_\alpha = 0$, and $\xi$ and $\omega_{\mu\nu}$ are constants. One should stress here that $\omega_{\mu\nu}$ is not necessarily equal to thermal vorticity $\varpi_{\mu\nu}$, although both are constant in the global equilibrium. The difference between $\omega_{\mu\nu}$ and $\varpi_{\mu\nu}$ in global equilibrium is understood on general grounds for conservative systems, provided energy–momentum tensor is asymmetric and spin tensor does not vanish.[229] As a matter of fact, the presence of nonlocal collisions may make the energy–momentum tensor asymmetric, giving rise to the spin–orbit coupling which in global equilibrium results in $\varpi_{\mu\nu} = \omega_{\mu\nu}$.[192,193,265]

In local equilibrium, only certain moments (integrals) in momentum space of Eqs. (137) are satisfied. In particular, following the standard hydrodynamic treatment of relativistic rarefied spinless gases,[229,267,288] one may show that the zeroth and first moments of the transport equations (137) (related to collisional invariants of baryon charge and four-momentum, respectively), lead to the net baryon current and energy–momentum tensor conservation laws. However, for spin-polarized particles considered in this paper, the precise form of the collisional kernel is so far unknown, hence, it is unclear which type of the moment one should consider to impose the angular momentum conservation in the collisions. Hence, in the following sections, we will follow a somewhat different methodology. We will use a so-called phenomenological approach to formulate conservation law for currents defined by the moments of the Wigner function. In particular, we will formulate the conservation equation for the spin current (the spin part of the total angular momentum rank-three tensor) which, as it was shown in Ref. 229, agrees with a certain moment of the equation for the axial coefficient of the equilibrium Wigner function.

*Net baryon current*

The equilibrium Wigner function, $W_{eq}(x,k)$, allows us to write the macroscopic form of net baryon current, with particle ($W_{eq}^+(x,k)$) and antiparticle ($W_{eq}^-(x,k)$) contributions, as a momentum integral[156]

$$N^\alpha(x) = \langle : \bar{\psi}\gamma^\alpha\psi : \rangle = \mathrm{tr} \int d^4k \gamma^\alpha (W_{eq}^+(x,k) - W_{eq}^-(x,k))$$

$$= \int d^4k (V_{eq}^{+,\alpha}(x,k) - V_{eq}^{-,\alpha}(x,k)), \tag{139}$$

where the right-hand side of the first line depicts the ensemble (statistical) average of the normal ordered, microscopic net baryon current, whereas tr (trace), in the second line, provides the expectation value (in other words, statistical average) of the microscopic quantity.

Using Eq. (133) in Eq. (139), one gets

$$N^\alpha(x) = \frac{1}{m} \int d^4k k^\alpha (F_{eq}^+(x,k) - F_{eq}^-(x,k)). \tag{140}$$

Substituting Eq. (128) into (140), we obtain the net baryon current as[86,225,229]

$$N^\alpha(x) = \frac{1}{m} \int d^4k \, k^\alpha (F_{eq}^+(x,k) - F_{eq}^-(x,k))$$

$$= 4\cosh(\zeta)\sinh(\xi) \int dP \, p^\alpha e^{-\beta \cdot p} = \mathcal{N} U^\alpha. \tag{141}$$

The quantity $\mathcal{N}$ in the above equation represents the net baryon density written as

$$\mathcal{N} = 4\cosh(\zeta)\sinh(\xi)\mathcal{N}_{(0)}(T), \tag{142}$$

where $\mathcal{N}_{(0)}(T)$ is the number density of spinless and neutral classical massive particles for an ideal relativistic Boltzmann gas defined as[50,86,229]

$$\mathcal{N}_{(0)}(T) = \int dP(U \cdot p) e^{-\beta \cdot p} = \frac{1}{2\pi^2} T^3 z^2 K_2(z). \tag{143}$$

Here, $z$ is defined as the ratio of the particle mass $m$ and the temperature $T$, $z \equiv m/T$,[qq] and $K_n$ denotes the modified Bessel function of the second kind[289]

$$K_n(z) = \frac{z^n}{(2n+1)!!} \int_1^\infty dx (x^2-1)^{n-1/2} e^{-xz}. \tag{144}$$

One can observe that the factor $\sinh(\xi) = (e^\xi - e^{-\xi})/2$ in Eq. (142) denotes the presence of both particles and antiparticles in the system, and the quantity $\cosh(\zeta) = (e^\zeta + e^{-\zeta})/2$ represents the presence of both spin-up and spin-down particles. Since the baryon current must be conserved, we write

$$\partial_\alpha N^\alpha(x) = 0. \tag{145}$$

*Energy–momentum tensor*

Similarly to the definition of the net baryon current, using Eq. (31), we can also define the macroscopic (noninteracting) energy–momentum tensor by taking the trace of the second moment of the equilibrium Wigner function for particle and antiparticle as, see p. 116 in Ref. 156,

$$T_{GLW}^{\mu\nu}(x) = \langle : \hat{T}_{GLW}^{\mu\nu} : \rangle = \frac{1}{m} \mathrm{tr} \int d^4k \, k^\mu k^\nu (W_{eq}^+(x,k) + W_{eq}^-(x,k))$$

$$= \frac{1}{m} \int d^4k \, k^\mu k^\nu (F_{eq}^+(x,k) + F_{eq}^-(x,k)). \tag{146}$$

where we use the assumption that $F^{(1)}(x,k) = 0$. Using Eq. (128) in Eq. (146), we obtain the perfect-fluid form of the energy–momentum tensor as[86,225,229]

$$T_{GLW}^{\mu\nu}(x) = \frac{1}{m} \int d^4k \, k^\mu k^\nu (F_{eq}^+(x,k) + F_{eq}^-(x,k))$$

$$= 4\cosh(\zeta)\cosh(\xi) \int dP \, p^\mu p^\nu e^{-\beta \cdot p},$$

$$= (\mathcal{E} + \mathcal{P}) U^\mu U^\nu - \mathcal{P} g^{\mu\nu}, \tag{147}$$

---

[qq]The variable "$z$" can also be interpreted as a controlling parameter for temperature, meaning that, for the fixed mass, temperature can be varied in order to go from relativistic (high temperature) limit to nonrelativistic (low temperature) limit.

where the energy density ($\mathcal{E}$) and pressure ($\mathcal{P}$) are, respectively, given as

$$\mathcal{E} = 4\cosh(\zeta)\cosh(\xi)\mathcal{E}_{(0)}(T), \quad \mathcal{P} = 4\cosh(\zeta)\cosh(\xi)\mathcal{P}_{(0)}(T). \tag{148}$$

Similar to the number density, we define the energy density ($\mathcal{E}_{(0)}(T)$) and pressure ($\mathcal{P}_{(0)}(T)$) for Boltzmann gas as[50,86,229]

$$\mathcal{E}_{(0)}(T) = \int d\mathrm{P}(U \cdot p)^2 e^{-\beta \cdot p} = \frac{1}{2\pi^2}T^4 z^2[zK_1(z) + 3K_2(z)], \tag{149}$$

$$\mathcal{P}_{(0)}(T) = -\frac{1}{3}\int d\mathrm{P}[p \cdot p - (U \cdot p)^2]e^{-\beta \cdot p} = \frac{1}{2\pi^2}T^4 z^2 K_2(z), \tag{150}$$

respectively. From Eqs. (150) and (143), we notice that

$$\mathcal{P}_{(0)}(T) = T\mathcal{N}_{(0)}(T), \tag{151}$$

which is the EoS for the ideal relativistic Boltzmann statistics. Again, since we assume the energy–momentum tensor to be conserved, we write

$$\partial_\alpha T^{\alpha\beta}_{\mathrm{GLW}}(x) = 0. \tag{152}$$

*Spin tensor*

The GLW form of the spin tensor is defined as the momentum average of the microscopic spin density (31)[156,rr]

$$
\begin{aligned}
S^{\alpha,\beta\gamma}_{\mathrm{GLW}} &= \langle : \hat{S}^{\alpha,\beta\gamma}_{\mathrm{GLW}} : \rangle \\
&= \frac{\hbar}{4}\int d^4 k \, \mathrm{tr}\left[\left(\{\sigma^{\beta\gamma}, \gamma^\alpha\} + \frac{2i}{m}(\gamma^{[\beta}k^{\gamma]}\gamma^\alpha - \gamma^\alpha\gamma^{[\beta}k^{\gamma]})\right) \right. \\
&\quad \left. \times (W^+_{\mathrm{eq}}(x,k) + W^-_{\mathrm{eq}}(x,k))\right].
\end{aligned}
\tag{153}
$$

Putting particle and antiparticle contributions from Eq. (127) in (153), performing the trace and carrying out the momentum integration, we get

$$S^{\alpha,\beta\gamma}_{\mathrm{GLW}} = \frac{\mathcal{C}}{m^2}\int d\mathrm{P}\, e^{-\beta \cdot p} p^\alpha (m^2\omega^{\beta\gamma} + 2p^\mu p^{[\beta}\omega^{\gamma]}_\mu) = S^{\alpha,\beta\gamma}_{\mathrm{PH}} + S^{\alpha,\beta\gamma}_\Delta, \tag{154}$$

where we identify[225,229]

$$S^{\alpha,\beta\gamma}_{\mathrm{PH}} = \mathcal{C}\mathcal{N}_{(0)}U^\alpha\omega^{\beta\gamma}, \tag{155}$$

$$S^{\alpha,\beta\gamma}_\Delta = \mathcal{C}[\mathcal{A}_{(0)}U^\alpha U^\delta U^{[\beta}\omega^{\gamma]}_\delta + \mathcal{B}_{(0)}(U^{[\beta}\Delta^{\alpha\delta}\omega^{\gamma]}_\delta + U^\alpha\Delta^{\delta[\beta}\omega^{\gamma]}_\delta + U^\delta\Delta^{\alpha[\beta}\omega^{\gamma]}_\delta)], \tag{156}$$

with the thermodynamic coefficients expressed as

$$\mathcal{B}_{(0)} = -\frac{2}{z^2}\frac{\mathcal{E}_{(0)} + \mathcal{P}_{(0)}}{T}, \quad \mathcal{A}_{(0)} = 2\mathcal{N}_{(0)} - 3\mathcal{B}_{(0)}. \tag{157}$$

---

[rr] Note that we explicitly placed the Planck constant $\hbar$ in (153) due to dimensional reason. Here, $\{A, B\} = AB + BA$ and $A^{[\mu}B^{\nu]} = \frac{1}{2}[A^\mu B^\nu - A^\nu B^\mu]$.

Here, we introduce $\mathcal{C} \equiv \hbar \cosh(\xi) \sinh(\zeta)/\zeta^{\text{ss}}$ and $\Delta^{\mu\nu} = g^{\mu\nu} - (U^\mu U^\nu)/(U \cdot U)$ which is the operator projecting onto the plane orthogonal to $U^\mu$. One may interpret $S_\Delta^{\alpha,\beta\gamma}$ as the correction to the phenomenological part ($S_{\text{PH}}^{\alpha,\beta\gamma}$).

We can write the spin tensor Eq. (154) in a more compact way as[238,241]

$$S_{\text{GLW}}^{\alpha,\beta\gamma} = U^\alpha(\mathcal{A}_1 \omega^{\beta\gamma} + \mathcal{A}_2 U^{[\beta} \omega_\delta^{\gamma]} U^\delta) + \mathcal{A}_3(U^{[\beta} \omega^{\gamma]\alpha} + g^{\alpha[\beta} \omega_\delta^{\gamma]} U^\delta), \tag{158}$$

with the thermodynamic coefficients having the forms

$$\mathcal{A}_1 = \mathcal{C}(\mathcal{N}_{(0)} - \mathcal{B}_{(0)}), \quad \mathcal{A}_2 = \mathcal{C}(\mathcal{A}_{(0)} - 3\mathcal{B}_{(0)}), \quad \mathcal{A}_3 = \mathcal{C}\mathcal{B}_{(0)}. \tag{159}$$

Since the energy–momentum tensor (147) is completely symmetric, the conservation of total angular momentum[156,180]

$$\partial_\alpha J^{\alpha,\beta\gamma} = T_{\text{GLW}}^{\beta\gamma} - T_{\text{GLW}}^{\gamma\beta} + \partial_\alpha S_{\text{GLW}}^{\alpha,\beta\gamma} = 0, \tag{160}$$

implies that the spin tensor is independently conserved[229]

$$\partial_\alpha S_{\text{GLW}}^{\alpha,\beta\gamma}(x) = T_{\text{GLW}}^{\gamma\beta} - T_{\text{GLW}}^{\beta\gamma} = 0. \tag{161}$$

Equations (145), (152) and (161), represent closed set of evolution equations constituting the formulation of perfect-fluid hydrodynamics with spin. Note that, in all above conservation laws, the small polarization limit ($\omega_{\alpha\beta} \ll 1$), which we consider throughout the remaining parts of the paper, is implemented by $\zeta \to 0$. Taking this limit prevents the coupling between the dynamics of the perfect-fluid background, represented through the energy–momentum and baryon number conservation laws, and spin. Thus, the evolution of spin polarization is determined solely through the conservation of total angular momentum (in practice, conservation of spin tensor).[tt]

## 3.2. *Classical approach*

In this section, we present a complementary method to derive the constitutive relations for the net baryon current (141), energy–momentum tensor (147) and spin tensor (154) using an approach based on the classical single-particle distribution function in a phase-space extended to spin. Such a distribution function may arise from the quantum kinetic considerations as discussed in Subsec. 2.4.

### 3.2.1. *Spin-dependent distribution function in equilibrium*

We consider classical single-particle distribution function for particles and antiparticles in a phase-space consisting of space–time position ($x^\mu$), four-momentum ($p^\mu$) and spin ($s^\mu$). Identifying the collisional invariants of the local Boltzmann equation, one finds[86]

$$f_{\text{eq}}^{\pm}(x, p, s) = \exp\left(-\beta(x) \cdot p \pm \xi(x)\right)\exp\left\{\frac{1}{2}\omega_{\mu\nu}(x)s^{\mu\nu}\right\}, \tag{162}$$

---

[ss]Note that $\mathcal{C}$ is not related to the collisional kernels discussed in the previous section.

[tt]Note that the spin tensor (154) diverges in the $m \to 0$ limit, breaking conformal symmetry explicitly, however, one may obtain approximate solutions for spin treating mass as a very small parameter. See Sec. 8 for extensive discussions on this matter.

where $s^{\mu\nu} = (1/m)\epsilon^{\mu\nu\alpha\beta}p_\alpha s_\beta$ is the internal angular momentum tensor.[290,179] In Eq. (162), the tensor $\omega_{\mu\nu}$ plays the role of the spin polarization tensor, encountered in the quantum statistical approach (121), and arises due to the spin angular momentum conservation considered herein.

Equation (162) gives the distribution function in terms of position $x$ and momentum $p$ coordinates only, once averaged over spin

$$\int \mathrm{dS} f_{\mathrm{eq}}^{\pm}(x,p,s) = f_{\mathrm{eq}}^{\pm}(x,p), \tag{163}$$

where the spin integration measure dS is defined as follows[86,uu]:

$$\mathrm{dS} = \frac{m}{\pi\mathfrak{s}} \mathrm{d}^4 s \delta(s \cdot s + \mathfrak{s}^2)\delta(p \cdot s). \tag{164}$$

Here, unlike in Eq. (110), $p$ is the on-shell kinetic four-momentum of the particle. For spin-half particles, the length of the spin vector is given by $\mathfrak{s}^2 = \frac{1}{2}(1 + \frac{1}{2}) = \frac{3}{4}$, which can also be related to the respective value of the Casimir operator.

In the remaining part of this section, we provide definitions of the conserved currents in terms of the moments of the distribution function (162), which ought to satisfy macroscopic balance equations describing conservation laws.

### 3.2.2. *Conservation laws*

*Net baryon current*

The net baryon current in equilibrium can be obtained using the formula[86]

$$N^\mu(x) = \int \mathrm{dP}\, \mathrm{dS}\, p^\mu [f_{\mathrm{eq}}^+(x,p,s) - f_{\mathrm{eq}}^-(x,p,s)]. \tag{165}$$

Putting the equilibrium function (162) in Eq. (165) leads to

$$N^\mu(x) = 2\sinh(\xi)\int \mathrm{dP}\, p^\mu e^{-p\cdot\beta}\int \mathrm{dS}\, \exp\left(\frac{1}{2}\omega_{\alpha\beta}s^{\alpha\beta}\right). \tag{166}$$

In our approach, we consider the case of small values of the spin polarization tensor $\omega_{\mu\nu}$, which allows us to expand the last exponential term up to the first order in $\omega$. Proceeding this way, we get

$$N^\mu = 2\sinh(\xi)\int \mathrm{dP}\, p^\mu e^{-p\cdot\beta}\int \mathrm{dS}\left(1 + \frac{1}{2}\omega_{\alpha\beta}s^{\alpha\beta}\right), \tag{167}$$

which, after performing the spin and momentum integration, yields[86]

$$N^\mu = \mathcal{N}U^\mu. \tag{168}$$

The resulting net baryon current matches exactly the one derived using the Wigner function approach, in the small polarization limit, see Eq. (141) in Subsec. 3.1.2.

---

[uu]Note that in the spin measure (164), we choose a different normalization of the spin four-vector as compared to (110) defined in Sec. 2. We note that the choice of the normalization convention in the spin measure definition does not affect our results, up to an overall constant factor.

*Energy–momentum tensor*

The symmetric energy–momentum tensor is defined by the following average[86]:

$$T^{\mu\nu} = \int \mathrm{dP}\,\mathrm{dS}\, p^\mu p^\nu [f_{\mathrm{eq}}^+(x,p,s) + f_{\mathrm{eq}}^-(x,p,s)]. \tag{169}$$

Substituting the distribution function (162) in (169), we obtain

$$T^{\mu\nu} = 2\cosh(\xi) \int \mathrm{dP}\, p^\mu p^\nu e^{-p\cdot\beta} \int \mathrm{dS} \exp\left(\frac{1}{2}\omega_{\alpha\beta}s^{\alpha\beta}\right), \tag{170}$$

where, again, employing the small polarization limit and performing the integrations, leads to the same formula as the one obtained in Subsec. 3.1.2, see Eq. (147).

*Spin tensor*

Spin tensor is defined as follows[86]:

$$S^{\lambda,\mu\nu} = \int \mathrm{dP}\,\mathrm{dS}\, p^\lambda s^{\mu\nu}[f_{\mathrm{eq}}^+(x,p,s) + f_{\mathrm{eq}}^-(x,p,s)], \tag{171}$$

$$= 2\cosh(\xi) \int \mathrm{dP}\, p^\lambda \exp\left(-p\cdot\beta\right) \int \mathrm{dS}\, s^{\mu\nu} \exp\left(\frac{1}{2}\omega_{\alpha\beta}s^{\alpha\beta}\right), \tag{172}$$

where, in the second line, we substituted the distribution function from Eq. (162). Performing the integral over the spin in Eq. (172), to the first order in $\omega_{\mu\nu}$, gives

$$\int \mathrm{dS}\, s^{\mu\nu} \exp\left(\frac{1}{2}\omega_{\alpha\beta}s^{\alpha\beta}\right) = \int \mathrm{dS}\, s^{\mu\nu}\left(1 + \frac{1}{2}\omega_{\alpha\beta}s^{\alpha\beta}\right)$$
$$= \frac{2}{3m^2}\mathfrak{s}^2(m^2\omega^{\mu\nu} + 2p^\alpha p^{[\mu}\omega^{\nu]}_{\ \alpha}). \tag{173}$$

Putting Eq. (173) back in Eq. (172), we obtain

$$S^{\lambda,\mu\nu} = \frac{4}{3m^2}\mathfrak{s}^2 \cosh(\xi) \int \mathrm{dP}\, p^\lambda e^{-p\cdot\beta}(m^2\omega^{\mu\nu} + 2p^\alpha p^{[\mu}\omega^{\nu]}_{\ \alpha}), \tag{174}$$

which after doing momentum integration leads to[86,229]

$$S^{\lambda,\mu\nu} = \mathcal{C}(\mathcal{N}_0(T)U^\lambda\omega^{\mu\nu} + S_\Delta^{\lambda,\mu\nu}). \tag{175}$$

The above result agrees with the spin tensor $S_{\mathrm{GLW}}^{\lambda,\mu\nu}$ derived in Subsec. 3.1.2, see Eq. (154). For the forthcoming discussions, it is useful to consider the spin tensor (154) in the large mass ($z \gg 1$ or $m \gg T$) regime, which is equivalent to neglecting the relativistic corrections in Eq. (154). This is particularly interesting in the context of low-energy heavy-ion experiments, where the mass of the particles ($\Lambda$ hyperons) is much higher than the system's temperature. Within this limit the thermodynamic coefficient $\mathcal{B}_{(0)} \sim -T^3 e^{-z}\sqrt{z}$ is small compared to $\mathcal{N}_{(0)} \sim T^3 e^{-z}z^{3/2}$ and thus can be neglected. Therefore, Eq. (154) takes the form[226,241]

$$S_{z\gg 1}^{\alpha,\beta\gamma} = \cosh(\xi)\mathcal{N}_{(0)}U^\alpha[\omega^{\beta\gamma} + 2U^\delta U^{[\beta}\omega^{\gamma]}_{\ \delta}]. \tag{176}$$

## 4. Propagation Properties of Spin Polarization

In this section, we study the propagation properties of the spin polarization components of spin hydrodynamics developed in the previous section through linear perturbation analysis. The idea behind this analysis is to observe the behavior of the spin components and their dependence on the thermodynamic parameters after they are perturbed in the longitudinal direction. We find that the longitudinal spin components do not propagate, however, the transverse spin components propagate in an analogous way to EM waves. Material presented here may be found in Refs. 241 and 248.

### 4.1. *Properties and parametrizations of the spin polarization tensor*

As the spin polarization tensor $\omega_{\mu\nu}$ is a second-rank antisymmetric tensor, it is convenient to decompose it with respect to fluid four-velocity $U^\mu$ in terms of two electric-like ($\kappa^\mu$) and magnetic-like ($\omega^\mu$) four-vectors as follows[225],[vv]:

$$\omega_{\mu\nu} = \kappa_\mu U_\nu - \kappa_\nu U_\mu + \epsilon_{\mu\nu\alpha\beta} U^\alpha \omega^\beta, \qquad (177)$$

with the following properties arising from it:

- Components of four-vectors $\kappa^\mu$ and $\omega^\mu$ parallel to $U^\mu$ have no contribution to the right-hand side of Eq. (177). Hence, $\kappa^\mu$ and $\omega^\mu$ satisfy the following constraints"

$$\kappa \cdot U = 0, \quad \omega \cdot U = 0. \qquad (178)$$

- One can obtain $\kappa_\mu$ and $\omega_\mu$ from $\omega_{\mu\nu}$ using the relations

$$\kappa_\mu = \omega_{\mu\alpha} U^\alpha, \quad \omega_\mu = \frac{1}{2} \epsilon_{\mu\alpha\beta\gamma} \omega^{\alpha\beta} U^\gamma. \qquad (179)$$

- Each of the four-vectors, $\kappa^\mu$ and $\omega^\mu$, has three independent components. Hence, they together constitute the same number of independent components as $\omega_{\mu\nu}$.

For mathematical convenience we introduce a basis $I \in \{U, X, Y, Z\}$ formed by a set of mutually orthogonal four-vectors: $U$, $X$, $Y$ and $Z$ satisfying the following normalization conditions:

$$U \cdot U = 1, \quad X \cdot X = Y \cdot Y = Z \cdot Z = -1. \qquad (180)$$

The four-vectors $X$, $Y$ and $Z$ span the space transverse to $U$ and may be obtained by canonical boost transformation with four-velocity $U$ of the LRF forms[292–294]

$$X^\alpha_{\text{LRF}} = (0, 1, 0, 0), \quad Y^\alpha_{\text{LRF}} = (0, 0, 1, 0), \quad Z^\alpha_{\text{LRF}} = (0, 0, 0, 1). \qquad (181)$$

---

[vv]We use terms "electric-like" and "magnetic-like" with respect to four-vectors $\kappa^\mu$ and $\omega^\mu$ to highlight their similarity to the components present in analogous decomposition of the Faraday tensor in electromagnetism[291]

$$F_{\mu\nu} = E_\mu U_\nu - E_\nu U_\mu + \epsilon_{\mu\nu\alpha\beta} U^\alpha B^\beta.$$

Using the four-vector basis, $\kappa^\mu$ and $\omega^\mu$ can be decomposed in terms of Lorentz-scalar spin coefficients $C_{\boldsymbol{\kappa}} = (C_{\kappa X}, C_{\kappa Y}, C_{\kappa Z})$ and $C_{\boldsymbol{\omega}} = (C_{\omega X}, C_{\omega Y}, C_{\omega Z})$[ww] as follows:

$$\kappa^\alpha = C_{\kappa X} X^\alpha + C_{\kappa Y} Y^\alpha + C_{\kappa Z} Z^\alpha, \quad \omega^\alpha = C_{\omega X} X^\alpha + C_{\omega Y} Y^\alpha + C_{\omega Z} Z^\alpha. \tag{182}$$

Decompositions (182) allow us to rewrite the spin polarization tensor equation (177) in the form

$$\begin{aligned} \omega_{\alpha\beta} = {} & 2(C_{\kappa X} X_{[\alpha} U_{\beta]} + C_{\kappa Y} Y_{[\alpha} U_{\beta]} + C_{\kappa Z} Z_{[\alpha} U_{\beta]}) \\ & + \epsilon_{\alpha\beta\gamma\delta} U^\gamma (C_{\omega X} X^\delta + C_{\omega Y} Y^\delta + C_{\omega Z} Z^\delta). \end{aligned} \tag{183}$$

In the laboratory (LAB) frame, one can have different parametrization[xx] of the spin polarization tensor consisting of electric-like, $\boldsymbol{e} = (e^1, e^2, e^3)$, and magnetic-like, $\boldsymbol{b} = (b^1, b^2, b^3)$, components as[226]

$$\omega_{\alpha\beta} = \begin{bmatrix} 0 & e^1 & e^2 & e^3 \\ -e^1 & 0 & -b^3 & b^2 \\ -e^2 & b^3 & 0 & -b^1 \\ -e^3 & -b^2 & b^1 & 0 \end{bmatrix}. \tag{184}$$

## 4.2. *Spin tensor for general statistics*

The spin tensor in Eq. (175) has been derived using the equilibrium distribution function (162) for the MJ statistics. In what follows, we extend the spin tensor to other statistics so that we can study the polarization of other species of particles produced in the experiments. For that purpose, we first extend the definition of the distribution function (162).

Let us start by considering the general distribution function

$$f_{\text{eq}}^\sigma \equiv f_{\text{eq}}^\sigma(y_\sigma), \tag{185}$$

with

$$y_\sigma = y_{\sigma;0} + y_{\text{spin}}, \quad y_{\sigma;0} = \beta p \cdot U - \sigma\xi, \quad y_{\text{spin}} = -\frac{1}{2}\omega^{\mu\nu} s_{\mu\nu}, \tag{186}$$

where $\beta$ is the inverse of temperature and $\sigma = +1(-1)$ represent particles (anti-particles).

Assuming $y_{\text{spin}} \ll y_{\sigma;0}$, we have

$$f_{\text{eq}}^\sigma(y_\sigma) = f_{\text{eq}}^\sigma(y_{\sigma;0}) + f_{\text{eq}}^{\sigma\prime}(y_{\sigma;0}) y_{\text{spin}} + \cdots, \tag{187}$$

where the derivative is done at vanishing spin polarization

$$f_{\text{eq}}^{\prime\sigma}(y_{\sigma;0}) = -\sigma\left(\frac{\partial f_{\text{eq}}^\sigma}{\partial \xi}\right)_\beta = \frac{1}{p \cdot U}\left(\frac{\partial f_{\text{eq}}^\sigma}{\partial \beta}\right)_\xi. \tag{188}$$

---

[ww]Note that all the spin components are dimensionless and are functions of space–time coordinates.
[xx]We followed the sign conventions of Ref. 291.

Thus, the spin tensor (174) can be written for an arbitrary statistics as

$$S^{\lambda,\mu\nu} = -\frac{2\mathfrak{s}^2}{3m^2}\sum_{\sigma=\pm}\int \mathrm{dP}\, p^\lambda f'^\sigma_{\mathrm{eq}}(m^2\omega^{\mu\nu} + 2p^\alpha p^{[\mu}\omega^{\nu]}_\alpha), \tag{189}$$

where we can write the integral of $p^\lambda f^{\sigma\prime}_{\mathrm{eq}}$ as

$$2\sum_{\sigma=\pm}\int \mathrm{dP}\, p^\lambda f^{\sigma\prime}_{\mathrm{eq}} = a_1 U^\lambda, \quad \text{with } a_1 = -\left(\frac{\partial \mathcal{N}}{\partial \xi}\right)_\beta. \tag{190}$$

Factor 2 appears due to spin degeneracy.

Furthermore, the integral of $p^\lambda p^\alpha p^\mu f'^\sigma_{\mathrm{eq}}$ can be decomposed as

$$2\sum_{\sigma=\pm}\int \mathrm{dP} f^{\sigma\prime}_{\mathrm{eq}} p^\lambda p^\alpha p^\mu = a_2 U^\lambda U^\alpha U^\mu + b_2(U^\lambda\Delta^{\alpha\mu} + U^\alpha\Delta^{\lambda\mu} + U^\mu\Delta^{\lambda\alpha}), \tag{191}$$

where thermodynamic coefficients $a_2$ and $b_2$ can be evaluated by contracting the above expression with $U_\lambda U_\alpha U_\mu$ and $U_\lambda g_{\alpha\mu}$. In this way, one gets

$$a_2 = \left(\frac{\partial \mathcal{E}}{\partial \beta}\right)_\xi, \quad a_2 + 3b_2 = -m^2\left(\frac{\partial \mathcal{N}}{\partial \xi}\right)_\beta, \tag{192}$$

respectively.

Using Eqs. (190) and (191) in Eq. (189) and comparing with Eq. (175) shows that $a_1 = -(3/\mathfrak{s}^2)(\mathcal{A}_1 + \mathcal{A}_3)$, $a_2 = (3m^2/2\mathfrak{s}^2)(\mathcal{A}_3 - 2\mathcal{A}_1)$ and $b_2 = -(3m^2/2\mathfrak{s}^2)\mathcal{A}_3$. Hence, one can have yet another decomposition of Eqs. (155) and (156) as

$$S^{\alpha,\beta\gamma}_{\mathrm{PH}} = (\mathcal{A}_1 + \mathcal{A}_3)U^\alpha\omega^{\beta\gamma}, \tag{193}$$

$$S^{\alpha,\beta\gamma}_\Delta = (2\mathcal{A}_1 - \mathcal{A}_3)U^\alpha U^\delta U^{[\beta}\omega^{\gamma]}_\delta + \mathcal{A}_3(\Delta^{\alpha\delta}U^{[\beta}\omega^{\gamma]}_\delta + U^\alpha\Delta^{\delta[\beta}\omega^{\gamma]}_\delta$$
$$+ U^\delta\Delta^{\alpha[\beta}\omega^{\gamma]}_\delta), \tag{194}$$

where $\mathcal{A}_1$ and $\mathcal{A}_3$ for general statistics are

$$\mathcal{A}_1 = \frac{\mathfrak{s}^2}{9}\left[\left(\frac{\partial \mathcal{N}}{\partial \xi}\right)_\beta - \frac{2}{m^2}\left(\frac{\partial \mathcal{E}}{\partial \beta}\right)_\xi\right], \quad \mathcal{A}_3 = \frac{2\mathfrak{s}^2}{9}\left[\left(\frac{\partial \mathcal{N}}{\partial \xi}\right)_\beta + \frac{1}{m^2}\left(\frac{\partial \mathcal{E}}{\partial \beta}\right)_\xi\right]. \tag{195}$$

One can check that above equations reduce to Eqs. (159) for MJ statistics.

### 4.3. *Dispersion relation of spin-wave velocity*

As mentioned in Sec. 3, due to the assumption of small polarization, conservation laws for net baryon current (145) as well as for energy and momentum (152) are independent of the spin dynamics,[229,86] hence, may be treated as a perfect-fluid background for the latter. As a result, the analysis of propagation of perturbations at the level of background will lead to a widely known spectrum of sound waves where the sound speed is defined as[157,149,249,250]

$$c^2_s = \left(\frac{\partial \mathcal{P}}{\partial \mathcal{E}}\right)_\mathcal{N} + \frac{\mathcal{N}}{\mathcal{E} + \mathcal{P}}\left(\frac{\partial \mathcal{P}}{\partial \mathcal{N}}\right)_\mathcal{E}, \tag{196}$$

In consequence, the propagation of perturbations at the level of spin tensor conservation law can be studied independently.

We will start with the spin tensor decomposition, derived for the general statistics, shown in Eqs. (193) and (194) along with the thermodynamic coefficients (195).

At the level of spin tensor, background fluid is at rest, hence, $U^\mu = g^{t\mu}$. In addition, we also consider that the system is homogeneous in the transverse plane. Due to the small polarization limit considered, we can also assume that the background fluid can be unpolarized.

Therefore, Eqs. (193) and (194) reduce to

$$
\begin{aligned}
S_{\rm PH}^{\alpha,\mu\nu} &= (\mathcal{A}_1 + \mathcal{A}_3)g^{t\alpha}\omega^{\mu\nu}, \\
S_{\Delta}^{\alpha,\mu\nu} &= 2(\mathcal{A}_1 - 2\mathcal{A}_3)g^{t\alpha}g^{t[\mu}\omega^{\nu]t} + \mathcal{A}_3(g^{t[\mu}\omega^{\nu]\alpha} + g^{\alpha[\mu}\omega^{\nu]t} - g^{t\alpha}\omega^{\mu\nu}),
\end{aligned}
\tag{197}
$$

and their divergence read

$$
\begin{aligned}
\partial_\alpha S_{\rm PH}^{\alpha,\mu\nu} &= (\mathcal{A}_1 + \mathcal{A}_3)\partial_t\omega^{\mu\nu}, \\
\partial_\alpha S_{\Delta}^{\alpha,\mu\nu} &= (2\mathcal{A}_1 - 3\mathcal{A}_3)g^{t[\mu}\partial_t\omega^{\nu]t} + \mathcal{A}_3(\partial^{[\mu}\omega^{\nu]t} - \partial_t\omega^{\mu\nu} + g^{t[\mu}\partial_z\omega^{[\nu]z}).
\end{aligned}
\tag{198}
$$

Considering cases such as $\{\mu = 0, \nu = i\}$ and $\{\mu = i, \nu = j\}$, and using the fact that $S_{\rm GLW}^{\alpha,\beta\gamma} = S_{\rm PH}^{\alpha,\beta\gamma} + S_{\Delta}^{\alpha,\beta\gamma}$, see Eq. (154), we obtain

$$
\begin{aligned}
\partial_\alpha S_{\rm GLW}^{\alpha,ti} &= \partial_\alpha S_{\rm PH}^{\alpha,ti} + \partial_\alpha S_{\Delta}^{\alpha,ti} = \mathcal{A}_3\left(\partial_t\omega^{ti} + \frac{1}{2}\partial_z\omega^{iz}\right), \\
\partial_\alpha S_{\rm GLW}^{\alpha,ij} &= \partial_\alpha S_{\rm PH}^{\alpha,ij} + \partial_\alpha S_{\Delta}^{\alpha,ij} = \mathcal{A}_1\partial_t\omega^{ij} + \mathcal{A}_3\partial^{[i}\omega^{j]t}.
\end{aligned}
\tag{199}
$$

Using Eq. (183), one can write the spin polarization tensor in the fluid rest frame as

$$
\omega^{ti} = -C_{\kappa i}, \quad \omega^{ij} = -\epsilon^{tijk}C_{\omega k},
\tag{200}
$$

and the conservation law for spin, $\partial_\alpha S_{\rm GLW}^{\alpha,\beta\gamma}(x) = 0$, gives

$$
\begin{aligned}
\partial_t C_{\kappa X} - \frac{1}{2}\partial_z C_{\omega Y} = 0, \quad \partial_t C_{\omega Y} + \frac{\mathcal{A}_3}{2\mathcal{A}_1}\partial_z C_{\kappa X} = 0, \\
\partial_t C_{\kappa Y} + \frac{1}{2}\partial_z C_{\omega X} = 0, \quad \partial_t C_{\omega X} - \frac{\mathcal{A}_3}{2\mathcal{A}_1}\partial_z C_{\kappa Y} = 0.
\end{aligned}
\tag{201}
$$

Since $\partial_t C_{\kappa Z} = \partial_t C_{\omega Z} = 0$, we find that the longitudinal spin components do not propagate. Hence, the spin degrees of freedom propagate as transverse waves analogous to EM waves.[291,248]

All four transverse spin components follow the same wave-like equation as

$$
\begin{aligned}
(\partial_t^2 - c_{\rm spin}^2\partial_z^2)C_{\kappa X} = 0, \quad (\partial_t^2 - c_{\rm spin}^2\partial_z^2)C_{\kappa Y} = 0, \\
(\partial_t^2 - c_{\rm spin}^2\partial_z^2)C_{\omega X} = 0, \quad (\partial_t^2 - c_{\rm spin}^2\partial_z^2)C_{\omega Y} = 0,
\end{aligned}
\tag{202}
$$

where $c_{\rm spin}$ is the speed of the spin wave expressed as

$$
c_{\rm spin}^2 = -\frac{1}{4}\frac{\mathcal{A}_3}{\mathcal{A}_1} = \frac{1}{4}\frac{(\partial\mathcal{E}/\partial T)_\xi - z^2(\partial\mathcal{N}/\partial\xi)_T}{(\partial\mathcal{E}/\partial T)_\xi + \frac{z^2}{2}(\partial\mathcal{N}/\partial\xi)_T},
\tag{203}
$$

which in the high temperature ($z \to 0$) limit gives $c_{\text{spin}} = 1/2$. We emphasize that the relation for $c_{\text{spin}}$ in Eq. (203) holds true for an arbitrary statistics.

### 4.3.1. *The case of MJ distribution*

We now move on to find the form of Eq. (203) for MJ statistics. In this case, the distribution reads

$$f_{\text{eq}}^{\sigma} = \frac{g_s}{(2\pi)^3} e^{-\beta U \cdot p + \sigma \xi}, \tag{204}$$

with $g_s = 2$ representing the spin degeneracy. The net baryon current ($N^{\mu}$) and energy–momentum tensor ($T^{\mu\nu}$) can be evaluated in the following way:

$$N^{\mu} = \sum_{\sigma} \sigma \int dP\, p^{\mu} f_{\text{eq}}^{\sigma}, \quad T^{\mu\nu} = \sum_{\sigma} \int dP\, p^{\mu} p^{\nu} f_{\text{eq}}^{\sigma}. \tag{205}$$

The above equations reproduce the relations given in Eqs. (168) and (170), respectively. The derivatives of net baryon density ($\mathcal{N}$) and energy density ($\mathcal{E}$) with respect to $\xi$ and $\beta$ give, respectively,

$$\begin{aligned}
\left(\frac{\partial \mathcal{N}}{\partial \xi}\right)_{\beta} &= \frac{2m^2 T}{\pi^2} \cosh(\xi) K_2(z), \\
\left(\frac{\partial \mathcal{E}}{\partial \beta}\right)_{\xi} &= -\frac{2m^3 T^2}{\pi^2} \cosh(\xi)[z K_2(z) + 3 K_3(z)].
\end{aligned} \tag{206}$$

Thus, $\mathcal{A}_1$ and $\mathcal{A}_3$ in Eq. (195) for the MJ statistics are

$$\mathcal{A}_1 = \frac{4\mathfrak{s}^2 m T^2}{3\pi^2} \cosh(\xi)\left[K_3(z) + \frac{z}{2} K_2(z)\right], \quad \mathcal{A}_3 = -\frac{4\mathfrak{s}^2 m T^2}{3\pi^2} \cosh(\xi) K_3(z), \tag{207}$$

in agreement with Eqs. (159) where $\mathfrak{s}^2 = 3/4$. Substituting the above equations in Eq. (203) gives $c_{\text{spin}}^2$ for (ideal) MJ gas

$$c_{\text{spin}}^2|_{\text{MJ}} = -\frac{1}{4} \frac{\mathcal{A}_3}{\mathcal{A}_1} = \frac{1}{4}\left[\frac{K_3(z)}{K_3(z) + \frac{z}{2} K_2(z)}\right], \tag{208}$$

which depends only on the parameter $z = m/T$, see Fig. 7.

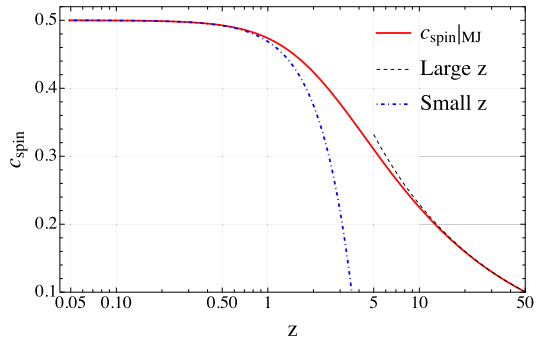

Fig. 7. The speed of the spin wave $c_{\text{spin}}$ for MJ statistics as a function of $z = m/T$ together with their respective asymptotic forms.

*Asymptotic limit*

To understand the asymptotic behavior of $c_{\text{spin}}|_{\text{MJ}}$, it is useful to discuss its asymptotic properties for the nonrelativistic ($z \gg 1$) and ultra-relativistic ($z \ll 1$) cases.

The modified Bessel functions can be expanded in the following way for large values of $z$[289]:

$$K_\nu(z) = \sqrt{\frac{\pi}{2z}} e^{-z} \sum_{k=0}^{\infty} \frac{a_k(\nu)}{z^k}, \quad \text{with } a_k(\nu) = \frac{(\frac{1}{2} - \nu)_k (\frac{1}{2} + \nu)_k}{(-2)^k k!}. \tag{209}$$

Thus

$$\begin{aligned}
K_2(z) &= \sqrt{\frac{\pi}{2z}} e^{-z} \left( 1 + \frac{15}{8z} + \frac{105}{128z^2} + \cdots \right), \\
K_3(z) &= \sqrt{\frac{\pi}{2z}} e^{-z} \left( 1 + \frac{35}{8z} + \frac{945}{128z^2} + \cdots \right).
\end{aligned} \tag{210}$$

Using Eqs. (210) in Eq. (208), we obtain $c_{\text{spin}}|_{\text{MJ}}$ for large values of $z$

$$c_{\text{spin}}|_{\text{MJ}(z \gg 1)} \simeq \frac{1}{\sqrt{2z}}. \tag{211}$$

Similarly, the modified Bessel functions of the second kind $K_n(z)$ can be expanded for small values of $z$ as[289]

$$\begin{aligned}
K_n(z) = \frac{1}{2} \left( \frac{z}{2} \right)^{-n} \sum_{k=0}^{n-1} \frac{(n-k-1)!}{k!} \left( -\frac{z^2}{4} \right)^k + (-1)^{n+1} \ln \left( \frac{z}{2} \right) I_n(z) \\
+ \frac{(-1)^n}{2} \left( \frac{z}{2} \right)^n \sum_{k=0}^{\infty} [\psi(k+1) + \psi(n+k+1)] \frac{(z^2/4)^k}{k!(n+k)!},
\end{aligned} \tag{212}$$

with $\psi(z) = \Gamma'(z)/\Gamma(z)$ being the digamma function, whereas $I_n(z)$ are the modified Bessel functions of the first kind

$$I_n(z) = \left( \frac{z}{2} \right)^n \sum_{k=0}^{\infty} \frac{(z^2/4)^k}{k!(n+k)!}. \tag{213}$$

Leading-order terms of $K_2(z)$ and $K_3(z)$ come from the first term of Eq. (212)

$$K_2(z) = \frac{2}{z^2} - \frac{1}{2} + O(z^2), \quad K_3(z) = \frac{8}{z^3} - \frac{1}{z} + O(z), \tag{214}$$

which, after using in Eq. (208), gives $c_{\text{spin}}|_{\text{MJ}}$ for small values of $z$

$$c_{\text{spin}}|_{\text{MJ}(z \ll 1)} = \frac{1}{2} \left[ 1 - \frac{z^2}{16} + O(z^4) \right]. \tag{215}$$

Figure 7 shows the comparison between the exact $c_{\text{spin}}|_{\text{MJ}}$ (208) and its asymptotic expressions (211) and (215). It can be observed, as expected, that at $z = 0$ the magnitude of $c_{\text{spin}} = 1/2$ which eventually vanishes for large values of $z$.

### 4.3.2. *The case of FD distribution*

FD gas is modeled using the distribution[295,296]

$$f_{\text{eq}}^\sigma = \frac{g_s}{8\pi^3} \frac{1}{e^{(\beta p \cdot U - \sigma \xi)} + 1}. \tag{216}$$

In this case, the net baryon density $\mathcal{N}$, energy density $\mathcal{E}$ and pressure $\mathcal{P}$ are evaluated, respectively, as[297]

$$\begin{pmatrix} \mathcal{N} \\ \mathcal{E} \\ \mathcal{P} \end{pmatrix} = \frac{1}{\pi^2} \sum_\sigma \int_m^\infty dE p \begin{pmatrix} \sigma E \\ E^2 \\ \frac{1}{3} p^2 \end{pmatrix} \frac{1}{e^{(\beta E - \sigma \xi)} + 1}, \tag{217}$$

where $p$ is the momentum of the particle. For $\xi < z = \beta m$, we can expand FD factor $[e^{(\beta E - \sigma \xi)} + 1]^{-1}$ as

$$\frac{1}{e^{(\beta E - \sigma \xi)} + 1} = \sum_{\ell=1}^\infty (-1)^{\ell+1} e^{-(\ell \beta E - \ell \sigma \xi)}, \tag{218}$$

which then allows $\mathcal{N}$ and $\mathcal{E}$ to have the forms

$$\begin{aligned}
\mathcal{N} &= \frac{2m^2 T}{\pi^2} \sum_{\ell=1}^\infty \frac{(-1)^{\ell+1}}{\ell} \sinh(\ell\xi) K_2(\ell z), \\
\mathcal{E} &= \frac{2m^2 T^2}{\pi^2} \sum_{\ell=1}^\infty \frac{(-1)^{\ell+1}}{\ell^2} \cosh(\ell\xi) [\ell z K_1(\ell z) + 3 K_2(\ell z)],
\end{aligned} \tag{219}$$

respectively. Putting $\ell = 1$ in Eq. (219) gives the results for the MJ statistics given in Eqs. (142) and (148). The derivatives of $\mathcal{E}$ with respect to $\beta$, and of $\mathcal{N}$ with respect to $\xi$ are

$$\begin{aligned}
\left( \frac{\partial \mathcal{E}}{\partial \beta} \right)_\xi &= -\frac{1}{\pi^2 \beta} \sum_\sigma \int_m^\infty dE E^2 \frac{3p + \frac{E^2}{p}}{e^{(\beta E - \sigma \xi)} + 1}, \\
\left( \frac{\partial \mathcal{N}}{\partial \xi} \right)_\beta &= \frac{1}{\pi^2 \beta} \sum_\sigma \int_m^\infty dE \frac{p + \frac{E^2}{p}}{e^{(\beta E - \sigma \xi)} + 1}.
\end{aligned} \tag{220}$$

However, for the case when $|\xi| < z$, the above integrals can be calculated using Eq. (218) allowing $\mathcal{A}_1$ and $\mathcal{A}_3$ to take the forms

$$\begin{aligned}
\mathcal{A}_1 &= \frac{4\mathfrak{s}^2 m T^2}{3\pi^2} \sum_{\ell=1}^\infty \frac{(-1)^{\ell+1}}{\ell} \cosh(\ell\xi) \left[ K_3(\ell z) + \frac{\ell z}{2} K_2(\ell z) \right], \\
\mathcal{A}_3 &= -\frac{4\mathfrak{s}^2 m T^2}{3\pi^2} \sum_{\ell=1}^\infty \frac{(-1)^{\ell+1}}{\ell} \cosh(\ell\xi) K_3(\ell z).
\end{aligned} \tag{221}$$

These results reduce to Eqs. (207) for $\ell = 1$. Therefore, for FD gas $c_{\mathrm{spin}}^2$ becomes[yy]

$$c_{\mathrm{spin}}^2|_{\mathrm{FD}} = -\frac{1}{4}\frac{\mathcal{A}_3}{\mathcal{A}_1} = \frac{1}{4}\frac{\sum_{\ell=1}^{\infty}\frac{(-1)^{\ell+1}}{\ell}\cosh(\ell\xi)K_3(\ell z)}{\sum_{\ell=1}^{\infty}\frac{(-1)^{\ell+1}}{\ell}\cosh(\ell\xi)\left[K_3(\ell z) + \frac{\ell z}{2}K_2(\ell z)\right]}. \tag{222}$$

*Asymptotic limit*

Now, let us derive the nonrelativistic and relativistic limits for $c_{\mathrm{spin}}|_{\mathrm{FD}}$. In the non-relativistic limit, we use large $z$ expansion (210) where the terms with $\ell > 1$ are replaced by $K_n(\ell z) \sim e^{-\ell z}/\sqrt{\ell z}$. Thus, for large $z$ values the expression of $c_{\mathrm{spin}}|_{\mathrm{FD}(z \gg 1)}$ agrees with $c_{\mathrm{spin}}|_{\mathrm{MJ}(z \gg 1)}$, see Eq. (211).

In the relativistic limit, we use Eq. (214). Using the notation $S_n = \sum_{\ell=1}^{\infty} \times \frac{(-1)^{\ell+1}}{\ell^n}\cosh(\ell\xi)$, the coefficients $\mathcal{A}_1$ and $\mathcal{A}_3$ can be expressed as

$$\mathcal{A}_1 = \frac{32\mathfrak{s}^2 T^4}{3\pi^2 m^2}[S_4 + O(z^4)], \quad \mathcal{A}_3 = -\frac{32\mathfrak{s}^2 T^4}{3\pi^2 m^2}\left[S_4 - \frac{z^2}{8}S_2 + O(z^4)\right]. \tag{223}$$

Now using $S_4 = (7\pi^4 + 30\pi^2\xi^2 + 15\xi^4)/720$ and $S_2 = (\pi^2 + 3\xi^2)/12$, we obtain $c_{\mathrm{spin}}^2$ for the relativistic case as

$$\begin{aligned}
c_{\mathrm{spin}}^2|_{\mathrm{FD}(z \ll 1)} &= \frac{1}{4}\left[1 - \frac{z^2}{8}\frac{S_2}{S_4} + O(z^4)\right] \\
&= \frac{1}{4}\left[1 - \frac{15z^2}{2}\frac{\pi^2 + 3\xi^2}{7\pi^4 + 30\pi^2\xi^2 + 15\xi^4} + O(z^4)\right]. \tag{224}
\end{aligned}$$

Figure 8 shows the comparison between the exact $c_{\mathrm{spin}}|_{\mathrm{FD}}$ (222) and its asymptotic expressions (211) and (224). Here, also at $z = 0$, $c_{\mathrm{spin}} = 1/2$ which vanishes for large $z$.

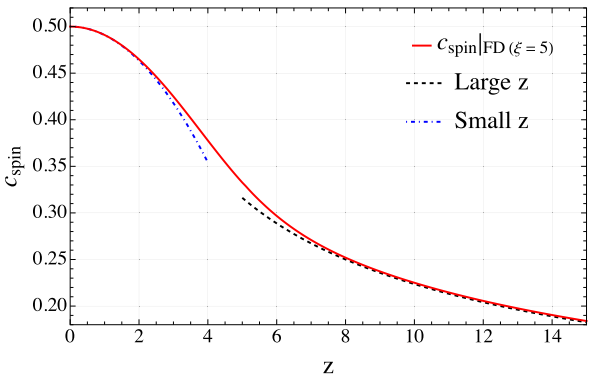

Fig. 8. The speed of the spin wave $c_{\mathrm{spin}}$ for FD statistics as a function of $z = m/T$ together with their respective asymptotic forms.

[yy]Equation (222) is true for $|\xi| < z$. For $|\xi| > z$, it diverges and then the asymptotic expression in Eq. (224), obtained in the following section, must be used.

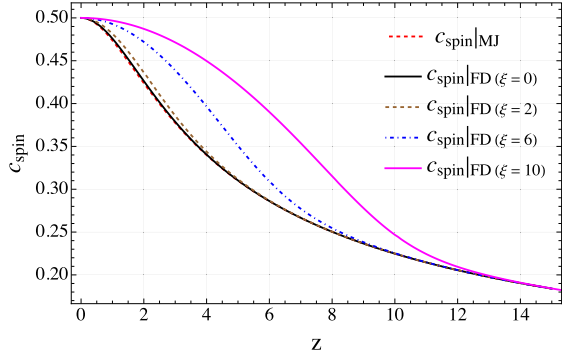

Fig. 9. Comparison between $c_{\mathrm{spin}}|_{\mathrm{MJ}}$ and $c_{\mathrm{spin}}|_{\mathrm{FD}}$ for various values of $\xi$. The MJ results are obtained using Eq. (208), while the FD results are obtained using Eqs. (222) and (224) when $z > |\xi|$ and $z < |\xi|$, respectively.

Figure 9 shows the comparison between the exact expressions of $c_{\mathrm{spin}}|_{\mathrm{MJ}}$ and $c_{\mathrm{spin}}|_{\mathrm{FD}}$ for different values of $\xi$. In summary, it is observed that $c_{\mathrm{spin}}$ is monotonically decreasing with $z$ from 0 to $\frac{1}{2}$ where the lower bound is for low-temperature limit (nonrelativistic limit) while the upper bound is for high temperatures.

*Degenerate limit*
Another interesting case which is relevant for the FD statistics is the degenerate limit, i.e. low temperature and high baryon density regime ($T \to 0$ and $\mu_B > m$), in which we have

$$\mathcal{N} = \frac{p_F^3}{3\pi^2}, \quad \mathcal{E} = \frac{1}{8\pi^2}\left[p_F \mu_B(p_F^2 + \mu_B^2) + m^4 \ln\left(\frac{m}{p_F + \mu_B}\right)\right], \qquad (225)$$

with $p_F = \sqrt{\mu_B^2 - m^2}$.

In this case $\mathcal{E} \equiv \mathcal{E}(\mu_B)$, and the derivatives of $\mathcal{E}$ and $\mathcal{N}$ with respect to $\beta$ and $\xi$, respectively, are

$$\begin{aligned}
\left(\frac{\partial \mathcal{E}}{\partial \beta}\right)_\xi &= -\frac{\mu_B}{\beta}\frac{\partial \mathcal{E}}{\partial \mu_B} = -\frac{\mu_B^3 \sqrt{\mu_B^2 - m^2}}{\pi^2 \beta}, \\
\left(\frac{\partial \mathcal{N}}{\partial \xi}\right)_\beta &= \frac{\mu_B \sqrt{\mu_B^2 - m^2}}{\pi^2 \beta} = -\frac{1}{\mu_B^2}\left(\frac{\partial \mathcal{E}}{\partial \beta}\right)_\xi.
\end{aligned} \qquad (226)$$

Hence, using the above equations in Eq. (203) leads to the expression of $c_{\mathrm{spin}}$ in the degenerate limit

$$c_{\mathrm{spin}}^2|_{\mathrm{FD(deg)}} = \frac{1}{4}\left(\frac{\xi^2 - z^2}{\xi^2 + z^2/2}\right), \qquad (227)$$

which is valid when $\xi \gg z$ or in other words $\mu_B \gg m$. Moreover, for the degenerate gas the expression of pressure can be written as

$$\mathcal{P} = \frac{1}{24\pi^2}\left[p_F \mu_B(5p_F^2 - 3\mu_B^2) - 3m^4 \ln\frac{m}{p_F + \mu_B}\right], \qquad (228)$$

which gives us the expression for the sound speed

$$c_s^2|_{\text{deg}} = \frac{\partial \mathcal{P}}{\partial \mathcal{E}} = \frac{1}{3}\left(\frac{\xi^2 - z^2}{\xi^2}\right). \tag{229}$$

Interestingly, we can establish a relationship between the spin-wave velocity (227) and the sound velocity (229) for the degenerate FD gas as

$$c_{\text{spin}}^2|_{\text{FD(deg)}} = \frac{1}{2}\left[\frac{c_s^2|_{\text{deg}}}{1 - c_s^2|_{\text{deg}}}\right]. \tag{230}$$

### 4.4. *Linear and circular polarization of spin waves*

For Eqs. (201), one can obtain the linearly polarized solutions for the components $C_{\boldsymbol{\kappa}}$ and $C_{\boldsymbol{\omega}}$ with $C_0$ as the real amplitude of the wave

$$
\begin{aligned}
C_{\boldsymbol{\kappa}} &= C_0 \text{Re}[e^{-ik(c_{\text{spin}}t-z)}](\hat{\boldsymbol{e}}_1 \cos(\theta) + \hat{\boldsymbol{e}}_2 \sin(\theta)), \\
C_{\boldsymbol{\omega}} &= 2c_{\text{spin}} C_0 \text{Re}[e^{-ik(c_{\text{spin}}t-z)}](\hat{\boldsymbol{e}}_1 \sin(\theta) - \hat{\boldsymbol{e}}_2 \cos(\theta)),
\end{aligned}
\tag{231}
$$

where $\theta$ is the inclination angle with respect to the $x$-axis. One can find the relation

$$C_{\boldsymbol{\omega}} = 2c_{\text{spin}} \hat{\boldsymbol{n}} \times C_{\boldsymbol{\kappa}}, \tag{232}$$

with $\hat{\boldsymbol{n}} = \hat{\boldsymbol{e}}_3$ being the direction of the wave propagation. One can observe that Eq. (232) is analogous to $\boldsymbol{H} = c\hat{\boldsymbol{n}} \times \boldsymbol{D}$ from electromagnetism[291] with $c$ denoting the speed of light.

Consequently, circularly polarized waves read

$$
\begin{aligned}
C_{\boldsymbol{\kappa};R/L} &= \frac{1}{\sqrt{2}} C_0 \text{Re}[e^{-ik(c_{\text{spin}}t-z)}(\hat{\boldsymbol{e}}_1 \cos(\theta) \pm i\hat{\boldsymbol{e}}_2 \sin(\theta))], \\
C_{\boldsymbol{\omega};R/L} &= \sqrt{2} C_0 c_{\text{spin}} \text{Re}[e^{-ik(c_{\text{spin}}t-z)}(\hat{\boldsymbol{e}}_1 \sin(\theta) \mp i\hat{\boldsymbol{e}}_2 \cos(\theta))],
\end{aligned}
\tag{233}
$$

which also satisfies Eq. (232).

## 5. Spin Polarization of Emitted Particles

In Sec. 3, we formulated the framework of relativistic perfect-fluid hydrodynamics with spin where the evolution of the background parameters is obtained through the conservation of net baryon current (145) and energy–momentum tensor (152) and the evolution of the spin polarization components is obtained through the conservation of spin tensor (161). Starting with certain initial conditions specified at a fixed space-like hypersurface, from the above hydrodynamic equations we may determine the dynamics of the thermodynamic and hydrodynamic fields in the future space–time region. However, it is known very well that when the system expands and dilutes, the mean free path of its constituents (particles) increases which gives rise to the decoupling of the particles. Eventually, the interactions are so weak that their momenta and spin get "frozen". This process is commonly known as *freeze-out*.

Using the values of hydrodynamic fields at the freeze-out, we can calculate the mean spin polarization per particle which can be compared with the experimental results.

In this section, we present details required to calculate the momentum-dependent and momentum-averaged mean spin polarization that will be used further for the numerical modeling of spin polarization of $\Lambda(\bar{\Lambda})$ hyperons in the subsequent sections. Parts presented in this section may be found in Ref. 241.

### 5.1. *PL four-vector*

To obtain the information about the spin of individual particles at the freeze-out we need to introduce PL four-vector, defined as[298,299]

$$\Pi_\mu = -\frac{1}{2}\epsilon_{\mu\nu\alpha\beta}J^{\nu\alpha}p^\beta, \tag{234}$$

where $p$ is the particle four-momentum and $J^{\nu\alpha}$ is total angular momentum which can be calculated by averaging the total angular momentum density over the three-dimensional volume

$$J^{\mu\nu} = \int d\Sigma_\lambda J^{\lambda,\mu\nu}. \tag{235}$$

Decomposing $J^{\nu\alpha}$ in Eq. (234) in terms of orbital $(x^\nu p^\alpha - x^\alpha p^\nu)$ and spin $(S^{\nu\alpha})$ parts gives

$$\begin{aligned}
\Pi_\mu &= -\frac{1}{2}\epsilon_{\mu\nu\alpha\beta}(x^\nu p^\alpha - x^\alpha p^\nu + S^{\nu\alpha})p^\beta \\
&= -\frac{1}{2}\epsilon_{\mu\nu\alpha\beta}(x^\nu p^\alpha - x^\alpha p^\nu)p^\beta - \frac{1}{2}\epsilon_{\mu\nu\alpha\beta}S^{\nu\alpha}p^\beta \\
&= -\frac{1}{2}\epsilon_{\mu\nu\alpha\beta}S^{\nu\alpha}p^\beta, \tag{236}
\end{aligned}$$

where, due to the presence of Levi-Civita symbol, orbital part vanishes and PL four-vector is linked directly to the spin angular momentum, describing spin states of the particle. Obviously, $p$ in Eq. (236) is orthogonal to $\Pi$.

### 5.2. *Phase-space density of the PL four-vector*

As the QGP is strongly coupled, we use relativistic hydrodynamics as an effective framework to describe its properties. However, due to expansion the interactions between the fluid constituents gradually cease, making the system weakly coupled. The adequate description of the subsequent dynamics of the particles is provided by transport theory. To make this transition possible, one is required to provide a prescription to describe the process of transforming the fluid elements to particles at the switching hypersurface $\Sigma$. A commonly used approach is the so-called Cooper–Frye formula,[300] which is based on the assumption of the continuity of the particle current at $\Sigma$, see, for instance, Ref. 301 for details.

Below we develop a respective freeze-out procedure for the spin of the particles using a similar reasoning to the one of Cooper–Frye approach. Using Eq. (234) we first calculate, using the invariant total angular momentum density ($J^{\lambda,\nu\alpha}$), the phase-space density of the PL four-vector in the volume $\Delta\Sigma$ and divide this quantity by the momentum density of all particles and antiparticles produced at the surface $\Delta\Sigma$. This ratio provides the information of average spin polarization per particle at the freeze-out hypersurface.

The phase-space density of PL four-vector in the volume $\Delta\Sigma$ is written as[159,226]

$$E_p \frac{d\Delta\Pi_\mu(x,p)}{d^3p} = -\frac{1}{2}\epsilon_{\mu\nu\alpha\beta}\Delta\Sigma_\lambda(x)E_p\frac{dJ^{\lambda,\nu\alpha}(x,p)}{d^3p}\frac{p^\beta}{m}. \tag{237}$$

Here, $J^{\lambda,\nu\alpha} = L^{\lambda,\nu\alpha} + S^{\lambda,\nu\alpha} = x^\nu T^{\lambda\alpha} - x^\alpha T^{\lambda\nu} + S^{\lambda,\nu\alpha}$. Levi-Civita symbol forces the orbital part ($L^{\lambda,\nu\alpha}$) to vanish, and we obtain

$$E_p \frac{d\Delta\Pi_\mu(x,p)}{d^3p} = -\frac{1}{2}\epsilon_{\mu\nu\alpha\beta}\Delta\Sigma_\lambda(x)E_p\frac{dS^{\lambda,\nu\alpha}(x,p)}{d^3p}\frac{p^\beta}{m}. \tag{238}$$

The term $E_p dS^{\lambda,\nu\alpha}/d^3p$ on the right-hand side of (238), within the GLW formulation, using Eq. (154), can be written as[229]

$$E_p \frac{dS^{\lambda,\nu\alpha}_{\text{GLW}}(x)}{d^3p} = \frac{\cosh(\xi)}{(2\pi)^3}e^{-\beta\cdot p}p^\lambda\left(\omega^{\nu\alpha} + \frac{2}{m^2}p^{[\nu}\omega^{\alpha]}_{\ \delta}p^\delta\right). \tag{239}$$

Note that $\zeta \to 0$ has been assumed herein and $\hbar = 1$. Putting Eq. (239) in Eq. (238), we get the phase-space density as

$$E_p \frac{d\Delta\Pi_\mu(x,p)}{d^3p} = -\frac{\cosh(\xi)}{2m(2\pi)^3}\epsilon_{\mu\nu\alpha\beta}\Delta\Sigma_\lambda p^\lambda e^{-\beta\cdot p}\left(\omega^{\nu\alpha}p^\beta + \frac{2}{m^2}p^{[\nu}\omega^{\alpha]}_{\ \delta}p^\delta p^\beta\right)$$

$$= -\frac{\cosh(\xi)}{(2\pi)^3 m}\Delta\Sigma_\lambda p^\lambda e^{-\beta\cdot p}\left(\omega^\star_{\mu\beta}p^\beta + \frac{1}{m^2}\epsilon_{\mu\nu\alpha\beta}p^{[\nu}\omega^{\alpha]}_{\ \delta}p^\delta p^\beta\right). \tag{240}$$

In the above equation, the second term in the bracket will vanish due to the Levi-Civita symbol, and we also used $\frac{1}{2}\epsilon_{\mu\nu\alpha\beta}\omega^{\nu\alpha} = \omega^\star_{\mu\beta}$. Thus, we obtain[zz]

$$E_p \frac{d\Pi^*_\mu(p)}{d^3p} = -\frac{1}{(2\pi)^3 m}\int\cosh(\xi)\Delta\Sigma_\lambda p^\lambda e^{-\beta\cdot p}(\omega^\star_{\mu\beta}p^\beta)^*, \tag{241}$$

which has to be integrated over volume $\Delta\Sigma_\lambda$ to obtain momentum density of the total value of the PL four-vector. The asterisk $()^*$ indicates that the quantity is boosted to the particle rest frame (PRF).[226,302,aaa]

---

[zz] The spin tensor (and total angular momentum) is a charge-conjugation even operator, therefore, in thermodynamic equilibrium, thermal effects cannot distinguish between the internal (spin) charge of the particles and the antiparticles. Hence, the direction of the spin polarization vector of particles and anti-particles will be the same, which is not the case if we introduce EM fields in the system.

[aaa] In the experiments, $\Lambda(\bar{\Lambda})$ hyperon spin polarization is measured in the rest frame of the decaying particle, hence, to have a comparison of our results with the experimental data, we must Lorentz transform the quantity $\omega^\star_{\mu\beta}p^\beta$ to the PRF.

The four-momentum $p^\alpha = (E_p, p_x, p_y, p_z)$ can be parametrized as

$$E_p = m_T \cosh(y_p), \quad p_x = p_T \cos(\phi_p), \quad p_y = p_T \sin(\phi_p), \quad p_z = m_T \sinh(y_p), \quad (242)$$

where $m_T$ is the transverse mass, $y_p$ is the momentum rapidity, $p_T$ is the transverse momentum and $\phi_p$ is the azimuthal angle. Using trigonometric formulae, from the momentum parametrization (242), we can get known relations for $p_T$ and $m_T$

$$p_T = \sqrt{p_x^2 + p_y^2}, \quad m_T = \sqrt{E_p^2 - p_z^2} = \sqrt{m^2 + p_T^2}. \quad (243)$$

## 5.3. *Mean spin polarization per particle*

To calculate the average spin polarization per particle, we first need to define the total particle number current as[226]

$$\mathcal{N}^\mu(x) = \int \mathrm{dP} \; \mathrm{dS} p^\mu [f_{\mathrm{eq}}^+(x,p,s) + f_{\mathrm{eq}}^-(x,p,s)], \quad (244)$$

where using the equilibrium function (162), after some algebraic manipulations, in the limit of $\omega_{\alpha\beta} \ll 1$ we get

$$\mathcal{N}^\mu(x) = 2 \cosh(\xi) \int \mathrm{dP} p^\mu e^{-p\cdot\beta} \int \mathrm{dS} \left(1 + \frac{1}{2}\omega_{\alpha\beta} s^{\alpha\beta}\right)$$
$$= 4 \cosh(\xi) \int \mathrm{dP} p^\mu e^{-p\cdot\beta}, \quad (245)$$

where we have used the identities $\int \mathrm{dS} = 2$ and $\int \mathrm{dS} s^{\alpha\beta} = 0$.[86] Note that Eq. (244) is the total particle current, i.e. sum of the number of particles and antiparticles, which is different from the net baryon current (165) that is coming from the subtraction of the number of baryons and number of antibaryons.

The momentum density of all particles and antiparticles being emitted from the hypersurface can be written as[229]

$$E_p \frac{d\mathcal{N}(p)}{d^3p} = \frac{4}{(2\pi)^3} \int \cosh(\xi) \Delta\Sigma_\lambda p^\lambda e^{-\beta\cdot p}. \quad (246)$$

Then, the ratio of the total PL four-vector (241) and the momentum density of particles and antiparticles (246) gives us the average spin polarization per particle as a function of momentum coordinates, defined as

$$\langle \pi_\mu \rangle_p = \frac{E_p \frac{d\Pi_\mu^*(p)}{d^3p}}{E_p \frac{d\mathcal{N}(p)}{d^3p}}. \quad (247)$$

It should be noted that $\langle \pi_\mu \rangle_p \langle \pi^\mu \rangle_p$ is a Lorentz-invariant quantity and the time component of $\langle \pi_\mu \rangle_p$ must vanish as in PRF, $\langle \pi_\mu \rangle_p p_*^\mu = \langle \pi_0 \rangle_p m = 0$. After integrating over momentum variables, we arrive at the momentum-averaged spin polarization as

$$\langle \pi_\mu \rangle = \frac{\int dP \langle \pi_\mu \rangle_p E_p \frac{d\mathcal{N}(p)}{d^3p}}{\int dP E_p \frac{d\mathcal{N}(p)}{d^3p}} \equiv \frac{\int d^3p \frac{d\Pi_\mu^*(p)}{d^3p}}{\int d^3p \frac{d\mathcal{N}(p)}{d^3p}}. \tag{248}$$

The second equality in (248) is obtained using Eqs. (246) and (247). Results coming from Eqs. (247) and (248) will be compared with the experimental data.

## 6. Modeling of the Spin Polarization Dynamics

After formulating relativistic perfect-fluid hydrodynamics with spin in Sec. 3, in this section, we will make use of it to study the dynamics of transversely homogeneous spin-polarizable matter with boost-invariant and boost-invariance-violating longitudinal (with respect to beam) expansion. For this purpose, we first tensor decompose the equations of motion for the background and spin. Subsequently, we use them to study different physical systems and draw some conclusions for spin polarization observables introduced in Sec. 5.

### 6.1. Basis-vector decomposition of evolution equations

In this section, we tensor decompose conservation laws for background and spin derived in Sec. 3 using the four-vector basis introduced in Subsec. 4.1. Resulting evolution equations will be subsequently used in the following sections to study spin polarization dynamics of systems respecting certain space–time symmetries. Details of this section are based mainly on Ref. 241.

6.1.1. (3+1)-dimensional parametrization of the four-vector basis

One can write a general (3+1)-dimensional parametrization of the fluid flow four-vector in Minkowski coordinates ($\mu$) in the following form[292–294]:

$$U^\mu = (U_0 \cosh(\Phi), U_x, U_y, U_0 \sinh(\Phi)), \tag{249}$$

with the quantities $U_0$, $U_x$ and $U_y$ defined as

$$U_0 = \cosh(\theta_\perp), \quad U_x = U_\perp \cos(\varphi), \quad U_y = U_\perp \sin(\varphi), \tag{250}$$

respectively. Here, $\Phi, \theta_\perp, \varphi$ are functions of $\tau, x, y, \eta$ in the Milne coordinate system, where $\tau = \sqrt{t^2 - z^2}$ is the longitudinal proper time and $\eta = \frac{1}{2} \ln[(t + z)/(t - z)]$ is the space–time rapidity. We adopt convention where the longitudinal fluid rapidity $\Phi = \vartheta(\tau, x, y, \eta) + \eta$ with $\vartheta$ denoting the deviation of longitudinal fluid rapidity from the boot-invariant one; we will come back to the discussion of the boot-invariant expansion in Subsec. 6.2. Our choice makes the parametrization (249) convenient for describing systems rapidly evolving along the longitudinal ($z$) direction. Such a situation takes place in ultra-relativistic heavy-ion collision experiments, where the particle production in central rapidity (also known as mid-rapidity) region is approximately boost-invariant along $z$ (beam) direction,[303] meaning $\vartheta \approx 0$.

Using trigonometric relations we find that $U_0 = \sqrt{1 + U_x^2 + U_y^2}$ and $U_\perp = \sqrt{U_x^2 + U_y^2} = \sinh(\theta_\perp)$. The remaining three basis vectors which span the space transverse to $U$ have the form

$$
\begin{aligned}
X^\mu &= \left( U_\perp \cosh(\Phi), \frac{U_0 U_x}{U_\perp}, \frac{U_0 U_y}{U_\perp}, U_\perp \sinh(\Phi) \right), \\
Y^\mu &= \left( 0, -\frac{U_y}{U_\perp}, \frac{U_x}{U_\perp}, 0 \right), \quad Z^\mu = (\sinh(\Phi), 0, 0, \cosh(\Phi)),
\end{aligned}
\tag{251}
$$

which, as mentioned in Sec. 4, result from canonical boost transformation $\Lambda_\nu^\mu(U^\lambda)$ (with $U$ given in (249)) applied to LRF forms (181).

Before moving ahead, it is also convenient to introduce some notation for the directional derivatives and divergences of the basis vectors, which will turn out handy in the subsequent sections. In particular, directional derivatives for the flow $U^\alpha$ (249) and space-like basis vectors $X^\alpha$, $Y^\alpha$ and $Z^\alpha$ (251) are denoted as

$$
\begin{aligned}
U^\alpha \partial_\alpha &= U \cdot \partial \equiv (\bullet), \quad X^\alpha \partial_\alpha = X \cdot \partial \equiv (\blacksquare), \\
Y^\alpha \partial_\alpha &= Y \cdot \partial \equiv (\square), \quad Z^\alpha \partial_\alpha = Z \cdot \partial \equiv (\circ),
\end{aligned}
\tag{252}
$$

while the divergences of the basis vectors are written as

$$
\begin{aligned}
\partial_\alpha U^\alpha &= \partial \cdot U \equiv \theta_U, \quad \partial_\alpha X^\alpha = \partial \cdot X \equiv \theta_X, \\
\partial_\alpha Y^\alpha &= \partial \cdot Y \equiv \theta_Y, \quad \partial_\alpha Z^\alpha = \partial \cdot Z \equiv \theta_Z.
\end{aligned}
\tag{253}
$$

Derivatives with respect to the Cartesian coordinates, $t$, $x$, $y$ and $z$, used above, are related to the ones with respect to Milne coordinates through the following matrix equation[238]:

$$
\begin{bmatrix} \partial_t \\ \partial_x \\ \partial_y \\ \partial_z \end{bmatrix} = \begin{bmatrix} \cosh(\eta) & 0 & 0 & -\sinh(\eta) \\ 0 & 1 & 0 & 0 \\ 0 & 0 & 1 & 0 \\ -\sinh(\eta) & 0 & 0 & \cosh(\eta) \end{bmatrix} \begin{bmatrix} \partial_\tau \\ \partial_x \\ \partial_y \\ \frac{1}{\tau} \partial_\eta \end{bmatrix}.
\tag{254}
$$

### 6.1.2. *Perfect-fluid background*

*Net baryon density conservation*

Substituting the expression for the net baryon current, see Eq. (141), in the net baryon density conservation law $\partial_\alpha N^\alpha(x) = 0$, see Eq. (145), we obtain the following equation of motion:

$$
U^\alpha \partial_\alpha \mathcal{N} + \mathcal{N} \partial_\alpha U^\alpha = \overset{\bullet}{\mathcal{N}} + \mathcal{N} \theta_U = 0,
\tag{255}
$$

where we used the notation listed in Eqs. (252) and (253). The net baryon density $\mathcal{N}$ is expressed by Eq. (142), and, in general, depends on $T$, $\mu_B$ and $m$ (recall that in the small polarization limit, which we will consider henceforth, we take $\zeta \to 0$).

*Energy and linear momentum conservation*

Contracting energy and linear momentum conservation law, $\partial_\alpha T_{\mathrm{GLW}}^{\alpha\beta}(x) = 0$, see Eq. (152), with fluid four-vector $U_\beta$ and then using perfect-fluid form of energy–momentum tensor ($T_{\mathrm{GLW}}^{\alpha\beta}(x) = (\mathcal{E} + \mathcal{P})U^\alpha U^\beta - \mathcal{P}g^{\alpha\beta}$), see Eq. (147), we obtain

$$U^\alpha \partial_\alpha \mathcal{E} + (\mathcal{E} + \mathcal{P})\partial_\alpha U^\alpha = \overset{\bullet}{\mathcal{E}} + (\mathcal{E} + \mathcal{P})\theta_U = 0, \qquad (256)$$

where the energy density $\mathcal{E}$ and pressure $\mathcal{P}$ are given by Eqs. (148). The implicit relation between equilibrium thermodynamic quantities ($\mathcal{N}, \mathcal{E}, \mathcal{P}$) expressed by their dependence on $T$, $\mu_B$ and $m$, defines the EoS for the system and employs the small polarization limit. It is equivalent to keeping terms in above equations of leading order in $\omega_{\mu\nu}$ or equivalently setting the limit $\zeta \to 0$. Effectively, it removes $\omega$ from the energy–momentum and net baryon density conservation laws and makes the background insensitive to the polarization dynamics.

Equation (256) has to be supplemented with its transverse counterpart; contracting Eq. (152) with the transverse projector ($\Delta_\beta^\mu = g_\beta^\mu - \frac{U^\mu U_\beta}{U \cdot U}$) and using Eq. (147) gives a set of three additional equations of motion

$$(\mathcal{E} + \mathcal{P})U^\beta \partial_\beta U^\alpha - \Delta^{\alpha\beta} \partial_\beta \mathcal{P} = (\mathcal{E} + \mathcal{P})\overset{\bullet}{U^\alpha} - (\partial^\alpha - U^\alpha U^\beta \partial_\beta)\mathcal{P} = 0, \qquad (257)$$

which, by taking nonrelativistic limit, may be interpreted as a relativistic generalization of the Euler equation.[142] In the subsequent sections, we will solve five partial differential equations given by Eqs. (255)–(257) for five unknowns, $T$, $\mu_B$ and three independent components of $U^\mu$, in order to determine a collective background on top of which we will study the dynamics of spin polarization. One should stress here, that, in practice, the small polarization limit makes Eqs. (255)–(257) to represent state-of-the-art formulation of relativistic perfect-fluid hydrodynamics for baryon-charged fluids.[301]

### 6.1.3. *Collective dynamics of spin polarization*

As discussed in Sec. 4, due to the antisymmetry of the spin polarization tensor $\omega$, it is convenient to work with its parametrization (177) in terms of electric-like and magnetic-like components, $C_{\boldsymbol{\kappa}}$ and $C_{\boldsymbol{\omega}}$, which behave as scalars under Lorentz boosts and rotations. To derive equations of motion for the components $C_{\boldsymbol{\kappa}}$ and $C_{\boldsymbol{\omega}}$, we will use the form of the spin tensor given by (158) with the general spin polarization tensor decomposition (183) in the angular momentum conservation law (161). The final equations of motion will be then obtained by projecting (161) on the basis tensors: $U_\beta X_\gamma$, $U_\beta Y_\gamma$, $U_\beta Z_\gamma$, $Y_\beta Z_\gamma$, $X_\beta Z_\gamma$ and $X_\beta Y_\gamma$.

The spin tensor (158) may be expressed in terms of basis four-vectors $I$, by contracting it with all their possible combinations. This leads to its following form:

$$S_{\text{GLW}}^{\alpha,\beta\gamma} = \sum_{i=x,y,z} S_{\alpha_i}^{\alpha,\beta\gamma} + S_{\beta_i}^{\alpha,\beta\gamma}, \tag{258}$$

where

$$S_{\alpha_x}^{\alpha,\beta\gamma} = 2\alpha_{x1}U^\alpha U^{[\beta}X^{\gamma]} + \alpha_{x2}Y^\alpha Y^{[\beta}X^{\gamma]} + \alpha_{x2}Z^\alpha Z^{[\beta}X^{\gamma]}, \tag{259}$$

$$S_{\alpha_y}^{\alpha,\beta\gamma} = 2\alpha_{y1}U^\alpha U^{[\beta}Y^{\gamma]} + \alpha_{y2}X^\alpha X^{[\beta}Y^{\gamma]} + \alpha_{y2}Z^\alpha Z^{[\beta}Y^{\gamma]}, \tag{260}$$

$$S_{\alpha_z}^{\alpha,\beta\gamma} = 2\alpha_{z1}U^\alpha U^{[\beta}Z^{\gamma]} + \alpha_{z2}X^\alpha X^{[\beta}Z^{\gamma]} + \alpha_{z2}Y^\alpha Y^{[\beta}Z^{\gamma]}, \tag{261}$$

$$S_{\beta_x}^{\alpha,\beta\gamma} = 2\beta_{x1}(Y^\alpha U^{[\beta}Z^{\gamma]} + Z^\alpha Y^{[\beta}U^{\gamma]}) - 2\beta_{x2}U^\alpha Y^{[\beta}Z^{\gamma]}, \tag{262}$$

$$S_{\beta_y}^{\alpha,\beta\gamma} = 2\beta_{y1}(Z^\alpha U^{[\beta}X^{\gamma]} + X^\alpha Z^{[\beta}U^{\gamma]}) - 2\beta_{y2}U^\alpha Z^{[\beta}X^{\gamma]}, \tag{263}$$

$$S_{\beta_z}^{\alpha,\beta\gamma} = 2\beta_{z1}(X^\alpha U^{[\beta}Y^{\gamma]} + Y^\alpha X^{[\beta}U^{\gamma]}) - 2\beta_{z2}U^\alpha X^{[\beta}Y^{\gamma]}, \tag{264}$$

and the coefficients $\alpha$ and $\beta$ are defined as

$$\alpha_{i1} = -\left(\mathcal{A}_1 - \frac{\mathcal{A}_2}{2} - \mathcal{A}_3\right)C_{\kappa i}, \quad \alpha_{i2} = -\mathcal{A}_3 C_{\kappa i},$$
$$\beta_{i1} = \frac{\mathcal{A}_3}{2}C_{\omega i}, \quad \beta_{i2} = \mathcal{A}_1 C_{\omega i}, \tag{265}$$

where, in the small polarization limit ($\zeta \to 0$)

$$\mathcal{A}_1 = \cosh(\xi)(\mathcal{N}_{(0)} - \mathcal{B}_{(0)}), \mathcal{A}_2 = \cosh(\xi)(\mathcal{A}_{(0)} - 3\mathcal{B}_{(0)}), \quad \text{and} \quad \mathcal{A}_3 = \cosh(\xi)\mathcal{B}_{(0)},$$

see Eqs. (159).

The divergence of Eq. (258) can be expressed as a sum

$$\partial_\alpha S_{\text{GLW}}^{\alpha,\beta\gamma} = \partial_\alpha S_{\alpha_x}^{\alpha,\beta\gamma} + \partial_\alpha S_{\alpha_y}^{\alpha,\beta\gamma} + \partial_\alpha S_{\alpha_z}^{\alpha,\beta\gamma} + \partial_\alpha S_{\beta_x}^{\alpha,\beta\gamma} + \partial_\alpha S_{\beta_y}^{\alpha,\beta\gamma} + \partial_\alpha S_{\beta_z}^{\alpha,\beta\gamma} = 0, \tag{266}$$

where the subsequent summands

$$\partial_\alpha S_{\alpha_x}^{\alpha,\beta\gamma} = 2\overset{\bullet}{\alpha_{x1}} U^{[\beta}X^{\gamma]} + \overset{\square}{\alpha_{x2}} Y^{[\beta}X^{\gamma]} + \overset{\circ}{\alpha_{x2}} Z^{[\beta}X^{\gamma]} + 2\alpha_{x1}[\theta_U U^{[\beta}X^{\gamma]}$$
$$+ \overset{\bullet}{U}{}^{[\beta} X^{\gamma]} + U^{[\beta} \overset{\bullet}{X}{}^{\gamma]}] + \alpha_{x2}[\theta_Y Y^{[\beta}X^{\gamma]} + \theta_Z Z^{[\beta}X^{\gamma]} + \overset{\square}{Y}{}^{[\beta} X^{\gamma]}$$
$$+ Y^{[\beta} \overset{\square}{X}{}^{\gamma]} + \overset{\circ}{Z}{}^{[\beta} X^{\gamma]} + Z^{[\beta} \overset{\circ}{X}{}^{\gamma]}], \tag{267}$$

$$\partial_\alpha S_{\alpha_y}^{\alpha,\beta\gamma} = 2\overset{\bullet}{\alpha_{y1}} U^{[\beta}Y^{\gamma]} + \overset{\blacksquare}{\alpha_{y2}} X^{[\beta}Y^{\gamma]} + \overset{\circ}{\alpha_{y2}} Z^{[\beta}Y^{\gamma]} + 2\alpha_{y1}[\theta_U U^{[\beta}Y^{\gamma]}$$
$$+ \overset{\bullet}{U}{}^{[\beta} Y^{\gamma]} + U^{[\beta} \overset{\bullet}{Y}{}^{\gamma]}] + \alpha_{y2}[\overset{\blacksquare}{X}{}^{[\beta} Y^{\gamma]} + X^{[\beta} \overset{\blacksquare}{Y}{}^{\gamma]} + \overset{\circ}{Z}{}^{[\beta} Y^{\gamma]} + Z^{[\beta} \overset{\circ}{Y}{}^{\gamma]}$$
$$+ \theta_X X^{[\beta}Y^{\gamma]} + \theta_Z Z^{[\beta}Y^{\gamma]}], \tag{268}$$

$$\partial_\alpha S^{\alpha,\beta\gamma}_{\alpha_z} = 2\,\overset{\bullet}{\alpha_{z1}}\,U^{[\beta}Z^{\gamma]} + \overset{\blacksquare}{\alpha_{z2}}\,X^{[\beta}Z^{\gamma]} + \overset{\square}{\alpha_{z2}}\,Y^{[\beta}Z^{\gamma]} + 2\alpha_{z1}[\theta_U U^{[\beta}Z^{\gamma]} + \overset{\bullet}{U}{}^{[\beta}Z^{\gamma]}$$
$$+ U^{[\beta}\overset{\bullet}{Z}{}^{\gamma]}] + \alpha_{z2}[\overset{\blacksquare}{X}{}^{[\beta}Z^{\gamma]} + X^{[\beta}\overset{\blacksquare}{Z}{}^{\gamma]} + \overset{\square}{Y}{}^{[\beta}Z^{\gamma]} + Y^{[\beta}\overset{\square}{Z}{}^{\gamma]}$$
$$+ \theta_X X^{[\beta}Z^{\gamma]} + \theta_Y Y^{[\beta}Z^{\gamma]}], \tag{269}$$

$$\partial_\alpha S^{\alpha,\beta\gamma}_{\beta_x} = 2[\overset{\square}{\beta_{x1}}\,U^{[\beta}Z^{\gamma]} + \overset{\circ}{\beta_{x1}}\,Y^{[\beta}U^{\gamma]} + \beta_{x1}(\theta_Y U^{[\beta}Z^{\gamma]} + \theta_Z Y^{[\beta}U^{\gamma]})$$
$$- \overset{\bullet}{\beta_{x2}}\,Y^{[\beta}Z^{\gamma]} - \beta_{x2}\theta_U Y^{[\beta}Z^{\gamma]}] + 2\beta_{x1}[\overset{\square}{U}{}^{[\beta}Z^{\gamma]} + \overset{\circ}{Y}{}^{[\beta}U^{\gamma]} + U^{[\beta}\overset{\square}{Z}{}^{\gamma]}$$
$$+ Y^{[\beta}\overset{\circ}{U}{}^{\gamma]}] - 2\beta_{x2}(\overset{\bullet}{Y}{}^{[\beta}Z^{\gamma]} + Y^{[\beta}\overset{\bullet}{Z}{}^{\gamma]}), \tag{270}$$

$$\partial_\alpha S^{\alpha,\beta\gamma}_{\beta_y} = 2[\overset{\circ}{\beta_{y1}}\,U^{[\beta}X^{\gamma]} + \overset{\blacksquare}{\beta_{y1}}\,Z^{[\beta}U^{\gamma]} + \beta_{y1}(\theta_Z U^{[\beta}X^{\gamma]} + \theta_X Z^{[\beta}U^{\gamma]})$$
$$- \overset{\bullet}{\beta_{y2}}\,Z^{[\beta}X^{\gamma]} - \beta_{y2}\theta_U Z^{[\beta}X^{\gamma]}] + 2\beta_{y1}[\overset{\circ}{U}{}^{[\beta}X^{\gamma]} + \overset{\blacksquare}{Z}{}^{[\beta}U^{\gamma]} + U^{[\beta}\overset{\circ}{X}{}^{\gamma]}$$
$$+ Z^{[\beta}\overset{\blacksquare}{U}{}^{\gamma]}] - 2\beta_{y2}(\overset{\bullet}{Z}{}^{[\beta}X^{\gamma]} + Z^{[\beta}\overset{\bullet}{X}{}^{\gamma]}), \tag{271}$$

$$\partial_\alpha S^{\alpha,\beta\gamma}_{\beta_z} = 2[\overset{\blacksquare}{\beta_{z1}}\,U^{[\beta}Y^{\gamma]} + \overset{\square}{\beta_{z1}}\,X^{[\beta}U^{\gamma]} + \beta_{z1}(\theta_X U^{[\beta}Y^{\gamma]} + \theta_Y X^{[\beta}U^{\gamma]})$$
$$- \overset{\bullet}{\beta_{z2}}\,X^{[\beta}Y^{\gamma]} - \beta_{z2}\theta_U X^{[\beta}Y^{\gamma]}] + 2\beta_{z1}[\overset{\blacksquare}{U}{}^{[\beta}Y^{\gamma]} + \overset{\square}{X}{}^{[\beta}U^{\gamma]} + U^{[\beta}\overset{\blacksquare}{Y}{}^{\gamma]}$$
$$+ X^{[\beta}\overset{\square}{U}{}^{\gamma]}] - 2\beta_{z2}(\overset{\bullet}{X}{}^{[\beta}Y^{\gamma]} + X^{[\beta}\overset{\bullet}{Y}{}^{\gamma]}), \tag{272}$$

were obtained from Eqs. (259)–(264), respectively.

Finally, to derive the evolution equations for the spin polarization components $C$ (recall that $\alpha$'s and $\beta$'s depend explicitly on $C$'s), we contract the partial differential equation (266) with $U_\beta X_\gamma$, $U_\beta Y_\gamma$, $U_\beta Z_\gamma$, $Y_\beta Z_\gamma$, $X_\beta Z_\gamma$ and $X_\beta Y_\gamma$, and obtain

$$\overset{\bullet}{\alpha_{x1}} + \alpha_{x1}\theta_U + \frac{\alpha_{x2}}{2}(U\overset{\square}{Y} + U\overset{\circ}{Z}) - \frac{\alpha_{y2}}{2}U\overset{\blacksquare}{Y} - \alpha_{y1}X\overset{\bullet}{Y} - \frac{\alpha_{z2}}{2}U\overset{\square}{Z} - \alpha_{z1}X\overset{\bullet}{Z}$$
$$- \beta_{x1}(X\overset{\circ}{Z} - X\overset{\square}{Y}) + \overset{\circ}{\beta_{y1}} + \beta_{y1}(\theta_Z + X\overset{\blacksquare}{Z}) - \beta_{y2}U\overset{\bullet}{Z} - \overset{\blacksquare}{\beta_{z1}} - \beta_{z1}(\theta_Y + X\overset{\square}{Y})$$
$$+ \beta_{z2}U\overset{\bullet}{Y} = 0, \tag{273}$$

$$\overset{\bullet}{\alpha_{y1}} + \alpha_{y1}\theta_U + \frac{\alpha_{y2}}{2}(U\overset{\blacksquare}{X} + U\overset{\circ}{Z}) - \frac{\alpha_{z2}}{2}U\overset{\square}{Z} - \alpha_{z1}Y\overset{\bullet}{Z} - \frac{\alpha_{x2}}{2}U\overset{\square}{X} - \alpha_{x1}Y\overset{\bullet}{X}$$
$$- \beta_{y1}(Y\overset{\circ}{X} - Y\overset{\blacksquare}{Z}) + \overset{\square}{\beta_{z1}} + \beta_{z1}(\theta_X + Y\overset{\square}{X}) - \beta_{z2}U\overset{\bullet}{X} - \overset{\circ}{\beta_{x1}} - \beta_{x1}(\theta_Z + Y\overset{\square}{Z})$$
$$+ \beta_{x2}U\overset{\bullet}{Z} = 0, \tag{274}$$

$$\overset{\bullet}{\alpha_{z1}} + \alpha_{z1}\theta_U + \frac{\alpha_{z2}}{2}(U\overset{\blacksquare}{X} + U\overset{\square}{Y}) - \frac{\alpha_{x2}}{2}U\overset{\circ}{X} - \alpha_{x1}Z\overset{\bullet}{X} - \frac{\alpha_{y2}}{2}U\overset{\circ}{Y} - \alpha_{y1}Z\overset{\bullet}{Y}$$
$$- \beta_{z1}(Z\overset{\blacksquare}{Y} - Z\overset{\square}{X}) + \overset{\square}{\beta_{x1}} + \beta_{x1}(\theta_Y + Z\overset{\circ}{Y}) - \beta_{x2}U\overset{\bullet}{Y} - \overset{\blacksquare}{\beta_{y1}} - \beta_{y1}(\theta_X + Z\overset{\circ}{X})$$
$$+ \beta_{y2}U\overset{\bullet}{X} = 0, \tag{275}$$

$$\frac{\overset{\circ}{\alpha}_{y2}}{2} + \frac{\alpha_{y2}}{2}(\theta_Z - Z\overset{\circ}{X}) - \alpha_{y1}Z\overset{\bullet}{U} - \frac{\overset{\Box}{\alpha}_{z2}}{2} - \frac{\alpha_{z2}}{2}(\theta_Y - Y\overset{\blacksquare}{X}) + \alpha_{z1}Y\overset{\bullet}{U}$$
$$- \frac{\alpha_{x2}}{2}(Y\overset{\circ}{X} - Z\overset{\Box}{X}) + \overset{\bullet}{\beta}_{x2} + \beta_{x2}\theta_U + \beta_{x1}(Y\overset{\Box}{U} + Z\overset{\circ}{U}) - \beta_{y1}Y\overset{\blacksquare}{U} + \beta_{y2}Y\overset{\bullet}{X}$$
$$- \beta_{z1}Z\overset{\bullet}{U} + \beta_{z2}Z\overset{\bullet}{X} = 0, \tag{276}$$

$$\frac{\overset{\blacksquare}{\alpha}_{z2}}{2} + \frac{\alpha_{z2}}{2}(\theta_X - X\overset{\Box}{Y}) - \alpha_{z1}X\overset{\bullet}{U} - \frac{\overset{\circ}{\alpha}_{x2}}{2} - \frac{\alpha_{x2}}{2}(\theta_Z - Z\overset{\Box}{Y}) + \alpha_{x1}Z\overset{\bullet}{U}$$
$$- \frac{\alpha_{y2}}{2}(Z\overset{\blacksquare}{Y} - X\overset{\circ}{Y}) + \overset{\bullet}{\beta}_{y2} + \beta_{y2}\theta_U + \beta_{y1}(Z\overset{\circ}{U} + X\overset{\blacksquare}{U}) - \beta_{z1}Z\overset{\Box}{U} + \beta_{z2}Z\overset{\bullet}{Y}$$
$$- \beta_{x1}X\overset{\Box}{U} + \beta_{x2}X\overset{\bullet}{Y} = 0, \tag{277}$$

$$\frac{\overset{\Box}{\alpha}_{x2}}{2} + \frac{\alpha_{x2}}{2}(\theta_Y - Y\overset{\circ}{Z}) - \alpha_{x1}Y\overset{\bullet}{U} - \frac{\overset{\blacksquare}{\alpha}_{y2}}{2} - \frac{\alpha_{y2}}{2}(\theta_X - X\overset{\circ}{Z}) + \alpha_{y1}X\overset{\bullet}{U}$$
$$- \frac{\alpha_{z2}}{2}(X\overset{\Box}{Z} - Y\overset{\blacksquare}{Z}) + \overset{\bullet}{\beta}_{z2} + \beta_{z2}\theta_U + \beta_{z1}(X\overset{\blacksquare}{U} + Y\overset{\Box}{U}) - \beta_{x1}X\overset{\circ}{U} + \beta_{x2}X\overset{\bullet}{Z}$$
$$- \beta_{y1}Y\overset{\circ}{U} + \beta_{y2}Y\overset{\bullet}{Z} = 0, \tag{278}$$

respectively.

As, in general, Eqs. (273)–(278) are quite complex, they need to be solved numerically to determine the dynamics of $C_{\boldsymbol{\kappa}}$ and $C_{\boldsymbol{\omega}}$. Moreover, one may observe that the evolution of the background enters Eqs. (273)–(278) solely through the thermodynamic factors $\mathcal{A}_1$–$\mathcal{A}_3$ and the flow vectors $I$, and can be entirely determined in advance; we will often exploit this property hereafter. Finally, as we will see in the following sections, while, in general, all spin components are entangled with each other, see Eqs. (273)–(278), in simple physics situations the dynamics of some components decouples from others, providing opportunity to get some insights on their collective properties more easily.

In the subsequent sections, we will study a number of special physics situations advocated by phenomenological observations, whose complexity will change with the change of symmetry constraints imposed on the system.

### 6.2. *Boost-invariant and transversely homogeneous background*

Experimental measurements show that in heavy-ion collisions, at top energies, the matter produced in the central (mid-rapidity) region exhibits invariance with respect to Lorentz boosts along the beam direction.[303] Moreover, as it takes finite time for the hydrodynamic gradients to build up, the transverse expansion of the system in the vicinity of the beam axis[bbb] may be treated initially as invariant with respect to translations in the transverse plane. These two approximate symmetries, namely

---

[bbb] By vicinity we mean that the distances from the $z$-axis are much smaller than the root-mean-square radius of the hard-sphere nuclear overlap region.

boost-invariance and transverse homogeneity, (together with the $\eta \to -\eta$ parity symmetry) constitute the so-called Bjorken symmetry.[ccc] This symmetry is employed commonly in the phenomenology of heavy-ion collisions, due to its advantage of decreasing dimensionality and reducing the number of independent parameters required to describe the dynamics of the system. In particular, one can show that all thermodynamic variables are Lorentz scalars which are functions of longitudinal proper time ($\tau$) only, while the flow pattern is completely fixed by symmetry. Indeed, in this case, the basis four-vectors (249) and (251) take the forms

$$\begin{aligned} U^\alpha &= (\cosh(\eta), 0, 0, \sinh(\eta)), \quad Z^\alpha = (\sinh(\eta), 0, 0, \cosh(\eta)), \\ X^\alpha &= (0, 1, 0, 0), \quad Y^\alpha = (0, 0, 1, 0). \end{aligned} \tag{279}$$

Moreover, the directional derivatives (252) and the divergences (253) take the forms $U \cdot \partial = \partial_\tau$, $Z \cdot \partial = \frac{1}{\tau}\partial_\eta$, $X \cdot \partial = \partial_x$, $Y \cdot \partial = \partial_y$, and

$$\partial \cdot U = \frac{1}{\tau}, \tag{280}$$

respectively. One can also check that $\partial \cdot X = \partial \cdot Y = \partial \cdot Z = 0$.

In the remaining part of Subsec. 6.2, we will make use of the above expressions to study polarization dynamics of Bjorken-expanding matter. The material discussed here is partially based on Ref. 238.

### 6.2.1. *Hydrodynamic evolution*

One can check that, due to the symmetries imposed, the relativistic Euler equation (257) is trivially satisfied. On the other hand, the energy equation (256) takes the form

$$\frac{\partial \mathcal{E}}{\partial \tau} + \frac{\mathcal{E} + \mathcal{P}}{\tau} = 0. \tag{281}$$

For baryon-charged matter, Eq. (281) has to be supplemented with the conservation law for net baryon density (255), which for Bjorken-expanding matter becomes

$$\frac{\partial \mathcal{N}}{\partial \tau} + \frac{\mathcal{N}}{\tau} = 0, \tag{282}$$

and has the solution

$$\mathcal{N} = \frac{\tau_0}{\tau}\mathcal{N}_0, \tag{283}$$

where $\mathcal{N}_0$ is the initial net baryon density at the initial proper time $\tau_0$, $\mathcal{N}_0 = \mathcal{N}(\tau_0)$.

We solve Eqs. (282) and (281) numerically by choosing the conditions aiming at mimicking the situation encountered in high-energy heavy-ion collisions, where the

---

[ccc] The concept of boost-invariance was first discussed by Feynman in 1969 within the context of hadron production at high energies.[304] In 1983, Bjorken implemented boost-invariance along with homogeneity in the transverse direction (now known as Bjorken symmetry) into the equations of hydrodynamics and estimated the values of initial energy densities in the collisions.[305]

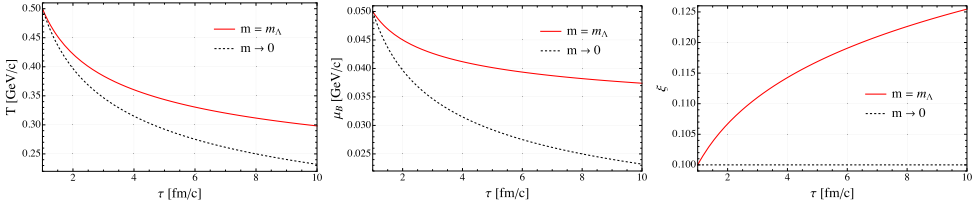

Fig. 10.  Evolution of $T$ (top panel), $\mu_B$ (middle panel) and $\xi = \mu_B/T$ (lower panel) as a function of $\tau$ for Bjorken background along with their respective massless case.

temperature is much higher than the baryon chemical potential. In such a case, we consider the initial baryon chemical potential and the initial temperature as $\mu_{B0} = \mu_B(\tau_0) = 0.05\,\text{GeV}$ and $T_0 = T(\tau_0) = 0.5\,\text{GeV}$, respectively, with $\tau_0 = 1\,\text{fm/c}$. For simplicity, we consider the system of $\Lambda$ hyperons and choose the particle mass as $m = m_\Lambda = 1.116\,\text{GeV}$.[306] The resulting evolution of the temperature $(T)$, baryon chemical potential $(\mu_B)$ and the ratio of baryon chemical potential over temperature $(\xi)$ as a function of $\tau$ is shown in Fig. 10. One can notice that, both, the temperature and baryon chemical potential, decrease with $\tau$ for massive $(m = m_\Lambda)$ as well as massless $(m \to 0)$ case. In the $m \to 0$ limit, $T$ and $\mu_B$ follow the analytical solutions $T_0(\tau_0/\tau)^{1/3}$ and $\mu_{B_0}(\tau_0/\tau)^{1/3}$, respectively, resulting in a constant $\xi$. In the massive case, the latter increases with $\tau$ as expected

$$\xi = \sinh^{-1}\left[\frac{\tau_0}{\tau}\frac{\mathcal{N}_{(0)}(\tau_0)}{\mathcal{N}_{(0)}}\sinh(\xi_0)\right], \qquad (284)$$

where $\xi_0$ and $\mathcal{N}_{(0)}(\tau_0)$ are the values of the parameters at $\tau_0$.

Due to the restrictive character of Bjorken expansion, the evolution equations for spin components (273)–(278) take the following simple form:

$$\begin{aligned}
\overset{\bullet}{\alpha}_{x1} &= -\alpha_{x1}\theta_U - \frac{\alpha_{x2}}{2}U\,\overset{\circ}{Z}, \\
\overset{\bullet}{\alpha}_{y1} &= -\alpha_{y1}\theta_U - \frac{\alpha_{y2}}{2}U\,\overset{\circ}{Z}, \\
\overset{\bullet}{\alpha}_{z1} &= -\alpha_{z1}\theta_U, \\
\overset{\bullet}{\beta}_{x2} &= -\beta_{x2}\theta_U - \beta_{x1}Z\,\overset{\circ}{U}, \\
\overset{\bullet}{\beta}_{y2} &= -\beta_{y2}\theta_U - \beta_{y1}Z\,\overset{\circ}{U}, \\
\overset{\bullet}{\beta}_{z2} &= -\beta_{z2}\theta_U,
\end{aligned} \qquad (285)$$

where one can check that $U\,\overset{\circ}{Z} = -Z\,\overset{\circ}{U} = 1/\tau$. As we observe from Eqs. (285), Bjorken symmetries prevent the spin components to couple with each other.

In Fig. 11, we show the evolution of the spin polarization components as a function of $\tau$ obtained by solving Eqs. (285) numerically. The initial value of all the components is chosen arbitrarily to be 0.1.[ddd] Transverse components $C_{\kappa X}$ and $C_{\omega X}$ decrease with, $\tau$ whereas the longitudinal components $C_{\kappa Z}$ and $C_{\omega Z}$ show opposite behavior. Due to the rotational invariance, the transverse components $C_{\kappa X}$ and $C_{\kappa Y}$ (as well as $C_{\omega X}$ and $C_{\omega Y}$) follow the same differential equations, hence, we refrain to show $C_{\kappa Y}$ and $C_{\omega Y}$ evolution in Fig. 11. We checked that the qualitative behavior of all the spin components does not change significantly with the change in the initial values of thermodynamic parameters such as $T_0$ or $\mu_{B0}$. Interestingly, in the large mass regime ($z \gg 1$), see the discussion in Subsec. 3.2.2, the quantity $\mathcal{B}_0$ can be neglected making only $\beta_{i2}$ remain nonvanishing. As a result, Eqs. (285) allow determining only the dynamics of magnetic-like components ($C_{\omega X}, C_{\omega Y}, C_{\omega Z}$), which follow the same dynamics given by equations

$$\overset{\bullet}{\beta_{i2}} = -\beta_{i2}\theta_U.$$ (286)

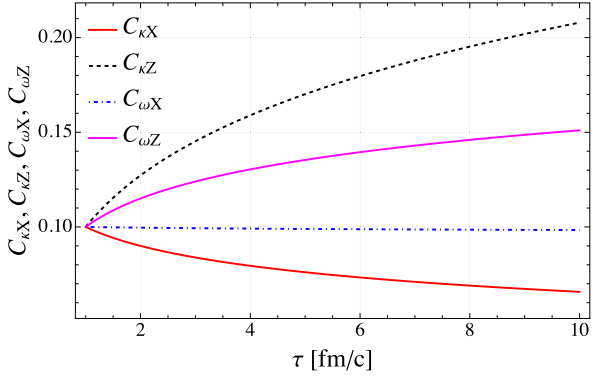

Fig. 11.   Time evolution of spin components.

The above equations have monotonically increasing solution

$$C_{\omega i} = C_{\omega i}^0 \frac{\tau_0}{\tau} \frac{\cosh(\xi_0)}{\cosh(\xi)} \frac{\mathcal{N}_{(0)}(\tau_0)}{\mathcal{N}_{(0)}},$$ (287)

where $C_{\omega i}^0$, $\xi_0$ and $\mathcal{N}_{(0)}(\tau_0)$ are the values of the parameters at $\tau_0$. The solution (287) is presented in Fig. 12. It shows that the qualitative behavior of $C_{\omega Z}$-component remains the same as in Fig. 11, whereas the behavior of $C_{\omega X}$ and $C_{\omega Y}$ becomes opposite in the $z \gg 1$ limit.

### 6.2.2.  *Angular momentum of a boost-invariant fire-cylinder*

Solutions obtained in the previous section acquire intuitive interpretation when one considers spin and orbital contributions to total angular momentum $J_{\text{FC}}^{\mu\nu} = L_{\text{FC}}^{\mu\nu} + S_{\text{FC}}^{\mu\nu}$ at the hypersurface of fixed longitudinal proper time we call a *fire-cylinder* (FC).

---

[ddd]We kept the initial value for all the spin components the same in order to observe their relative behavior with respect to each other, as well as much smaller than 1 to respect the assumed small polarization limit.

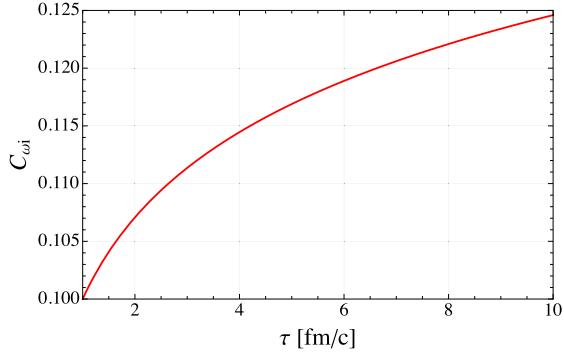

Fig. 12.   Evolution of $C_{\omega i}$ components in the $z \gg 1$ limit.

Using the fact that the total angular momentum is conserved, see Eq. (160), this will help us with initialization of the spin components in numerical simulations. For this purpose, let us consider a boost-invariant three-dimensional hypersurface defined by $\tau = \tau_{\text{FC}} = \text{const.}$ and contained within the space–time region defined by conditions $-\eta_{\text{FC}}/2 \leq \eta \leq \eta_{\text{FC}}/2$ and $(x^2 + y^2)^{1/2} \leq R_{\text{FC}}$, with constant radius $R_{\text{FC}}$, see Fig. 13. The respective hypersurface element is defined as

$$\Delta\Sigma_\mu = \tau_{\text{FC}} U_\mu dx dy d\eta, \tag{288}$$

with the flow vector given by Eq. (279).

The orbital contribution for the GLW pseudogauge can be calculated as follows:

$$
\begin{aligned}
L_{\text{FC}}^{\mu\nu} &= \int \Delta\Sigma_\lambda L^{\lambda,\mu\nu} = \int \Delta\Sigma_\lambda (x^\mu T_{\text{GLW}}^{\lambda\nu} - x^\nu T_{\text{GLW}}^{\lambda\mu}) \\
&= \int \tau_{\text{FC}} U_\lambda dx dy d\eta (x^\mu T_{\text{GLW}}^{\lambda\nu} - x^\nu T_{\text{GLW}}^{\lambda\mu}) \\
&= \tau_{\text{FC}} \pi R_{\text{FC}}^2 \int_{-\eta_{\text{FC}}/2}^{\eta_{\text{FC}}/2} \mathcal{E} d\eta (x^\mu U^\nu - x^\nu U^\mu),
\end{aligned}
\tag{289}
$$

where, in the last line, we used the fact that $U^\mu$ is the eigenvector of the energy–momentum tensor with $\mathcal{E}$ as an eigenvalue

$$T_{\text{GLW}}^{\lambda\nu} U_\lambda = \mathcal{E} U^\nu. \tag{290}$$

Putting the value of $U^\mu$ from Eq. (279) in Eq. (289) we find that $L_{\text{FC}}^{\mu\nu} = 0$. Hence, the orbital part does not contribute to the total angular momentum for Bjorken-expanding matter.

The respective contribution from the spin part can be calculated as

$$
\begin{aligned}
S_{\text{FC}}^{\mu\nu} &= \int \Delta\Sigma_\lambda S_{\text{GLW}}^{\lambda,\mu\nu} = \tau_{\text{FC}} \int dx dy \int_{-\eta_{\text{FC}}/2}^{\eta_{\text{FC}}/2} d\eta U_\lambda S_{\text{GLW}}^{\lambda,\mu\nu} \\
&= \tau_{\text{FC}} \pi R_{\text{FC}}^2 \int_{-\eta_{\text{FC}}/2}^{\eta_{\text{FC}}/2} d\eta U_\lambda S_{\text{GLW}}^{\lambda,\mu\nu}.
\end{aligned}
\tag{291}
$$

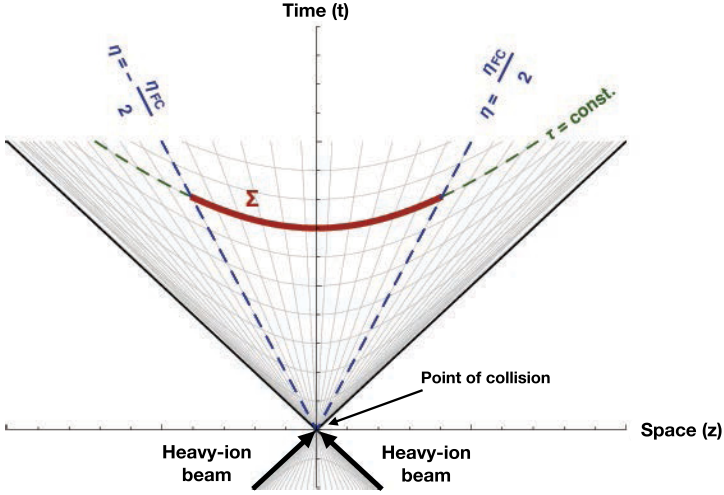

Fig. 13. Schematic diagram of the boost-invariant FC with $\Sigma$ denoting the hypersurface.

Using in Eq. (291) the relation $U_\lambda S_{\text{GLW}}^{\lambda,\mu\nu} = \mathcal{A}_3(\kappa^\mu U^\nu - \kappa^\nu U^\mu) + \mathcal{A}_1 \epsilon^{\mu\nu\beta\gamma} U_\beta \omega_\gamma$, see Eq. (158), and integrating over $\eta$ the spin angular momentum tensor takes the form

$$S_{\text{FC}}^{\mu\nu} = -S_{\text{FC}}^{\nu\mu} = \begin{bmatrix} 0 & S_{\text{FC}}^{01} & S_{\text{FC}}^{02} & S_{\text{FC}}^{03} \\ -S_{\text{FC}}^{01} & 0 & S_{\text{FC}}^{12} & S_{\text{FC}}^{13} \\ -S_{\text{FC}}^{02} & -S_{\text{FC}}^{12} & 0 & S_{\text{FC}}^{23} \\ -S_{\text{FC}}^{03} & -S_{\text{FC}}^{13} & -S_{\text{FC}}^{23} & 0 \end{bmatrix}, \tag{292}$$

with its components expressed by

$$\begin{aligned} S_{01}^{\text{FC}} &= -S_{10}^{\text{FC}} = 2\pi R_{\text{FC}}^2 \tau_{\text{FC}} \mathcal{A}_3 C_{\kappa X} \sinh(\eta_{\text{FC}}/2), \\ S_{02}^{\text{FC}} &= -S_{20}^{\text{FC}} = 2\pi R_{\text{FC}}^2 \tau_{\text{FC}} \mathcal{A}_3 C_{\kappa Y} \sinh(\eta_{\text{FC}}/2), \\ S_{03}^{\text{FC}} &= -S_{30}^{\text{FC}} = \pi R_{\text{FC}}^2 \tau_{\text{FC}} \mathcal{A}_3 C_{\kappa Z} \eta_{\text{FC}}, \\ S_{23}^{\text{FC}} &= S_{32}^{\text{FC}} = -2\pi R_{\text{FC}}^2 \tau_{\text{FC}} \mathcal{A}_1 C_{\omega X} \sinh(\eta_{\text{FC}}/2), \\ S_{13}^{\text{FC}} &= S_{31}^{\text{FC}} = 2\pi R_{\text{FC}}^2 \tau_{\text{FC}} \mathcal{A}_1 C_{\omega Y} \sinh(\eta_{\text{FC}}/2), \\ S_{12}^{\text{FC}} &= S_{21}^{\text{FC}} = -\pi R_{\text{FC}}^2 \tau_{\text{FC}} \mathcal{A}_1 C_{\omega Z} \eta_{\text{FC}}. \end{aligned} \tag{293}$$

We observe that in the boost-invariant and transversely homogeneous system different spin components $C$ are directly related to different components of spin angular momentum tensor $S_{\text{FC}}^{\mu\nu}$, which, due to $L_{\text{FC}}^{\mu\nu} = 0$, are conserved during the evolution.

At this point, it is important to emphasize the physics picture for the spin polarization evolution we have in mind, which will be followed in the analyses presented in this section. We assume that during the initial stage of the noncentral relativistic heavy-ion collision, the total angular momentum ($\boldsymbol{J}$) consists of only orbital angular momentum ($\boldsymbol{L}$), the direction of which is perpendicular to the reaction ($x - z$) plane, i.e. along the $-y$-axis,[85,91,125,133] see Fig. 2.

This derives its motivation from the experimental measurements of the spin polarization of $\Lambda(\bar\Lambda)$ hyperons where the average direction of the spin is along $-y$-axis. After the collision some fraction of the orbital angular momentum is transferred, due to initial scatterings, to the initial spin angular momentum $(\boldsymbol{S})$[307] which is, on average, along the same direction as the original total angular momentum

$$\boldsymbol{J}_{\text{initial}} = \boldsymbol{L}_{\text{initial}} = \boldsymbol{L}_{\text{final}} + \boldsymbol{S}_{\text{final}}. \tag{294}$$

This is reflected by the nonvanishing $xz$-component of spin angular momentum.[308] One may notice from Eq. (293) that the spin component $C_{\omega Y}$ is directly linked to the $xz$-component of the spin angular momentum ($S_{13}^{\text{FC}}$). Hence, the situation discussed here can be reproduced in the numerical simulation of Bjorken-expanding system by assuming nonvanishing and positive component $C_{\omega Y}(\tau_{\text{FC}} = \tau_0)$ and keeping all the other spin components zero.

### 6.2.3. *Spin polarization at freeze-out*

Having the collective dynamics of the system determined in previous sections, we now use it to calculate the spin polarization of particles emitted from the fluid at the freeze-out. Due to simplifications resulting from the Bjorken symmetries, the phase-space density of the PL four-vector (241) as well as the momentum density of all particles (246) can be calculated analytically allowing us to write the following expressions of average spin polarization (247) as a function of particle momentum, in PRF:

$$\langle \pi_\mu \rangle_p = \frac{1}{8m} \begin{bmatrix} 0 \\ \chi C_{\omega X} m_T{}^2 - 2C_{\kappa Z}p_y - \left(\dfrac{p_x p_z}{E_p+m}\right)[\chi(C_{\kappa X}p_y - C_{\kappa Y}p_x) + 2C_{\omega Z}] \\ \qquad - \dfrac{1}{E_p+m}(\chi p_x E_p(C_{\omega X}p_x + C_{\omega Y}p_y)) \\ \chi C_{\omega Y} m_T{}^2 + 2C_{\kappa Z}p_x - \left(\dfrac{p_y p_z}{E_p+m}\right)[\chi(C_{\kappa X}p_y - C_{\kappa Y}p_x) + 2C_{\omega Z}] \\ \qquad - \dfrac{1}{E_p+m}(\chi p_y E_p(C_{\omega X}p_x + C_{\omega Y}p_y)) \\ \left(\dfrac{mE_p+m_T{}^2}{E_p+m}\right)[\chi(C_{\kappa X}p_y - C_{\kappa Y}p_x) + 2C_{\omega Z}] + \dfrac{\chi m p_z(C_{\omega X}p_x + C_{\omega Y}p_y)}{E_p+m} \end{bmatrix}, \tag{295}$$

where $\chi = (K_0(\hat{m}_T) + K_2(\hat{m}_T))/(m_T K_1(\hat{m}_T))$ and $\hat{m}_T$ is the ratio of the transverse mass over temperature, $\hat{m}_T = m_T/T$. Note that the time component of $\langle \pi_\mu \rangle_p$ vanishes. Indeed, since $\langle \pi_\mu \rangle_p \langle \pi^\mu \rangle_p$ is a Lorentz-invariant quantity, one can write $\langle \pi_\mu \rangle_p p_*^\mu = \langle \pi_0 \rangle_p m = 0$. Using the numerical results of the thermodynamic parameters from Eqs. (281) and (282), and the spin components from Eqs. (285), from Eq. (295) we can obtain $x$-, $y$-and $z$-components of $\langle \pi_\mu \rangle_p$ as a function of $p_x$ and $p_y$ at mid-rapidity ($y_p = 0$). In Fig. 14, we show plots for $\langle \pi_x \rangle_p$ (top panel) and $\langle \pi_y \rangle_p$ (bottom panel) at mid-rapidity (one can check that at mid-rapidity, $\langle \pi_z \rangle_p$ vanishes). The component $\langle \pi_y \rangle_p$ has negative magnitude reflecting the initial direction of the spin in the system, whereas the component $\langle \pi_x \rangle_p$ exhibits a quadrupole structure depicting nontrivial momentum dependence of Eq. (295).

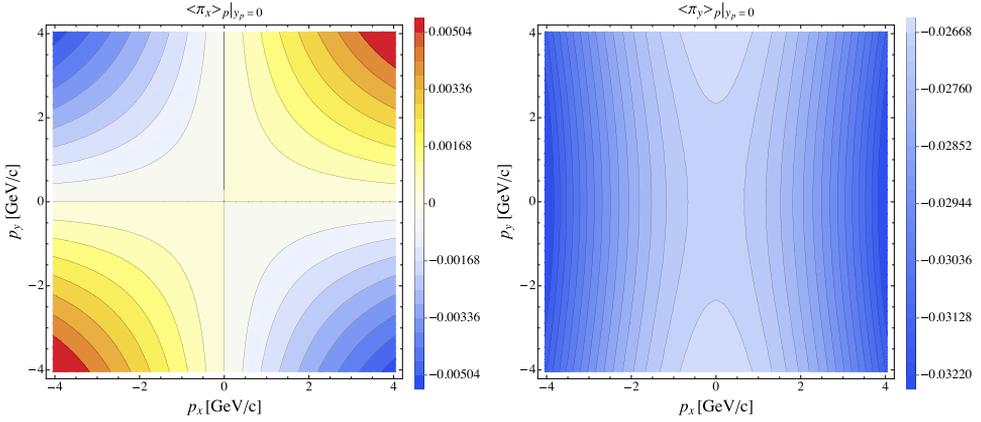

Fig. 14. The $x$-component (top panel) and $y$-component (bottom panel) of the mean spin polarization per particle at mid-rapidity as a function of $p_x$ and $p_y$ for the Bjorken-expanding matter.

We note that the results presented here do not reproduce the experimentally observed polarization measurements, which is primarily due to the symmetries involved in the Bjorken expansion. In particular, our results do not capture the quadrupole structure of the longitudinal spin polarization $\langle \pi_z \rangle_p$ at mid-rapidity,[125] which is expected to arise due to the flow gradients in the transverse plane (anisotropic flow resulting from the elliptic deformation of the system).[124] Obviously, such mechanism may be addressed only in the framework which relaxes assumption of transverse homogeneity.[eee]

## 6.3. *Nonboost-invariant and transversely homogeneous background*

The experimental results from the STAR Collaboration clearly show the decrease of the amplitude of global polarization as a function of center-of-mass energy, with its values approaching zero at top RHIC and LHC energies, see Fig. 4. This makes the low-and mid-energy collisions most interesting from the point of view of polarization phenomenology. On the other hand, the Bjorken model is known for being too restrictive on its assumptions in this region. In this section, we try to, partially, address this issue. In what follows, we extend the study done in Subsec. 6.2 by relaxing the symmetry of boost-invariance while keeping the assumption of homogeneity of the system in the transverse plane. Details of the study presented in this section may be found in Ref. 241.

---

[eee]We note that in the spin-thermal-based models quadrupole structure in the longitudinal spin polarization component, comes, albeit with an opposite sign, from the imposed coupling between vorticity and polarization.[127]

### 6.3.1. *Four-vector basis and spin polarization components*

By relaxing the boost-invariance of the produced matter in the beam direction, we need to account for possible flow gradients which may build up along the beam direction. This is done by introducing a deviation in the flow vector (279) in the following way:

$$U^\alpha = (\cosh(\Phi), 0, 0, \sinh(\Phi)), \tag{296}$$

where, as mentioned before, $\Phi = \vartheta(\tau, \eta) + \eta$, and we kept transverse components zero due to homogeneity in $x$–$y$ plane. The remaining basis vectors, see Eqs. (251), take the forms

$$X^\alpha = (0, 1, 0, 0), \quad Y^\alpha = (0, 0, 1, 0), \quad Z^\alpha = (\sinh(\Phi), 0, 0, \cosh(\Phi)). \tag{297}$$

The directional derivatives (252) and the divergences (253) read

$$U \cdot \partial = \cosh(\vartheta)\partial_\tau + \frac{\sinh(\vartheta)}{\tau}\partial_\eta, \quad Z \cdot \partial = \sinh(\vartheta)\partial_\tau + \frac{\cosh(\vartheta)}{\tau}\partial_\eta,$$
$$X \cdot \partial = \partial_x, \quad Y \cdot \partial = \partial_y, \tag{298}$$

$$\partial_\alpha U^\alpha = \frac{\cosh(\vartheta)}{\tau} + \overset{\circ}{\vartheta}, \quad \partial_\alpha Z^\alpha = \frac{\sinh(\vartheta)}{\tau} + \overset{\bullet}{\vartheta}, \tag{299}$$

respectively. Since all the scalar functions must depend now on $\tau$ and $\eta$, one has $\partial \cdot X = \partial \cdot Y = 0$.

### 6.3.2. *Background dynamics*

In the case of the nonboost-invariant expansion, equations resulting from baryon density conservation and energy–momentum conservation, (255) and (256)–(257), respectively, are used together with the expressions listed in Subsec. 6.3.1. However, unlike in the case of Bjorken expansion, see Subsec. 6.2.1, this time the relativistic Euler equation (257) has one nontrivial ($z$) component

$$(\mathcal{E} + \mathcal{P})U^\beta \partial_\beta \sinh(\Phi) - (\partial^z - \sinh(\Phi)U^\beta\partial_\beta)\mathcal{P} = 0. \tag{300}$$

As a result, we are left with three partial differential equations in $\tau$–$\eta$ space to be solved for temperature, baryon chemical potential and longitudinal fluid rapidity correction.

To model the nontrivial rapidity dependence of the initial energy density $\mathcal{E}_0(\eta) = \mathcal{E}(\tau_0, \eta)$ deposited in the collision, we introduce the profile

$$\mathcal{E}_0(\eta) = \frac{\mathcal{E}_0^c}{2}[\Theta(\eta)(\tanh(a - \eta b) + 1) + \Theta(-\eta)(\tanh(a + \eta b) + 1)], \tag{301}$$

with $a = 6.2$, $b = 1.9$ and $\Theta$ denoting the Heaviside step function.[309] Here, $\mathcal{E}_0^c = \mathcal{E}(T_0^c, \mu_{B0}^c)$ is the initial energy density at the center ($\eta = 0$) calculated from Eq. (148) at the initial central temperature $T_0^c$ and baryon chemical potential $\mu_{B0}^c$ (note that we assume $\zeta \to 0$ herein). In numerical simulations, we use $T_0^c = T(\tau_0, \eta = 0) = 0.26\,\text{GeV}$.

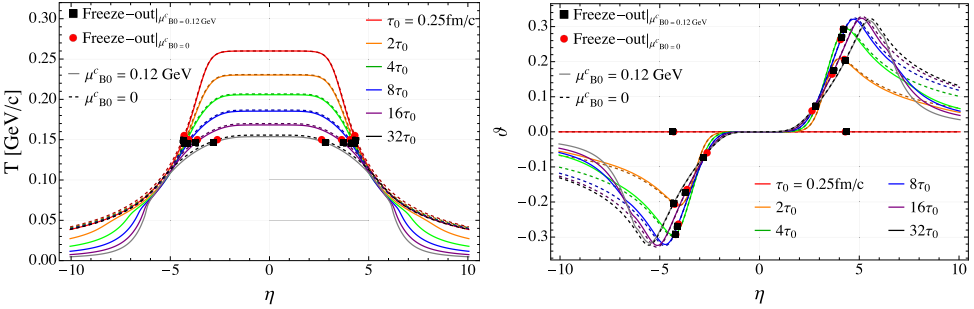

Fig. 15.   (Color online) Temperature (top panel) and fluid rapidity (bottom panel) evolution in $\eta$ at various $\tau$. Solid and dashed lines represent nonvanishing and vanishing baryon chemical potential, respectively. The black and red symbols denote freeze-out points at different times.

In the case of baryon chemical potential profile, we consider $\mu_{B0}(\eta) = \mu_{B0}^c = $ const. with two possibilities: either we choose vanishing baryon chemical potential $\mu_{B0}^c = 0$ (resembling conditions present at the high-energy experiments), or we pick $\mu_{B0}^c = 0.12\,\mathrm{GeV}$ (to address possible effects coming from baryon chemical potential present at lower energies). Throughout, we assume the initial longitudinal flow profile to have the Bjorken form $\Phi_0(\eta) = \eta$.

Figure 15 shows the evolution for $T$ (top panel) and $\vartheta$ (bottom panel) as a function of space–time rapidity for different longitudinal proper times $\tau$ with initial $\tau_0 = 0.25\,\mathrm{fm}/c$. The dashed lines represent $\mu_{B0}^c = 0$ and solid lines denote $\mu_{B0}^c = 0.12\,\mathrm{GeV}$. We find that the temperature evolution is $\eta$-even similar to the evolution of baryon chemical potential, see Fig. 16, while the evolution of the fluid rapidity is $\eta$-odd. At mid-rapidity ($\eta = 0$), temperature decreases in $\tau$, similar to the Bjorken case, see Fig. 10. Moreover, we observe that at large, $|\eta|$ the energy density gradients lead to the buildup of fluid velocity gradients. One may also notice the steep decay in $T$ and $\vartheta$ evolution at $\eta \approx \pm 5$ in the case of nonzero baryon chemical potential. Up to this effect, the baryon chemical potential has no significant effect on the evolution of the background parameters.

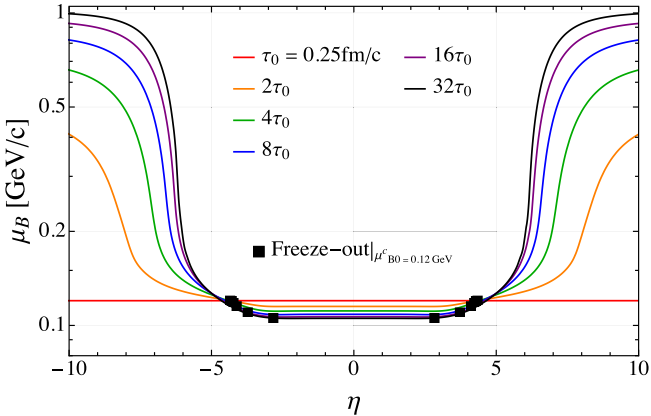

Fig. 16.   Baryon chemical potential evolution in $\eta$ at various $\tau$.

### 6.3.3. *Spin dynamics*

The dynamics in the spin sector follows from Eqs. (273)–(278) which in the boost-invariance-breaking case reduce to

$$\overset{\bullet}{\alpha}_{x1} + \overset{\circ}{\beta}_{y1} = -\alpha_{x1}\theta_U - \frac{\alpha_{x2}U\overset{\circ}{Z}}{2} - \beta_{y1}\theta_Z + \beta_{y2}U\overset{\bullet}{Z}, \tag{302}$$

$$\overset{\bullet}{\alpha}_{y1} - \overset{\circ}{\beta}_{x1} = -\alpha_{y1}\theta_U - \frac{\alpha_{y2}U\overset{\circ}{Z}}{2} + \beta_{x1}\theta_Z - \beta_{x2}U\overset{\bullet}{Z}, \tag{303}$$

$$\overset{\bullet}{\alpha}_{z1} = -\alpha_{z1}\theta_U, \tag{304}$$

$$\frac{\overset{\circ}{\alpha}_{y2}}{2} + \overset{\bullet}{\beta}_{x2} = -\frac{\alpha_{y2}\theta_Z}{2} + \alpha_{y1}Z\overset{\bullet}{U} - \beta_{x2}\theta_U - \beta_{x1}Z\overset{\circ}{U}, \tag{305}$$

$$\frac{\overset{\circ}{\alpha}_{x2}}{2} - \overset{\bullet}{\beta}_{y2} = -\frac{\alpha_{x2}\theta_Z}{2} + \alpha_{x1}Z\overset{\bullet}{U} + \beta_{y2}\theta_U + \beta_{y1}Z\overset{\circ}{U}, \tag{306}$$

$$\overset{\bullet}{\beta}_{z2} = -\beta_{z2}\theta_U, \tag{307}$$

where

$$U\overset{\circ}{Z} = -Z\overset{\circ}{U} = \cosh(\vartheta)(1 + \partial\vartheta/\partial\eta)/\tau + \sinh(\vartheta)\partial\vartheta/\partial\tau,$$

$$U\overset{\bullet}{Z} = -Z\overset{\bullet}{U} = \sinh(\vartheta)(1 + \partial\vartheta/\partial\eta)/\tau + \cosh(\vartheta)\partial\vartheta/\partial\tau.$$

Note that, in contrast to the case of Bjorken expansion, in the present case some spin components in the above equations are coupled, cf. Eqs. (285). In particular, from Eqs. (302) and (306) one observes that this is the case for $C_{\kappa X}$ and $C_{\omega Y}$. Similarly, the coupling concerns $C_{\kappa Y}$ and $C_{\omega X}$, which is evident from Eqs. (303) and (305). This does not apply to longitudinal spin components $C_{\kappa Z}$ and $C_{\omega Z}$ which evolve independently of others.

In the numerical simulations, we follow the initialization scheme of the spin components, $C$ which results from the physical considerations discussed in Subsec. 6.2.2. Since at the initial time $\vartheta_0(\eta) = 0$ (representing Bjorken flow profile), the nonvanishing $y$-component of the spin angular momentum at the initial time is related to the component $C_{\omega Y}$ and it requires $C_{\omega Y}$ being symmetric in $\eta$, see Ref. 241 for more details. Hence, it is enough, at the initial time, if we choose

$$C_{\omega Y}^0(\eta) = C_{\omega Y}(\tau_0, \eta) = \frac{d}{\cosh(\eta)}, \tag{308}$$

where $d = 0.1$, and keep all other spin components zero to obtain the $y$-component of $S_{\rm FC}^{\mu\nu}$. With the help of the following relations between the parametrizations (183) and (184) of the spin polarization tensor:

$$C_{\kappa X} = e^1 \cosh(\Phi) - b^2 \sinh(\Phi), \quad e^1 = C_{\kappa X} \cosh(\Phi) + C_{\omega Y} \sinh(\Phi), \tag{309}$$

$$C_{\kappa Y} = e^2 \cosh(\Phi) + b^1 \sinh(\Phi), \quad e^2 = C_{\kappa Y} \cosh(\Phi) - C_{\omega X} \sinh(\Phi), \tag{310}$$

$$C_{\kappa Z} = e^3, \tag{311}$$

$$C_{\omega X} = b^1 \cosh(\Phi) + e^2 \sinh(\Phi), \quad b^1 = C_{\omega X} \cosh(\Phi) - C_{\kappa Y} \sinh(\Phi), \tag{312}$$

$$C_{\omega Y} = b^2 \cosh(\Phi) - e^1 \sinh(\Phi), \quad b^2 = C_{\kappa X} \sinh(\Phi) + C_{\omega Y} \cosh(\Phi), \tag{313}$$

$$C_{\omega Z} = b^3, \tag{314}$$

the respective initial LAB frame spin components $e^1$ and $b^2$ are initialized as

$$e_0^1(\eta) = e^1(\tau_0, \eta) = d \tanh(\eta), \quad b_0^2(\eta) = b^2(\tau_0, \eta) = d. \tag{315}$$

In Fig. 17, we present results of numerical simulations for spin components $C_{\kappa X}$ and $C_{\omega Y}$. As discussed above, although initially chosen to be zero, $C_{\kappa X}$-component undergoes nontrivial evolution due to its coupling to $C_{\omega Y}$ through Eqs. (302) and (306) (obviously, all other spin components remain zero). The symmetry in $\eta$ of $C_{\kappa X}$ and $C_{\omega Y}$ remains intact throughout the evolution, which is governed by the evolution equations (302) and (306), along with the initial condition (308). One can observe effects of background evolution on spin, appearing in spin equations of motion through the thermodynamic coefficients (265). Similar to temperature, see Fig. 15, the coefficient $C_{\omega Y}$ also decreases in time, at the center, reproducing the Bjorken behavior observed in Fig. 11. However, at the edges ($\eta \approx \pm 5$), as the system enters the large mass limit regime, the spin dynamics gets reversed, which means that the magnitude of $C_{\omega Y}$ starts to increase with $\tau$. This is also the case for the spin components in the Bjorken expansion, see Fig. 12. In Fig. 17, we also observe that the presence of homogeneous nonzero, but small, baryon chemical potential is almost irrelevant for spin dynamics, which is understood since baryon chemical potential enters only through $\cosh(\xi)$ in the thermodynamic coefficients (265).

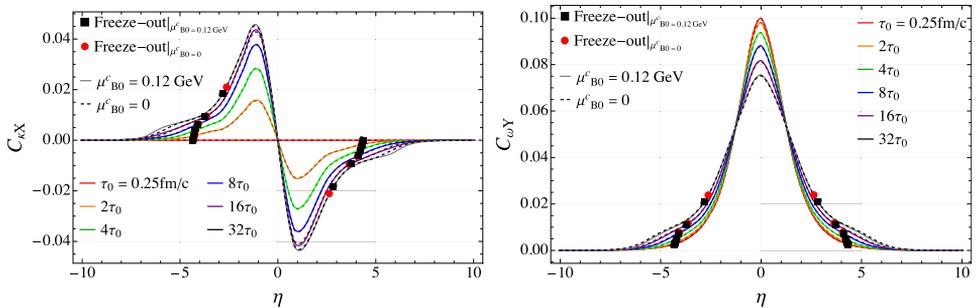

Fig. 17. Evolution of the coefficients $C_{\kappa X}$ (top panel) and $C_{\omega Y}$ (bottom panel). Solid and dashed lines represent nonvanishing and vanishing baryon chemical potential, respectively.

### 6.3.4. *Spin polarization at freeze-out*

After exploring the collective dynamics of the spin polarization components, we calculate the average spin polarization of the particles emitted at the freeze-out to assess implications of the dynamics for the measured observables. As the freeze-out times ($\tau_{\text{FO}}$) depend on the space–time rapidity of the fluid cells, see black and red symbols in Fig. 15, in the current analysis, first we need to define the freeze-out hypersurface ($\Sigma_\mu$) appearing in the phase-space density of the PL four-vector (241) accordingly. In consequence, the resulting factor $\Delta\Sigma \cdot p$ is

$$\Delta\Sigma \cdot p = m_T[\tau_{\text{FO}}(\eta)\cosh(y_p - \eta) - \tau'_{\text{FO}}(\eta)\sinh(y_p - \eta)]dxdyd\eta, \tag{316}$$

while the thermal factor has the form

$$\beta \cdot p = \frac{U \cdot p}{T} = \frac{m_T}{T}\cosh(y_p - \Phi). \tag{317}$$

Furthermore, the product of the dual polarization tensor $\omega^\star_{\mu\beta}$ with $p^\beta$ is

$$\omega^\star_{\mu\beta}p^\beta = \begin{bmatrix} (C_{\kappa X}p_y - C_{\kappa Y}p_x)\sinh(\Phi) + (C_{\omega X}p_x + C_{\omega Y}p_y)\cosh(\Phi) + C_{\omega Z}m_T\sinh(y_p) \\ -C_{\kappa Y}m_T\sinh(y_p - \Phi) - C_{\omega X}m_T\cosh(y_p - \Phi) + C_{\kappa Z}p_y \\ C_{\kappa X}m_T\sinh(y_p - \Phi) - C_{\omega Y}m_T\cosh(y_p - \Phi) - C_{\kappa Z}p_x - (C_{\kappa X}p_y - C_{\kappa Y}p_x) \\ \times\cosh(\Phi) - (C_{\omega X}p_x + C_{\omega Y}p_y)\sinh(\Phi) - C_{\omega Z}m_T\cosh(y_p) \end{bmatrix}, \tag{318}$$

which after Lorentz transformation to the PRF becomes

$$(\omega^\star_{\mu\beta}p^\beta)^* = \begin{bmatrix} 0 \\ m\alpha_p p_x p_y[C_{\kappa X}\sinh(\Phi) + C_{\omega Y}\cosh(\Phi)] \\ m\alpha_p p_y^2[C_{\kappa X}\sinh(\Phi) + C_{\omega Y}\cosh(\Phi)] \\ -m_T[C_{\kappa X}\sinh(\Phi - y_p) + C_{\omega Y}\cosh(\Phi - y_p)] \\ -m\alpha_p p_y[m_T(C_{\kappa X}\cosh(\Phi - y_p) + C_{\omega Y}\sinh(\Phi - y_p)) \\ +m(C_{\kappa X}\cosh(\Phi) + C_{\omega Y}\sinh(\Phi))] \end{bmatrix}, \tag{319}$$

where $\alpha_p \equiv 1/(m^2 + mE_p)$.[226] Note that, in the above equation, we have kept only $C_{\kappa X}$ and $C_{\omega Y}$ as these are the ones nonvanishing for the current numerical analysis.

Putting Eq. (319) in Eq. (247), we calculate the average spin polarization per particle as a function of momentum coordinates. The results of this procedure for the case of vanishing baryon chemical potential at mid-rapidity and forward rapidity are shown in Figs. 18–20. As we have shown above, see Fig. 17, the nonvanishing baryon chemical potential has very little effect on the spin dynamics. We verified that the same applies to momentum-dependent polarization, hence, we restrain from showing these results here.

Figure 18 shows the $\langle\pi_x\rangle_p$ component of the mean polarization vector which has quadrupole structure with the sign changing sequentially through the subsequent quadrants. Similar results were found in the case with Bjorken expansion, see Fig. 14 (top panel). One can notice that this quadrupole structure is arising because of the $p_x p_y$ term in the $x$-component of $(\omega^{\star}_{\mu\beta} p^{\beta})^*$, see Eq. (319). One observes that the magnitude of this component decreases with rapidity.

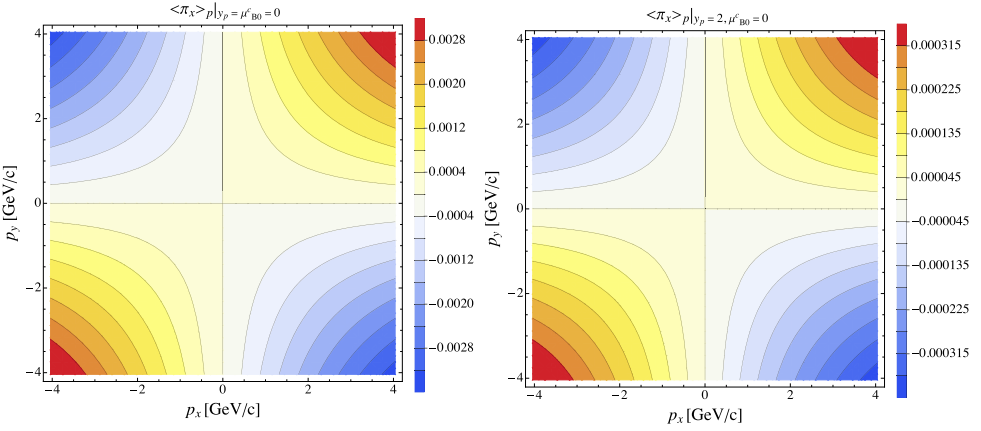

Fig. 18. The $x$-component of the mean spin polarization per particle (247) at mid-rapidity (top panel) and forward rapidity (bottom panel) as a function of $p_x$ and $p_y$ for the nonboost-invariant system with vanishing baryon chemical potential.

The $y$-component of polarization vector $\langle \pi_y \rangle_p$, shown in Fig. 19, is negative, depicting the direction of the spin angular momentum assumed in the initialization of the hydrodynamic equations. Its magnitude decreases with increasing rapidity and eventually becomes independent of $\phi_p$.

From the experimental point of view the most interesting is the polarization measured along the beam ($z$) direction, i.e. longitudinal spin polarization[125,134] which is still awaiting clear explanation. This component is given by the quantity $\langle \pi_z \rangle_p$. Using symmetry arguments in Eqs. (241) and (319), one can immediately obtain important information about the $\langle \pi_z \rangle_p$ quantity. Note that the symmetric integration range in $\eta$ makes only the contribution of $\eta$-even integrands relevant. Thus, due to the assumption of $C_{\kappa X}$ and $C_{\omega Y}$ being odd and even function of $\eta$, respectively, makes $\langle \pi_z \rangle_p$ vanish at mid-rapidity ($y_p = 0$). On the other hand, at forward rapidity we observe nontrivial longitudinal polarization pattern, see Fig. 20. We note that, similar to the Bjorken expansion case, the results shown here do not reproduce the quadrupole structure of the longitudinal spin polarization seen in the experiments, primarily due to the homogeneity in the transverse plane.

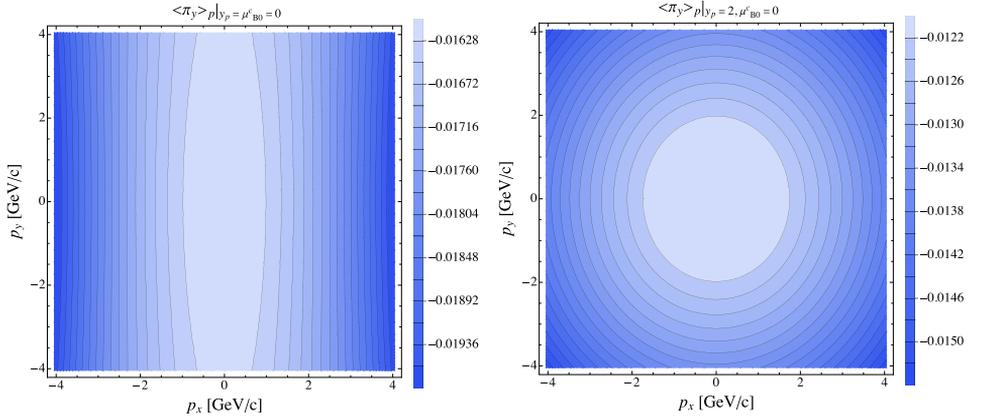

Fig. 19. Same as Fig. 18 but for $y$-component.

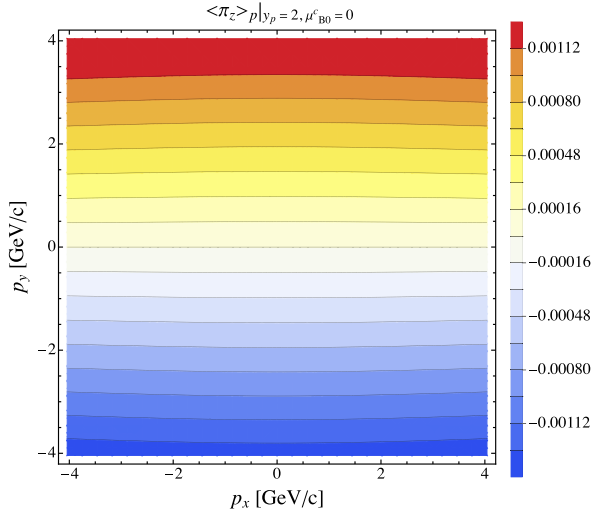

Fig. 20. The $z$-component of the mean spin polarization per particle (247) at forward rapidity as a function of $p_x$ and $p_y$ for the nonboost-invariant background with vanishing baryon chemical potential.

From the experimental viewpoint, apart from calculating multiple-differential spin polarization vector it is useful to consider integrated $\langle \pi_\mu \rangle$ components (note missing index $p$). In particular, in order to get the spin polarization $\langle \pi_\mu \rangle$, we integrate (247) over momentum coordinates.

We find that the initialization of the parameters that we adopted and the symmetry properties of the spin polarization components allow only $y$-component of $\langle \pi_\mu \rangle$ to be nonvanishing (and negative).

Figure 21 (top panel) shows the behavior of the global polarization as a function of rapidity. At mid-rapidity the magnitude of $\langle \pi_y \rangle$ component is maximal, and then it decreases in forward rapidities indicating that most of the hyperon polarization is coming from the mid-rapidity region. This behavior is qualitatively similar to other models[165] and is a subject of future experiments.[310] One can notice that the magnitude of $\langle \pi_y \rangle$ at $y_p = 0$ is qualitatively similar to the one obtained in the global polarization measurements,[85] see Fig. 4.

We also study the relation between the final polarization ($\langle \pi_y \rangle$) and the initial value of the spin component ($b_0^2$) at mid-rapidity and forward rapidity, which we show in Fig. 21 (bottom panel). We observe that $\langle \pi_y \rangle$ depends linearly on $b_0^2$ for, both, vanishing and nonvanishing baryon chemical potential. Note that for $b_0^2 = 0.1$, as considered in the current analysis, the magnitude of $\langle \pi_y \rangle$ is about 1.6% ($y_p = 0$) as depicted also in the top panel of Fig. 21, however, interestingly for $b_0^2 = 0.3$, magnitude of $\langle \pi_y \rangle$ is about 5% at mid-rapidity which, qualitatively, agrees with the magnitude of spin polarization of $\Lambda$ hyperons observed in the low-energy heavy-ion collisions.[132]

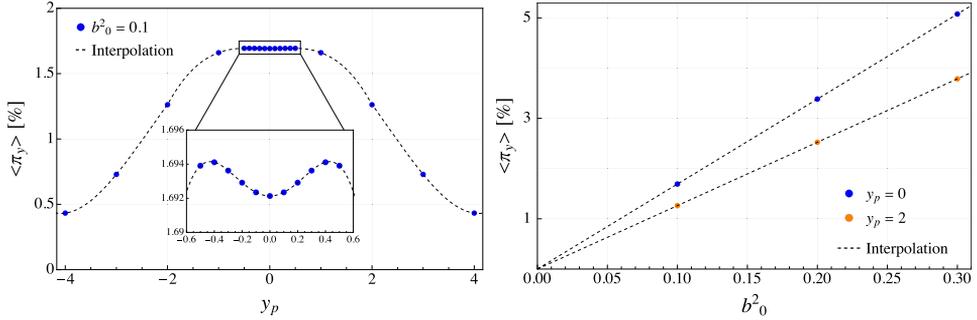

Fig. 21. Global spin polarization, $\langle \pi_y \rangle$, as a function of rapidity (top panel) and spin coefficient $b_0^2$ (bottom panel).

It is also interesting to find how spin polarization behaves with respect to transverse momentum ($p_T$) and azimuthal angle ($\phi_p$). This may provide important information about the dynamics of spin polarization in the transverse-momentum plane. Integrating Eq. (247) over transverse momentum and azimuthal angle gives global polarization as a function of $\phi_p$ and $p_T$, respectively, which can be calculated as follows[176]:

$$\langle \pi_\mu(\phi_p) \rangle = \frac{\int p_T dp_T E_p \frac{d\Pi_\mu^*(p)}{d^3 p}}{\int d\phi_p p_T dp_T E_p \frac{d\mathcal{N}(p)}{d^3 p}}, \quad \langle \pi_\mu(p_T) \rangle = \frac{\frac{1}{2\pi} \int d\phi_p \sin(2\phi_p) E_p \frac{d\Pi_\mu^*(p)}{d^3 p}}{\int d\phi_p E_p \frac{d\mathcal{N}(p)}{d^3 p}}. \quad (320)$$

Figure 22 shows the behavior of $\langle \pi_y \rangle$ with respect to $p_T$ (top panel) and $\phi_p$ (bottom panel). The $p_T$ dependence is found to be strong as compared to other models[165] and the experiments[91] which may be because we assume nonzero initial spin polarization which then evolves in time. On the other hand, if $p_T$ dependence is weak then it probably means that polarization is arising due to spin–orbit coupling, which is not present in the current formulation of our framework. The behavior of $\phi_p$ dependence of polarization is more pronounced at mid-rapidity, which, within the range of $0 < \phi_p < \pi/2$, is qualitatively similar to the polarization behavior in the experiments.[91,165] In general, we find that the effect of nonzero baryon chemical potential on $\langle \pi_y \rangle$ is small, and has opposite effect in forward rapidity and mid-rapidity.

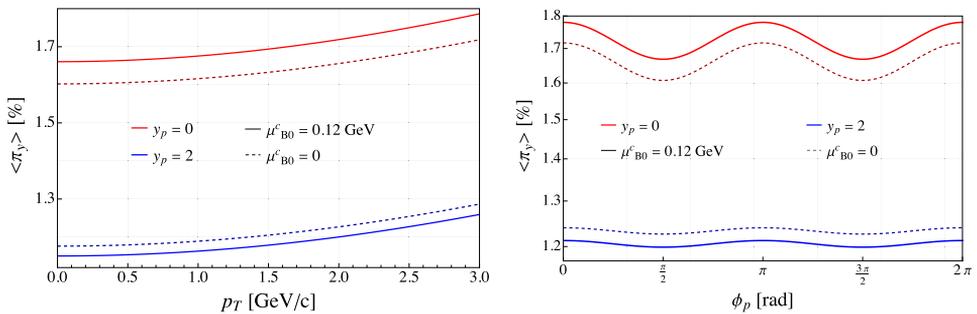

Fig. 22. Component $\langle \pi_y \rangle$ of momentum-averaged polarization (248) as a function of $p_T$ (top panel) and $\phi_p$ (bottom panel) with dashed and solid lines representing vanishing and nonvanishing baryon chemical potential, respectively.

## 7. Bjorken-Expanding Spin-Polarized System in External Electric Field

In the heavy-ion collisions, it is expected that large EM fields are produced[114,311,312] with the strength of the order of $eE/m_\pi^2 \sim eB/m_\pi^2 \sim \mathcal{O}(1)$, where $m_\pi$ is the mass of the pion and $e$ is the electric charge. Understanding the dynamics of the EM fields is still one of the key goals in the physics of heavy-ion collisions, for recent studies, see Refs. 313–322. These fields may have an effect on the behavior of spin polarization of particles, for instance, the splitting of $\Lambda$ and $\bar{\Lambda}$ global polarization, see Fig. 4, which is still an open question. Thus, in this section, we incorporate electric field into our formalism and study its effects on the evolution of the spin components in the Bjorken-expanding background. Details of this section may be found in Ref. 240.

### 7.1. *Stationary solution to the Boltzmann–Vlasov equation*

For the sake of simplicity, we consider the fluid in equilibrium that, at the microscopic scale, is composed of quark-like quasi-particles of $N_f$ flavors which can be described by Boltzmann–Vlasov (BV) equation. With this assumption, we are not taking into consideration any direct interaction (coupling) between spin and EM fields. In Ref. 280, a (stationary) solution to the BV equation was presented as the zeroth-order expansion in $\hbar$ where modified chemical potential allows electric field to be nonvanishing in equilibrium.[323] It is possible that in the event-by-event averaging[311] electric field may vanish, however, its effects on the thermodynamics may survive. We first obtain the modification of the equilibrium hydrodynamic variables, using the stationary solution to the BV equation, which are then inserted into the equations of magnetohydrodynamics (MHD) along with solutions of Maxwell equations derived for the case of Bjorken flow.[317] Subsequently, we obtain the evolution of temperature and chemical potential and plug them into the conservation law for spin to find the evolution of spin components. We observe that the dynamics of the temperature and chemical potential is affected by the Joule heating (JH) term and modified thermodynamic parameters at the macroscopic and microscopic level, respectively.[fff]

Due to the presence of EM fields, we use the stationary solution to the BV equation[156] where the relativistic collisionless BV equation takes the form

$$p^\mu \partial_\mu f_i^\pm \pm q_i F^{\mu\nu} p_\nu \partial_\mu^p f_i^\pm = 0, \tag{321}$$

---

[fff]Our approach is sometimes dubbed a strong electric field regime.[323]

where $i = 1, \ldots, N_f$ represents quark flavor number, and $q_i(-q_i)$ denotes the (anti) particle electric charge for each $i$th flavor. The field strength tensor $F_{\mu\nu}$ can be written in terms of gauge field $A_\mu{}^{\text{ggg}}$ as

$$F_{\mu\nu} = \partial_\mu A_\nu - \partial_\nu A_\mu. \tag{322}$$

One can also decompose $F_{\mu\nu}$ in terms of electric $(E_\mu)$ and magnetic $(B_\mu)$ components as[324]

$$F_{\mu\nu} = E_\mu U_\nu - E_\nu U_\mu + \epsilon_{\mu\nu\alpha\beta} U^\alpha B^\beta, \tag{323}$$

where

$$E^\mu \equiv F^{\mu\nu} U_\nu, \quad B^\mu \equiv \frac{1}{2} \epsilon^{\mu\nu\alpha\beta} F_{\nu\alpha} U_\beta. \tag{324}$$

The equilibrium distribution function (162) for the system with noninteracting quark-like quasi-particles with spin, having the same baryon number but different electric charges, becomes

$$f_{\text{eq},i}^{\pm}(x, p, s) = f_{\text{eq},i}^{\pm}(x, p) \exp\left[\frac{1}{2} \omega_{\mu\nu}(x) s^{\mu\nu}\right], \tag{325}$$

where

$$f_{\text{eq},i}^{\pm}(x, p) = \exp\left[-\beta^\mu(p_\mu \pm q_i A_\mu) \pm \frac{\xi(x)}{3}\right], \tag{326}$$

is the stationary solution to Eq. (321). Note that we have divided $\xi$ by the number of quarks (3) present in a baryon to get the quark chemical potential $(\mu_B/3)$. Thus, $\xi/3 = \mu_B/(3T) = \mu_Q/T$ where $\mu_Q$ is the quark chemical potential.

Putting Eq. (326) into (321) gives

$$\frac{1}{2} p^\mu p^\nu \mathcal{L}_\beta g_{\mu\nu} \pm q_i p^\mu \mathcal{L}_\beta A_\mu = 0, \tag{327}$$

where $\mathcal{L}_\beta X$ is the Lie derivative of a tensor $X$ with respect to $\beta$[325]

$$\mathcal{L}_\beta A_\mu = \beta^\nu \partial_\nu A_\mu + A_\nu \partial_\mu \beta^\nu, \quad \mathcal{L}_\beta g_{\mu\nu} = \partial_\mu \beta_\nu + \partial_\nu \beta_\mu. \tag{328}$$

In the global equilibrium Eq. (327) is satisfied, thus we get

$$\mathcal{L}_\beta g_{\mu\nu} = 0, \quad \mathcal{L}_\beta A_\mu = 0, \tag{329}$$

where, using Eq. (328), one gets

$$\begin{aligned} \mathcal{L}_\beta A_\nu &= \beta^\mu(\partial_\mu A_\nu - \partial_\nu A_\mu) + \beta^\mu \partial_\nu A_\mu + A_\mu \partial_\nu \beta^\mu \\ &= \beta^\mu F_{\mu\nu} + \partial_\nu(\beta \cdot A) = 0. \end{aligned} \tag{330}$$

---

ggg Note that $A_\mu$, defined here, have completely different meaning than the definition of axial-vector component of the Wigner function used in the previous sections.

Using (323) in (330), we obtain

$$\frac{E_\mu}{T} = \partial_\mu (\beta \cdot A). \tag{331}$$

Assuming $E_\mu$ and $T$ being slowly varying functions at the microscopic scale, the integral of the above equation gives

$$\beta \cdot A = \frac{E_\mu}{T} \int dx^\mu, \tag{332}$$

up to a gauge transformation. This is immersed in the quark chemical potential.[326] Thus, Eq. (325) is expressed, in the small polarization limit ($\omega_{\mu\nu} \ll 1$), as

$$f_{\text{eq},i}^\pm (x, p, s) = f_{\text{eq},i}^{pm}(x, p) \left[ 1 + \frac{1}{2} \omega_{\mu\nu}(x) s^{\mu\nu} \right], \tag{333}$$

with

$$f_{\text{eq},i}^\pm (x, p) = \exp\left( \pm \xi_i - \beta^\mu p_\mu \right), \quad \text{where} \quad \xi_i = \xi - q_i \frac{E_\mu}{T} \int dx^\mu. \tag{334}$$

The interpretation of Eq. (334) is that, though the event-by-event average of the electric field vanishes, its traces in the distribution function may survive.[311]

### 7.2. *Baryon and electric charges*

In this study, the fluid has two charge currents: net baryon current ($N^\alpha$) and electric charge current ($J^\alpha$). The net baryon current is defined as

$$N^\alpha = \sum_i^{N_f} \int d\text{P} \ d\text{S} p^\alpha [f_{\text{eq},i}^+(x, p, s) - f_{\text{eq},i}^-(x, p, s)], \tag{335}$$

which reduces, after plugging Eq. (333), to

$$N^\alpha = \sum_i^{N_f} \mathcal{N}_i U^\alpha, \tag{336}$$

with

$$\mathcal{N}_i = 4 \sinh(\xi_i) \mathcal{N}_{(0),i}(T) \quad \text{and} \quad \mathcal{N}_{(0),i}(T) = \frac{T^3}{2\pi^2} z_i^2 K_2(z_i), \tag{337}$$

where $z_i \equiv m_i/T$ is the ratio between $i$th flavor mass and temperature.

The inhomogeneous Maxwell equation

$$\partial_\mu F^{\mu\nu} = J^\nu = \rho_e U^\nu + \Delta^{\nu\rho} J_\rho, \tag{338}$$

with $\rho_e$ being local electric charge density

$$\rho_e = \sum_i^{N_f} \int d\text{P} \ d\text{S}(p \cdot U) q_i [f_{\text{eq},i}^+(x, p, s) - f_{\text{eq},i}^-(x, p, s)] = \sum_i^{N_f} q_i \mathcal{N}_i, \tag{339}$$

implies electric current conservation. Since we assume Bjorken–expanding resistive MHD, this requires electric neutrality.[317] In this case, it is possible for the fluid to have a net baryon density $\mathcal{N}$, and vanishing electric charge density $\rho_e$.[hhh] Away from equilibrium, the (dissipative) electric current reads

$$J^\mu = \sigma_e E^\mu, \tag{340}$$

with $\sigma_e$ being the electric conductivity which, in general, can depend on $T$ and $\mu_B$.

### 7.3. *Energy–momentum conservation*

The energy–momentum tensor of the fluid is defined, in equilibrium, as

$$T^{\mu\nu} = \sum_i^{N_f} \int \mathrm{dP}\,\mathrm{dS} p^\mu p^\nu [f_{\mathrm{eq},i}^+(x,p,s) + f_{\mathrm{eq},i}^-(x,p,s)]. \tag{341}$$

After plugging Eq. (333) in the above equation, we obtain

$$\partial_\mu T^{\mu\nu} = F^{\nu\rho} J_\rho, \tag{342}$$

where $T^{\mu\nu}$ has the same form as Eq. (147) with the energy density and pressure expressed as

$$\mathcal{E} = 4 \sum_i^{N_f} \cosh(\xi_i) \mathcal{E}_{(0),i}(T), \quad \mathcal{P} = 4 \sum_i^{N_f} \cosh(\xi_i) \mathcal{P}_{(0),i}(T), \tag{343}$$

respectively. In above equations

$$\mathcal{E}_{(0),i}(T) = \frac{1}{2\pi^2} T^4 z_i^2 [z_i K_1(z_i) + 3K_2(z_i)], \quad \mathcal{P}_{(0),i}(T) = T\mathcal{N}_{(0),i}(T). \tag{344}$$

### 7.4. *Entropy conservation*

It is important to comment on the conservation of entropy in the presence of external electric field in the global equilibrium. The definition of the entropy current is written as[86,156]

$$\begin{aligned} H^\mu = -\sum_i^{N_f} \int \mathrm{dP}\ \mathrm{dS} p^\mu \{ f_{\mathrm{eq},i}^+(x,p,s)[\ln f_{\mathrm{eq},i}^+(x,p,s) - 1] \\ + f_{\mathrm{eq},i}^-(x,p,s)[\ln f_{\mathrm{eq},i}^-(x,p,s) - 1] \}. \end{aligned} \tag{345}$$

Putting Eq. (325) in Eq. (345), after some straightforward calculations we have

$$H^\mu = \mathcal{P}\beta^\mu + \beta_\alpha T^{\mu\alpha} - \sum_i^{N_f} \xi_i(x) N_i^\mu, \tag{346}$$

---

[hhh] This kind of setup, although greatly simplified, may hold at later stages of the evolution of the QGP.

whereas, in the global equilibrium, using (342), we obtain

$$
\begin{aligned}
\partial_\mu H^\mu &= \partial_\mu \left( \mathcal{P} \beta^\mu + \beta_\alpha T^{\mu\alpha} - \sum_i^{N_f} \xi_i(x) N_i^\mu \right) \\
&= \beta_\alpha \partial_\mu T^{\mu\alpha} - \sum_i^{N_f} N_i^\mu \partial_\mu \xi_i(x) = \beta_\alpha F^{\alpha\beta} J_\beta - \sum_i^{N_f} N_i^\mu \partial_\mu \xi_i(x) \\
&= -\sum_i^{N_f} N_i^\mu \partial_\mu (\xi_i(x) + q_i E_\mu) = -\mathcal{N} T \beta^\mu \partial_\mu \xi = -\mathcal{N} T \mathcal{L}_\beta \xi = 0. \quad (347)
\end{aligned}
$$

Hence, we can safely say that electric field does not produce entropy in global equilibrium with the condition that chemical potential follows Eq. (334). We emphasize that Eq. (345) also includes terms of spin polarization tensor ($\omega_{\mu\nu}$), however, those contributions come at quadratic order which we do not consider in our analysis.

The current study assumes resistive MHD equations where the source of dissipation is the electrical conductivity. For out of equilibrium system, the divergence of entropy reads

$$
\partial_\mu H^\mu = \frac{\sigma_e}{T} E^2, \quad \text{where } E \equiv \sqrt{-E^\mu E_\mu}. \quad (348)
$$

## 7.5. *Background dynamics*

The form of the net baryon density conservation (282) and the energy equation (281) in the presence of the external electric field and Bjorken expansion become

$$
\frac{\partial \mathcal{N}}{\partial \tau} + \frac{\mathcal{N}}{\tau} = 0, \quad \frac{\partial \mathcal{E}}{\partial \tau} + \frac{\mathcal{E} + \mathcal{P}}{\tau} = \sigma_e E^2, \quad (349)
$$

respectively. Along with the above equations, we must also solve the Maxwell equations to obtain the evolution of the EM fields, which, in general, is quite difficult. Nevertheless, here we assume maximally boost-invariant or nonrotating solution as[317]

$$
B^\mu = B_0 \frac{\tau_0}{\tau} Y^\mu, \quad E^\mu = \ell E_0 \frac{\tau_0}{\tau} e^{-\sigma_e(\tau - \tau_0)} Y^\mu, \quad (350)
$$

with $E_0$ and $B_0$ being the electric and magnetic field values at the initial proper time $\tau_0$, respectively. The variable $\ell \equiv \frac{\boldsymbol{B} \cdot \boldsymbol{E}}{BE} = \pm 1$ represents the parallel ($+1$) and antiparallel ($-1$) field configurations.

One obtains solution (350) to the resistive MHD equations as follows. Bjorken symmetries prevent magnetic field to exist in the longitudinal ($z$) direction together with $E_z$-component and electric charge density $\rho_e$,[319] hence, allowing electric and magnetic fields to be nonvanishing only in the transverse ($x$–$y$) plane. Moreover, electric and magnetic fields can only be parallel $0(\ell = +1)$ or antiparallel $\pi(\ell = -1)$ to each other in order to preserve the Bjorken flow. We also consider that the direction of the fields is boost-invariant. This allows us to obtain the solution (350) from the Maxwell's equations.[iii] Using (350) in (334), we have

$$\xi_i = \xi - \ell q_i R_{\text{RMS}} \frac{E_0}{T} \left(\frac{\tau_0}{\tau}\right) e^{-\sigma_e(\tau - \tau_0)},$$

(351)

with $R_{\text{RMS}}$ being the nucleon root-mean-square charge radius.

Another important point is of charge neutrality, which is not satisfied at the initial time. However, considering $\rho_e$ at the initial time for $N_f = 3$, charge neutrality forces the initial number densities to follow

$$2\mathcal{N}_u(\tau_0) - \mathcal{N}_d(\tau_0) - \mathcal{N}_s(\tau_0) = 0,$$

(352)

which, in realistic situations, is not satisfied by the fluid evolution. For our analysis, local electric charge density takes very small values within a fraction of a Fermi, hence can be neglected and neutrality is achieved approximately.

For the numerical modeling of the background, we initialize the parameters at the initial proper time $\tau_0 = 1$ fm as

$$T_0 = 0.6 \, \text{GeV}, \quad \mu_{B0} = 0.05 \, \text{GeV},$$

(353)

which corresponds to $\sqrt{s_{\text{NN}}} = 200 \, \text{GeV}$ collision energy.[327] We use the masses of constituent quarks such as up, down and strange at $\Lambda$'s mass scale for $N_f = 3$[328]

$$m_u = m_d = 0.382 \, \text{GeV}, \quad m_s = 0.537 \, \text{GeV},$$

(354)

whereas $R_{\text{RMS}} = 4.3$ fm.[329] The electrical conductivity follows the relation up to second order in $\xi = \mu_B/T$ as[330]

$$\sigma_e(T, \mu_B) = 0.37 \, Q_e T \left[1 + 0.15 \left(\frac{\mu_B}{T}\right)^2\right],$$

(355)

and $Q_e = (6/9)e^2$ is the sum of the square of up $(2e/3)$, down $(-e/3)$ and strange $(-e/3)$ quark electric charges. We also introduce $\alpha$ parameter to employ different initial values of the electric field $(E_0)$ for the analysis

$$\alpha \equiv \ell \frac{e E_0}{m_\pi^2},$$

(356)

where $e E_0 = m_\pi^2$. $\alpha$ is our only free variable which we can tune to change the value of the initial electric field. In order to better understand the evolution of temperature and baryon chemical potential, we first write the energy equation in Eq. (349) as

$$\tau \frac{d\mathcal{E}}{d\tau} = - \underbrace{w}_{\text{expansion term}} + \overbrace{\tau \sigma_e E^2}^{\text{Joule heating term}},$$

(357)

---

[iii]Note that the presence of EM fields can, in general, break the boost-invariance of the system due to a nonvanishing value of the Poynting vector. In the presence of EM field, the most general form of the energy–momentum tensor can have terms of the form $U^\mu \epsilon^{\nu\lambda\alpha\beta} E_\lambda B_\alpha U_\beta$, which can be argued to break the boost-invariant flow. Thus, to preserve the Bjorken symmetry in the transverse boost-invariant MHD, the direction of the electric field can either be parallel or antiparallel to the direction of the magnetic field.

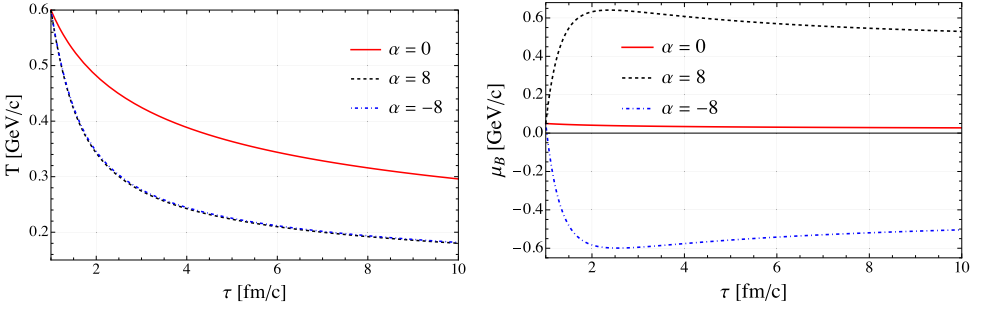

Fig. 23. Evolution of $T$ (top panel) and $\mu_B$ (bottom panel) as a function of proper time $\tau$ with initial values $T_0 = 0.6$ GeV, and $\mu_{B0} = 0.05$ GeV for Bjorken background in the presence of external electric field. $\alpha = 0$ corresponds to no electric field in the system.

with $w = \mathcal{E} + \mathcal{P}$ being the enthalpy density. As observed from Eq. (357), the proper-time evolution of the energy density depends on the expansion and JH terms. JH increases the temperature with the increase in the values of $E^2$ and $\sigma_e \tau$, thus producing heating effect. Equation (355) tells us that $\sigma_e \tau$ is a small factor and its increase will decrease the electric field in the hydrodynamic evolution, see (350). However, the JH term can be dominating over the expansion term in the early time for large initial value of electric field, which may induce reheating effect.[317] Nonetheless, in the current setup, fluid evolution gets modified through both JH term and EoS which prevents reheating and increases the enthalpy density. This in turn means that temperature is decreasing due to electric field ($|\alpha| \neq 0$) and makes the fluid elements heavier, which can be seen from Fig. 23 (upper panel). The temperature behavior for both negative and positive values of $\alpha$ is almost similar, however, for large values of $\alpha$ it is more pronounced.

Evolution of the baryon chemical potential, as seen from Fig. 23 (lower panel), is more sensitive to the presence of the electric field in the system. With the increase in proper time, the absolute value of $\mu_B$ starts to decrease, specifically at $\tau = 3$ fm, because electric field starts to dominate over $\mu_B$, and the sign of $\mu_B$ depends on the sign of $\alpha$.

## 7.6. *Spin dynamics*

The spin tensor (158) for quark-like quasi-particles takes the form

$$S_{\text{GLW}}^{\alpha,\beta\gamma} = \sum_i^{N_f} \cosh(\xi_i)[U^\alpha(\mathcal{A}_{1,i}\omega^{\beta\gamma} + \mathcal{A}_{2,i}U^{[\beta}\omega_\delta^{\gamma]}U^\delta) + \mathcal{A}_{3,i}(U^{[\beta}\omega^{\gamma]\alpha} + g^{\alpha[\beta}\omega_\delta^{\gamma]}U^\delta)],$$

(358)

with the thermodynamic coefficients

$$\mathcal{A}_{1,i} = \mathcal{N}_{(0),i} - \mathcal{B}_{(0),i}, \quad \mathcal{A}_{2,i} = \mathcal{A}_{(0),i} - 3\mathcal{B}_{(0),i}, \quad \mathcal{A}_{3,i} = \mathcal{B}_{(0),i}.$$

(359)

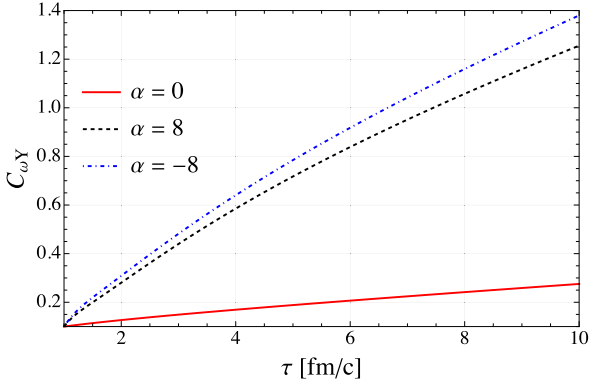

Fig. 24.   Proper-time evolution of spin polarization component $C_{\omega Y}$ with initial value $C_{\omega Y_0} = 0.1$ in the presence of external electric field.

We observe that the forms of the spin equations of motion remain the same as Eqs. (285) even in the presence of the electric field, and find no coupling between the spin components. The evolution of the spin components is qualitatively similar to the Bjorken-expanding system without mean fields, see Subsec. 6.2. However, we notice that electric field enhances the spin dynamics, as depicted in Fig. 24. We only show $C_{\omega Y}$ evolution as this component is relevant keeping in mind the physics situation of heavy-ion collisions, see the discussion in Subsec. 6.2.2, and its enhancement suggests that electric field may play an important role in the heavy-ions collision experiments. As the spin component behavior is similar to the Bjorken-expanding matter, we find that the dynamics of mean spin polarization is also qualitatively similar, as shown in Fig. 14 in Subsec. 6.2.3, hence, we do not show those plots here.

## 8. Conformal Symmetry of Perfect-Fluid Hydrodynamics with Spin

In Subsecs. 6.2 and 6.3, we have considered the produced matter to be homogeneous in the direction transverse to the beam, and we focused on its dynamics in the longitudinal direction. However, due to the finite size of the colliding nuclei, the dynamics of the produced matter in the transverse plane is inhomogeneous, with energy density strongly decreasing at the radial edge. To address such a possibility, in this section, we consider the simplest nontrivial extension of Bjorken expansion assuming that the system is cylindrically symmetric with respect to the beam direction. Such conditions arise in perfectly central relativistic collisions of heavy ions or smaller systems. Although, in general, to study such configuration, it is necessary to perform full numerical simulations, in what follows, we choose a different strategy leading to semi-analytic results. For this purpose, we use a generalization of the Bjorken flow known as Gubser flow.[331,332] The latter introduces the flow component in the transverse direction and can be shown invariant for boost-invariant and cylindrically symmetric systems under conformal symmetry group. $SO(3)_q \otimes SO(1,1) \otimes \mathbf{Z}_2$. This special symmetry, known as "Gubser's symmetry", is a generalized version of the Bjorken symmetry $ISO(2) \otimes SO(1,1) \otimes \mathbf{Z}_2$[iii] and has been used in various studies of hydrodynamics.[333–341]

In this section, we study the dynamics of spin components in the Gubser-expanding perfect-fluid background. For this purpose, we first introduce certain aspects of conformal mapping to the de Sitter space required to make the Gubser's symmetry manifest. We find transformation rules which the conservation laws must follow for the dynamics to respect conformal invariance. While the resulting well-known tracelessness of the energy–momentum tensor is a feasible condition, the additional condition for the spin tensor is not satisfied by the form we use (154), thus breaking conformal symmetry explicitly. However, as our formulation prevents any back-reaction from the spin evolution to the perfect-fluid background evolution, the invariance of the background is not spoiled by the conformal symmetry breaking at the level of the spin tensor. This motivates us to find nontrivial approximate solutions for the spin keeping a small but finite mass. The results presented in this section may be found in Ref. 239.

## 8.1. *Boost-invariant and cylindrically symmetric expansion*

For boost-invariant and cylindrically symmetric (with respect to the beam direction) systems respecting Gubser symmetry it is convenient to introduce the polar-Milne coordinates $x^\mu = (\tau, r, \phi, \eta)$ with the line element given as

$$ds^2 = -d\tau^2 + dr^2 + r^2 d\phi^2 + \tau^2 d\eta^2,$$

where $\tau$ and $\eta$, as defined in Subsec. 6.1.1, are longitudinal proper time and longitudinal space–time rapidity, while $r = \sqrt{x^2 + y^2}$ and $\phi = \tan^{-1}(y/x)$ are the radial distance and the azimuthal angle, respectively, which parametrize the plane transverse to the beam direction.

The four-vector basis for boost-invariant and cylindrically symmetric system can be written in the form (note in LRF, $U$, $X$, $Y$ and $Z$ are the unit vectors in the directions of $\tau$, $r$, $\phi$ and $\eta$, respectively)

$$U^\mu = (\cosh(\vartheta), \sinh(\vartheta), 0, 0), \quad X^\mu = (\sinh(\vartheta), \cosh(\vartheta), 0, 0),$$
$$Y^\mu = (0, 0, 1/r, 0), \quad Z^\mu = (0, 0, 0, 1/\tau), \tag{360}$$

which follows

$$U \cdot U = -1, \quad X \cdot X = 1, \quad Y \cdot Y = 1, \quad Z \cdot Z = 1, \tag{361}$$

and allows polar-hyperbolic metric tensor, $g^{\mu\nu} = \text{diag}(-1, 1, 1/r^2, 1/\tau^2)$ (note opposite signature as compared to previous sections), to be written as

$$g^{\mu\nu} = -U^\mu U^\nu + X^\mu X^\nu + Y^\mu Y^\nu + Z^\mu Z^\nu. \tag{362}$$

---

[iii] $SO(3)_q$ is a special orthogonal rotation group in three dimensions characterized by a length scale $q$, $SO(1,1)$ is a special orthogonal indefinite group which ensures boost-invariance symmetry in the system and $\mathbf{Z}_2$ symmetry is generated by the reflections in the $r$–$\phi$ plane. $ISO(2)$ is a Euclidean group with dimension 3 ensuring translational and rotational invariance in the transverse plane. In the limit $q \to 0$, $SO(3)_q$ reduces to $ISO(2)$ and Gubser flow takes the form of Bjorken flow.

R. Singh

## 8.2. *Conformal symmetry and conformal mapping to de Sitter space*

The invariance of the flow profile with respect to the Gubser's symmetry is manifest if one uses the conformal mapping to the curved (de Sitter) space–time $dS_3 \otimes R$ defined by three-dimensional de Sitter space and a line. This is performed by Weyl rescaling of the metric

$$ds^2 \to \frac{ds^2}{\Omega_{\rm cf}^2} = \frac{-d\tau^2 + dr^2 + r^2 d\phi^2}{\tau^2} + d\eta^2, \tag{363}$$

where $\Omega_{\rm cf} = \tau$ is the conformal factor, and the coordinate transformation from polar Milne coordinates to de Sitter coordinates $\hat{x}^\mu = (\rho, \theta, \phi, \eta)$[kkk] with the help of the relations

$$\sinh(\rho(\tau, r)) = -\frac{1 - (q\tau)^2 + (qr)^2}{2q\tau}, \quad \tan(\theta(\tau, r)) = \frac{2qr}{1 + (q\tau)^2 - (qr)^2}. \tag{364}$$

The rescaled line element in the de Sitter space with the metric

$$\hat{g}_{\mu\nu} = \text{diag}(-1, (\cosh(\rho))^2, (\cosh(\rho))^2(\sin(\theta))^2, 1),$$

takes the form

$$d\hat{s}^2 = -d\rho^2 + (\cosh(\rho))^2(d\theta^2 + (\sin(\theta))^2 d\phi^2) + d\eta^2. \tag{365}$$

From Eq. (365), we observe that the Weyl scaling (363) along with the change of the coordinates through the relations (364) makes the $(SO(3)_q)$ conformal isometry to become a manifest isometry $(SO(3))$ in de Sitter coordinates. For a system to respect conformal symmetry it is required that $(m_A, n_A)$ tensors transform homogeneously under Weyl rescaling, namely[331,332,342–344]

$$A^{\mu_1 \ldots \mu_m}_{\nu_1 \ldots \nu_n}(x) \to e^{(-\varphi(x))\Delta_A} A^{\mu_1 \ldots \mu_m}_{\nu_1 \ldots \nu_n}(x) \equiv \Omega_{\rm cf}^{\Delta_A} A^{\mu_1 \ldots \mu_m}_{\nu_1 \ldots \nu_n}(x), \tag{366}$$

where $\varphi(x)$ depends on the space–time coordinates, while $\Delta_A = [A] + m_A - n_A$ indicates the conformal weight of $A$ with $[A]$ representing the mass dimension of $A$, and $m_A$ and $n_A$ are the number of contravariant and covariant indices, respectively, of the quantity $A$.

Take for example the metric tensor $g_{\mu\nu}$: $g_{\mu\nu}$ is a second-rank $(0,2)$ tensor which is dimensionless, thus $[g_{\mu\nu}] = 0$ with $m_A = 0$ and $n_A = 2$, hence, $\Delta_{g_{\mu\nu}} = -2$, which implies that $g_{\mu\nu}$ transforms homogeneously under Weyl rescaling as[332,342]

$$g_{\mu\nu} \to \Omega_{\rm cf}^{-2} g_{\mu\nu}. \tag{367}$$

Similarly, conformal weight of $g^{\mu\nu}$ is $\Delta_{g^{\mu\nu}} = 2$.

---

[kkk] In the following, the quantities with a double hat on top indicate that they are defined in the de Sitter space.

With the use of Eq. (367) one can obtain the relation between $R^3 \otimes R$ and $dS_3 \otimes R$ as

$$\hat{g}_{\mu\nu} = \frac{1}{\tau^2} \frac{\partial x^\alpha}{\partial \hat{x}^\mu} \frac{\partial x^\beta}{\partial \hat{x}^\nu} g_{\alpha\beta}. \tag{368}$$

With the knowledge of $\Delta_{g_{\mu\nu}} = -2$ and fluid flow normalization $U \cdot U = -1$, the conformal weight of contravariant fluid flow vector can be shown to be $\Delta_{U^\mu} = 1$, where the transformation rule is

$$\hat{U}_\nu = \frac{1}{\tau} \frac{\partial x^\mu}{\partial \hat{x}^\nu} U_\mu. \tag{369}$$

The advantage of going to de Sitter space–time is that the flow profile (360) under (369) becomes static

$$\hat{U}^\mu = (1, 0, 0, 0). \tag{370}$$

Hence, it respects Gubser's symmetry with the transverse rapidity of the form[331,332]

$$\vartheta(\tau, r) = \tanh^{-1}\left(\frac{2q\tau qr}{1 + (q\tau)^2 + (qr)^2}\right). \tag{371}$$

Similarly, other basis vectors (360) in the de Sitter space–time become

$$\hat{X}^\mu = (0, (\cosh(\rho))^{-1}, 0, 0), \quad \hat{Y}^\mu = (0, 0, (\cosh(\rho)\sin(\theta))^{-1}, 0),$$
$$\hat{Z}^\mu = (0, 0, 0, 1). \tag{372}$$

Using basis vectors, the metric $\hat{g}^{\mu\nu}$ reads

$$\hat{g}^{\mu\nu} = -\hat{U}^\mu \hat{U}^\nu + \hat{X}^\mu \hat{X}^\nu + \hat{Y}^\mu \hat{Y}^\nu + \hat{Z}^\mu \hat{Z}^\nu, \tag{373}$$

with its determinant given by

$$\hat{g} \equiv \det(\hat{g}_{\mu\nu}) = -(\cosh(\rho))^4 (\sin(\theta))^2. \tag{374}$$

From Eq. (148), we find that the mass dimension of the energy density and the pressure is $[\varepsilon] \equiv [\mathcal{P}] = 4$, thus their conformal weight is $\Delta_\varepsilon = \Delta_\mathcal{P} = 4$. Similarly, for the net baryon density (142) one arrives at $[\mathcal{N}] = 3$ with $\Delta_\mathcal{N} = 3$. For the temperature and baryon chemical potential, it is easy to observe that their conformal weight is $\Delta_T = \Delta_{\mu_B} = 1$ since their mass dimension is $[T] = [\mu_B] = 1$.

Using the method described above, see Eq. (369), transformation rules for other quantities can be found

$$U_\mu(\tau, r) = \tau \frac{\partial \hat{x}^\nu}{\partial x^\mu} \hat{U}_\nu(\rho), \quad T(\tau, r) = \frac{\hat{T}(\rho)}{\tau}, \quad \mu_B(\tau, r) = \frac{\hat{\mu}_B(\rho)}{\tau},$$
$$\varepsilon(\tau, r) = \frac{\hat{\varepsilon}(\rho)}{\tau^4}, \quad \mathcal{P}(\tau, r) = \frac{\hat{\mathcal{P}}(\rho)}{\tau^4}, \quad \mathcal{N}(\tau, r) = \frac{\hat{\mathcal{N}}(\rho)}{\tau^3}. \tag{375}$$

For the purpose of the following analysis, it is particularly important to find conformal weights of the energy–momentum and the spin tensors. As mentioned before, see Eq. (19), total angular momentum has both orbital and spin contributions

$$J^{\lambda,\mu\nu} = L^{\lambda,\mu\nu} + S^{\lambda,\mu\nu} = x^\mu T^{\lambda\nu} - x^\nu T^{\lambda\mu} + S^{\lambda,\mu\nu}. \tag{376}$$

As the mass dimension of $x^\mu$ is $[x^\mu] = -1$, one has $\Delta_{x^\mu} = 0$. From the knowledge of the conformal weights of $\varepsilon$ and $U^\alpha$, and using Eq. (147), we find that the conformal weight of the energy–momentum tensor is $\Delta_{T^{\alpha\beta}} = 6$ (note that the left-hand side of Eq. (147) must have the same conformal weight as each term on the right-hand side). From the reasoning that each term in Eq. (376) should have the same conformal weight one can then find that the spin tensor has the conformal weight $\Delta_{S^{\alpha\beta\gamma}} = 6$. Therefore, the GLW spin tensor in our formalism should respect $\Delta_{S^{\alpha\beta\gamma}_{\mathrm{GLW}}} = 6$.[lll]

In a similar way, using the information of the conformal weights of net baryon density and fluid flow, we also obtain the conformal weight of the net baryon current (141) as $\Delta_{N^\alpha} = 4$. From Eq. (155), we find, using the conformal weight of the spin tensor, that the conformal weight of the spin polarization tensor is $\Delta_{\omega^{\alpha\beta}} = 2$ which gives the conformal weights of $\kappa_\alpha$ and $\omega_\alpha$ to be $\Delta_{\kappa^\alpha} = 1$ and $\Delta_{\omega^\alpha} = 1$. Here, we have used the fact that $\epsilon^{\alpha\beta\gamma\delta}$ has no mass dimension and has four contravariant indices resulting in $\Delta_{\epsilon^{\alpha\beta\gamma\delta}} = 4$, see Eq. (179). Using the information of how the spin polarization tensor transforms under Weyl scaling we find that spin polarization components $C_\kappa$ and $C_\omega$ have conformal weights $\Delta_{C_\kappa} = \Delta_{C_\omega} = 0$, which makes them conformally invariant (note that $C_\kappa$ and $C_\omega$ are dimensionless scalars).

To summarize, transformation rules equation (366) for the net baryon current, the energy–momentum tensor and the spin tensor for four-dimensional space–time read

$$N^\alpha \to \Omega_{\mathrm{cf}}^4 N^\alpha, \quad T^{\alpha\beta} \to \Omega_{\mathrm{cf}}^6 T^{\alpha\beta}, \quad S^{\alpha\beta\gamma} \to \Omega_{\mathrm{cf}}^6 S^{\alpha\beta\gamma}. \tag{377}$$

It can be easily noticed that the transformation rules for the covariant quantities can be found using the information that raising (lowering) the Lorentz index with the metric tensor changes the conformal weight by $2(-2)$.

### 8.3. *Conformal invariance of conservation equations*

In this section, we derive conformal transformation rules for the conservation laws for $N^\alpha$, $T^{\alpha\beta}$ and $S^{\alpha\beta\gamma}$, which, in curved space–time, read

$$d_\alpha N^\alpha(x) = \partial_\alpha N^\alpha + \Gamma_{\alpha\beta}^\alpha N^\beta = 0, \tag{378}$$

$$d_\alpha T^{\alpha\beta}(x) = \partial_\alpha T^{\alpha\beta} + \Gamma_{\alpha\lambda}^\alpha T^{\lambda\beta} + \Gamma_{\alpha\lambda}^\beta T^{\alpha\lambda} = 0, \tag{379}$$

$$d_\alpha S^{\alpha\beta\gamma}(x) = \partial_\alpha S^{\alpha\beta\gamma} + \Gamma_{\alpha\lambda}^\alpha S^{\lambda\beta\gamma} + \Gamma_{\alpha\lambda}^\beta S^{\alpha\lambda\gamma} + \Gamma_{\alpha\lambda}^\gamma S^{\alpha\beta\lambda} = 0, \tag{380}$$

---

[lll]Using the information that the conformal weight of the Dirac spinor $\psi$ and Dirac dual spinor $\bar\psi$ is $\Delta_\psi = \Delta_{\bar\psi} = \frac{3}{2}$ and Dirac gamma matrix has the conformal weight $\Delta_{\gamma^\mu} = 1$,[345,346] we can find that the canonical spin tensor (26) and the HW spin tensor (30) have the same conformal weight as the GLW spin tensor.

with $\Gamma^{\nu}_{\mu\lambda}$ being the Christoffel symbol defined as[347–349]

$$\Gamma^{\nu}_{\mu\lambda} \equiv \Gamma^{\nu}_{\lambda\mu} = \frac{1}{2}g^{\nu\sigma}(\partial_{\mu}g_{\sigma\lambda} + \partial_{\lambda}g_{\sigma\mu} - \partial_{\sigma}g_{\mu\lambda}). \tag{381}$$

Note that $d_{\alpha}$ in Eqs. (378)–(380) is the covariant derivative, which reduces to partial derivative in flat space–time. To obtain conformal transformation of the conservation equations (378)–(380) we require Christoffel symbols (381) to transform accordingly. Using Eq. (367), we get[344,350,351]

$$\hat{\Gamma}^{\nu}_{\mu\lambda} = \Gamma^{\nu}_{\mu\lambda} - \delta^{\nu}_{\lambda}\partial_{\mu}\varphi - \delta^{\nu}_{\mu}\partial_{\lambda}\varphi + g^{\nu\sigma}g_{\mu\lambda}\partial_{\sigma}\varphi, \tag{382}$$

with $\delta^{\beta}_{\lambda}$ being the Kronecker delta function. However, one may notice that when $\varphi$ is constant then $\partial_{\sigma}\varphi = 0$ in Eq. (382) which makes $\hat{\Gamma}^{\nu}_{\mu\lambda} = \Gamma^{\nu}_{\mu\lambda}$.

One can easily show that the net baryon current conservation (378) is conserved in Minkowski and de Sitter space–times which means that it is conformal-frame-independent[342–344] resulting in

$$d_{\alpha}N^{\alpha} = \Omega^{4}_{\text{cf}}\hat{d}_{\alpha}\hat{N}^{\alpha}. \tag{383}$$

Plugging (382) in (379), we arrive at the conformal transformation of energy and linear momentum conservation[343,344,350]

$$d_{\alpha}T^{\alpha\beta} = \Omega^{6}_{\text{cf}}[\hat{d}_{\alpha}\hat{T}^{\alpha\beta} - \hat{T}^{\lambda}_{\lambda}\hat{g}^{\beta\delta}\partial_{\delta}\varphi], \tag{384}$$

which indicates that $\hat{T}^{\alpha\beta}$ has to be traceless ($\hat{T}^{\alpha}_{\alpha} = 0$) to be conserved in de Sitter space–time.[352–355] Finally, using Eqs. (377) and (382) in (380), we obtain the conformal transformation of the spin conservation law

$$d_{\alpha}S^{\alpha\beta\gamma} = \Omega^{6}_{\text{cf}}[\hat{d}_{\alpha}\hat{S}^{\alpha\beta\gamma} - (\hat{S}^{\lambda\gamma}_{\lambda}\hat{g}^{\beta\sigma} + \hat{S}^{\alpha\beta}_{\alpha}\hat{g}^{\sigma\gamma})\partial_{\sigma}\varphi], \tag{385}$$

indicating that the spin tensor needs to fulfill the condition $\hat{S}^{\alpha\beta}_{\alpha} = 0$ for the spin conservation to be conformally invariant. We also observe from Eqs. (383)–(385) that when $\varphi$ is constant, all the conservation laws automatically become conformally invariant.

## 8.4. *Background dynamics*

We will now explore the Gubser-expanding background dynamics which will then be used in the next section to obtain the evolution of the spin polarization components. Using Eq. (375), our results can be transformed back to the Minkowski space–time.

We find that the GLW energy–momentum tensor (147) satisfies the traceless condition only in the conformal limit where energy density (148), pressure (148) and net baryon density (142) are expressed as

$$\hat{\varepsilon} = \frac{12}{\pi^2}\cosh(\xi)\hat{T}^4, \quad \hat{\mathcal{P}} = \frac{4}{\pi^2}\cosh(\xi)\hat{T}^4, \quad \hat{\mathcal{N}} = \frac{4}{\pi^2}\sinh(\xi)\hat{T}^3, \tag{386}$$

respectively. One finds that $\hat{\hat{\varepsilon}} = 3\hat{\hat{\mathcal{P}}}$, which makes the energy–momentum traceless, as required in Eq. (384). Moreover, we observe that the GLW (158) and HW (30) definitions of spin tensor do not satisfy the condition of conformal invariance (385), which is not the case for canonical spin tensor (26).

Since, as before, we assume small spin polarization limit which makes spin dynamics decouple from the background dynamics, in the following, we study the dynamics of the spin components on top of the Gubser flow background without spoiling its conformal invariance. Considering evolution of the spin components in the de Sitter coordinates we keep finite mass in the spin tensor (158).

The conservation law for net baryon density (255) can be written in the de Sitter coordinates as

$$\hat{\hat{U}}^\alpha \partial_\alpha \hat{\hat{\mathcal{N}}} + \hat{\hat{\mathcal{N}}} \partial_\alpha \hat{\hat{U}}^\alpha + \hat{\hat{\mathcal{N}}} \hat{\hat{U}}^\alpha \frac{\partial_\alpha \sqrt{-\hat{\hat{g}}}}{\sqrt{-\hat{\hat{g}}}} = \partial_\rho \hat{\hat{\mathcal{N}}} + 2\hat{\hat{\mathcal{N}}} \tanh(\rho) = 0, \qquad (387)$$

where $\hat{\hat{g}}$ is the value of the determinant of de Sitter metric (374). Similarly, the conservation of energy (256) gives

$$\partial_\rho \hat{\hat{\varepsilon}} + 2(\hat{\hat{\varepsilon}} + \hat{\hat{\mathcal{P}}}) \tanh(\rho) = 0, \qquad (388)$$

whereas Euler equation (257) is satisfied trivially. The solutions of the above equations of motion can be obtained as[331,332]

$$\hat{\hat{\varepsilon}} = \hat{\hat{\varepsilon}}_0 \left(\frac{\cosh(\rho_0)}{\cosh(\rho)}\right)^{8/3}, \quad \hat{\hat{\mathcal{N}}} = \hat{\hat{\mathcal{N}}}_0 \left(\frac{\cosh(\rho_0)}{\cosh(\rho)}\right)^2, \qquad (389)$$

respectively, where $\hat{\hat{\varepsilon}}_0 \equiv \hat{\hat{\varepsilon}}(\rho_0)$ and $\hat{\hat{\mathcal{N}}}_0 \equiv \hat{\hat{\mathcal{N}}}(\rho_0)$ are constants of integration at the initial de Sitter time ($\rho_0$). The respective solutions for temperature and baryon chemical potential in de Sitter space can be obtained using Eqs. (386)

$$\hat{\hat{T}} = \hat{\hat{T}}_0 \left(\frac{\cosh(\rho_0)}{\cosh(\rho)}\right)^{2/3}, \quad \hat{\hat{\mu}}_B = \hat{\hat{\mu}}_{B0} \left(\frac{\cosh(\rho_0)}{\cosh(\rho)}\right)^{2/3}, \qquad (390)$$

where $\hat{\hat{T}}_0 \equiv \hat{\hat{T}}(\rho_0)$ and $\hat{\hat{\mu}}_{B0} \equiv \hat{\hat{\mu}}_B(\rho_0)$ are the integration constants. This result implies that $\hat{\hat{\xi}} = \hat{\hat{\mu}}_B/\hat{\hat{T}}$ does not depend on $\rho$.[mmm]

Figure 25 shows strongly correlated behavior of temperature (390) with the flow-vector components $(U^\tau, U^r)/\sqrt{(U^\tau)^2 + (U^r)^2}$, where the initial temperature is $\hat{\hat{T}}_0 \equiv \hat{\hat{T}}(\rho_0) = 1.2$ at $\rho_0 = 0$ such that considering $q = 1\,\text{fm}^{-1}$ gives $T(\tau_0 = 1\,\text{fm}, r = 0) = 1.2\,\text{fm}^{-1} = 0.24\,\text{GeV}$.[335]

## 8.5. *Spin dynamics*

In this section, we study the evolution of the spin polarization components. For simplicity, we continue on using the de Sitter coordinates instead of polar Milne

---

[mmm] We observe here that the dynamics of the system, due to Gubser symmetry, depends only on $\rho$.[331,332]

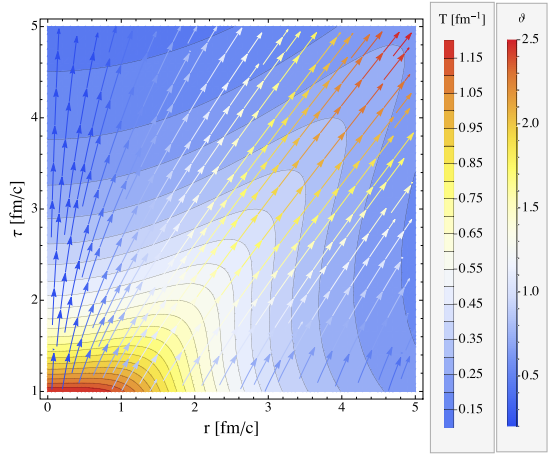

Fig. 25.  Temperature (contours) and flow-vector components $(U^\tau, U^r)/\sqrt{(U^\tau)^2 + (U^r)^2}$ (streamlines — the coloring of arrows are indicated by the fluid rapidity $\vartheta$) as functions of longitudinal proper time ($\tau$) and radial distance ($r$).

coordinates, although, as we noted above, the considered GLW form of the spin tensor leads to the conformal symmetry breaking.

The spin equations (273)–(278) reduce, in this case, to

$$\hat{\hat{\mathcal{Q}}}\dot{\hat{\hat{C}}}_{\kappa X} = -\hat{\hat{C}}_{\kappa X}\left[\dot{\hat{\hat{\mathcal{Q}}}} + \frac{5}{2}\hat{\hat{\mathcal{Q}}}\tanh(\rho)\right], \tag{391}$$

$$\hat{\hat{\mathcal{Q}}}\dot{\hat{\hat{C}}}_{\kappa Y} + \frac{\hat{\hat{\mathcal{Q}}}}{2}\cosh(\rho)\sin(\theta)\overset{\circ}{\hat{\hat{C}}}_{\omega Z}$$

$$= -\hat{\hat{C}}_{\kappa Y}\left[\dot{\hat{\hat{\mathcal{Q}}}} + \frac{5}{2}\hat{\hat{\mathcal{Q}}}\tanh(\rho)\right] - \hat{\hat{C}}_{\omega Z}\frac{\hat{\hat{\mathcal{Q}}}}{2}\cosh(\rho)\cos(\theta), \tag{392}$$

$$\hat{\hat{\mathcal{Q}}}\dot{\hat{\hat{C}}}_{\kappa Z} - \frac{\hat{\hat{\mathcal{Q}}}}{2}\cosh(\rho)\sin(\theta)\overset{\circ}{\hat{\hat{C}}}_{\omega Y}$$

$$= -\hat{\hat{C}}_{\kappa Z}\left[\dot{\hat{\hat{\mathcal{Q}}}} + 3\hat{\hat{\mathcal{Q}}}\tanh(\rho)\right] + \hat{\hat{C}}_{\omega Y}\hat{\hat{\mathcal{Q}}}\cosh(\rho)\cos(\theta), \tag{393}$$

$$(\hat{\hat{\mathcal{Q}}} - \hat{\hat{\mathcal{R}}})\dot{\hat{\hat{C}}}_{\omega X} = -\hat{\hat{C}}_{\omega X}\left[\dot{\hat{\hat{\mathcal{Q}}}} - \dot{\hat{\hat{\mathcal{R}}}} + \left(\frac{9\hat{\hat{\mathcal{Q}}}}{2} - 4\hat{\hat{\mathcal{R}}}\right)\tanh(\rho)\right], \tag{394}$$

$$(\hat{\hat{\mathcal{Q}}} - \hat{\hat{\mathcal{R}}})\dot{\hat{\hat{C}}}_{\omega Y} - \frac{\hat{\hat{\mathcal{Q}}}}{2\sin(\theta)}\frac{1}{(\cosh(\rho))^3}\overset{\circ}{\hat{\hat{C}}}_{\kappa Z}$$

$$= -\hat{\hat{C}}_{\omega Y}\left[\dot{\hat{\hat{\mathcal{Q}}}} - \dot{\hat{\hat{\mathcal{R}}}} + \left(\frac{9\hat{\hat{\mathcal{Q}}}}{2} - 4\hat{\hat{\mathcal{R}}}\right)\tanh(\rho)\right], \tag{395}$$

$$(\hat{\hat{\mathcal{Q}}} - \hat{\hat{\mathcal{R}}})\dot{\hat{C}}_{\omega Z} + \frac{\hat{\hat{\mathcal{Q}}}}{2\sin(\theta)}\frac{1}{(\cosh(\rho))^3}\overset{\circ}{\hat{C}}_{\kappa Y}$$

$$= -\hat{C}_{\omega Z}[\dot{\hat{\mathcal{Q}}} - \dot{\hat{\mathcal{R}}} + (5\hat{\hat{\mathcal{Q}}} - 4\hat{\hat{\mathcal{R}}})\tanh(\rho)] - \frac{\hat{\hat{\mathcal{Q}}}}{2\sin(\theta)}\frac{\cot(\theta)}{(\cosh(\rho))^3}\hat{C}_{\kappa Y}, \quad (396)$$

where $(\cdot) \equiv \partial/\partial\rho$, $(\circ) \equiv \partial/\partial\theta$ with $\hat{\hat{\mathcal{Q}}} = \cosh(\hat{\hat{\xi}})\hat{\mathcal{B}}_{(0)}$ and $\hat{\hat{\mathcal{R}}} = \cosh(\hat{\hat{\xi}})\hat{\mathcal{N}}_{(0)}$.

In contrast to the Bjorken-expanding system (285), where all spin components evolve independently and to the case of nonboost-invariant system, see Eqs. (302)–(307), where the spin components $C_{\kappa X}$, $C_{\omega Y}$ and $C_{\kappa Y}$, $C_{\omega X}$ are coupled with each other, respectively, here we find that $C_{\kappa X}$ and $C_{\omega X}$ evolve independently whereas $C_{\kappa Y}$, $C_{\omega Z}$ and $C_{\kappa Z}$, $C_{\omega Y}$ are coupled with each other, respectively. We observe that these couplings arise due to the breaking of the conformal symmetry in the spin tensor through the dependence of the spin components on the $\theta$ coordinate.

For the conformal EoS, the solutions to Eqs. (391) and (394) can be obtained analytically as

$$\hat{C}_{\kappa X} = \hat{C}_{\kappa X}^0\left(\frac{\cosh(\rho)}{\cosh(\rho_0)}\right)^{5/6}, \quad \hat{C}_{\omega X} = \hat{C}_{\omega X}^0\left(\frac{\cosh(\rho_0)}{\cosh(\rho)}\right)^{7/6}, \quad (397)$$

where $\hat{C}_{\kappa X}^0 \equiv \hat{C}_{\kappa X}(\rho_0)$ and $\hat{C}_{\omega X}^0 \equiv \hat{C}_{\omega X}(\rho_0)$ are the initial values of the spin components. Interestingly, the behavior (concave function of $\rho$) of $\hat{C}_{\omega X}$ is qualitatively similar to the evolution of temperature and baryon chemical potential (390), whereas dynamics of $\hat{C}_{\kappa X}$ is given by a convex function of de Sitter time $\rho$.

Figure 26 shows the evolution of $C_{\kappa X}$ (top panel) and $C_{\omega X}$ (bottom panel) in the polar Milne coordinates $(\tau, r)$ with the initial values of the spin components as $\hat{\hat{C}}^0_{\kappa X} = \hat{\hat{C}}^0_{\omega X} = 0.1$ implying $C_{\kappa X}(\tau_0 = 1\,\text{fm}, r = 0) = C_{\omega X}(\tau_0 = 1\,\text{fm}, r = 0) = 0.1$ and the mass we considered in the calculations is $m = 0.5 \hat{T}_0\,\text{fm}^{-1}$.

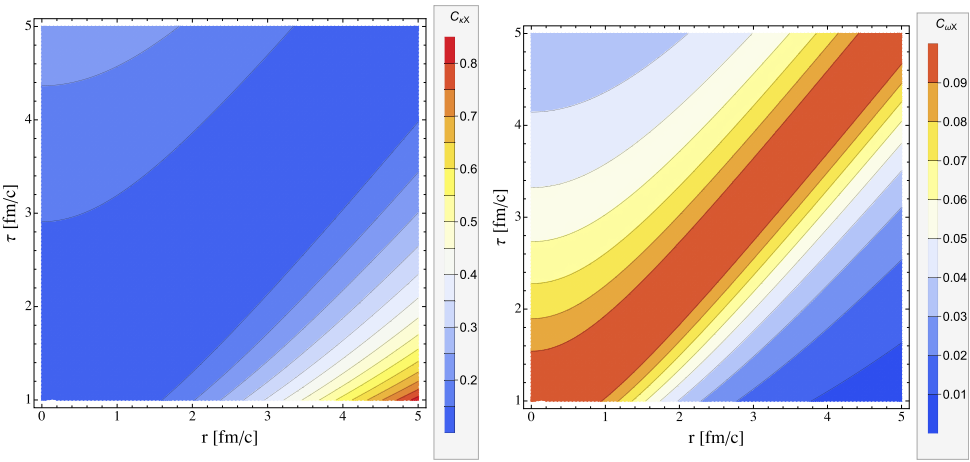

Fig. 26. Dynamics of coefficients $C_{\kappa X}$ (top panel) and $C_{\omega X}$ (bottom panel) in $\tau$–$r$ plane. Note different color scaling for the two panels.

The dynamics of the components $\hat{\hat{C}}_{\omega Y}$ and $\hat{\hat{C}}_{\kappa Z}$ is quite complicated due to their coupling through Eqs. (393) and (395). However, when $\hat{\hat{C}}_{\omega Y}$ is kept initially negligible, then it can be shown that $\hat{\hat{C}}_{\kappa Z}$ is approximately $\theta$-independent

$$\hat{\hat{C}}_{\kappa Z}|_{\hat{\hat{C}}_{\omega Y} = 0} \approx \hat{\hat{C}}^0_{\kappa Z}\left(\frac{\cosh(\rho)}{\cosh(\rho_0)}\right)^{1/3}, \tag{398}$$

where $\hat{\hat{C}}^0_{\kappa Z} \equiv \hat{\hat{C}}_{\kappa Z}(\rho_0)$ and $\hat{\hat{C}}_{\kappa Z}(\rho) \sim 1/\sqrt{\hat{T}_{(\rho)}}$.

Furthermore, when $\hat{\hat{C}}_{\omega Y} \neq 0$, one may notice that $\hat{\hat{C}}_{\kappa Z}$ varies slowly with $\theta$ and the second term on the left-hand side of (395) can be considered negligible. In this case, the solution to $\hat{\hat{C}}_{\omega Y}$-component takes the form

$$\hat{\hat{C}}_{\omega Y} \approx \hat{\hat{C}}^0_{\omega Y}\left(\frac{\cosh(\rho_0)}{\cosh(\rho)}\right)^{7/6}, \tag{399}$$

where $\hat{\hat{C}}^0_{\omega Y}$ is the initial value of $\hat{\hat{C}}_{\omega Y}$ at $\rho_0$. Numerically, it has checked that $\hat{\hat{C}}_{\omega Y}$ depends weakly on $\theta$ and thus approximately proportional to $\hat{\hat{C}}_{\omega X}$.

From Eqs. (392) and (396), we observe that the solutions for $\hat{\hat{C}}_{\kappa Y}$ and $\hat{\hat{C}}_{\omega Z}$ can be obtained only numerically. Nevertheless, some special solutions can be obtained if we assume that $\theta$ terms vanish which will make $\hat{\hat{C}}_{\kappa Y}$ and $\hat{\hat{C}}_{\omega Z}$ decouple from each other

$$\hat{\hat{C}}_{\kappa Y} \approx \hat{\hat{C}}^0_{\kappa Y}\left(\frac{\cosh(\rho)}{\cosh(\rho_0)}\right)^{5/6}\csc\theta, \quad \hat{\hat{C}}_{\omega Z} \approx \hat{\hat{C}}^0_{\omega Z}\left(\frac{\cosh(\rho_0)}{\cosh(\rho)}\right)^{5/3}\csc\theta, \tag{400}$$

where $\hat{\hat{C}}^0_{\kappa Y} \equiv \hat{\hat{C}}_{\kappa Y}(\rho_0)$ and $\hat{\hat{C}}^0_{\omega Z} \equiv \hat{\hat{C}}_{\omega Z}(\rho_0)$ and thus $\hat{\hat{C}}_{\kappa Y} \sim \hat{\hat{C}}_{\kappa X}$. All the spin components tend to exhibit the behavior $(\cosh\rho)^c$, with $c$ being positive (or negative) constant.

## 9. Summary

In this paper, we developed a quantum kinetic theory formalism for the Wigner function for spin-half massive particles considering local and nonlocal collisional effects, which we used to derive the Boltzmann-like kinetic equation for classical distribution function in phase-space extended to spin. In this process, we assumed no extra constraints on the collisional kernels and considered that the spin polarization may have, both, classical and quantum origins (zeroth and the first order, in $\hbar$). In the case of local equilibrium where the collisional kernels vanish using the developed quantum kinetic theory we derived the perfect-fluid hydrodynamics with spin. We obtained them from the conservation laws for the net baryon current, the energy–momentum tensor and the spin tensor using GLW definitions of the currents. These relations agree with the conservation equations derived using the classical approach in the small polarization limit. We then studied the propagation properties of the spin polarization components using the general form of spin tensor and derived the spin-wave velocity for the MJ and FD statistics. We analyzed the wave spectrum of the spin polarization components and found that only the transverse spin components propagate, as in the case of EM waves. Finally, we studied the space–time evolution of the spin polarization components using the formalism of perfect-fluid hydrodynamics with spin and obtained some novel numerical and analytic results using various hydrodynamic backgrounds and/or external electric field. We found qualitative agreement between our spin polarization results with other models and experimental observations.

## Acknowledgments

I am very grateful to V. E. Ambrus, A. Das, W. Florkowski, A. Kumar, R. Ryblewski, M. Shokri, G. Sophys, S. M. A. Tabatabaee Mehr for their collaborations and inspiring discussions. I would also like to acknowledge the funding resources which contributed partly to the completion of this work:

- Institute of Nuclear Physics Polish Academy of Sciences.
- Polish National Science Centre Research Grant Nos. 2016/23/B/ST2/00717 and 2018/30/E/ST2/00432.
- NAWA PROM PPI/PRO/2019/1/00016/U/001.
- Polish NAWA Bekker Program No. BPN/BEK/2021/1/00342.

I would also like to thank following people for their interesting comments: S. Bhadury, S. Bhattacharyya, K. G. Biernat, W. Broniowski, A. Dash, A. Jaiswal, I. Karpenko, A. Khuntia, A. Palermo, D. H. Rischke, N. Sadooghi, D. Sénéchal, E. Speranza, D. Wagner, Y. Wang, N. Weickgenannt and K. Zalewski. I also thank the Institute for Nuclear Theory at the University of Washington for its kind hospitality and stimulating research environment. This research was supported in part by the INT's U.S. Department of Energy grant No. DE-FG02-00ER41132.

## References

1. G. Lemaître, *Ann. Soc. Sci. Brux.* **47**, 49 (1927).
2. A. Friedmann, *Gen. Relativ. Gravit.* **31**, 1991 (1999), https://doi.org/10.1023/A:1026751225741.
3. H. Nussbaumer, L. Bieri and A. Sandage, *Discovering the Expanding Universe* (Cambridge University Press, 2009).
4. K. Lundmark, *Mon. Not. R. Astron. Soc.* **84**, 747 (1924).
5. H. P. Robertson, *Astrophys. J.* **82**, 284 (1935).
6. A. G. Walker, *Proc. Lond. Math. Soc.* **s2-42**, 90 (1937).
7. E. Hubble, *Proc. Natl. Acad. Sci.* **15**, 168 (1929).
8. A. A. Penzias and R. W. Wilson, *Astrophys. J.* **142**, 419 (1965).
9. R. A. Alpher, H. Bethe and G. Gamow, *Phys. Rev.* **73**, 803 (1948).
10. H. R. Schmidt and J. Schukraft, *J. Phys. G* **19**, 1705 (1993).
11. J. Schukraft and R. Stock, *Adv. Ser. Direct. High Energy Phys.* **23**, 61 (2015), arXiv:1505.06853 [nucl-ex].

12. E. W. Kolb and M. S. Turner, *The Early Universe* (CRC Press, 1990).
13. A. D. Linde, *Rep. Prog. Phys.* **42**, 389 (1979).
14. D. Boyanovsky, H. J. de Vega and D. J. Schwarz, *Annu. Rev. Nucl. Part. Sci.* **56**, 441 (2006), arXiv:hep-ph/0602002.
15. H. Satz, *Statistical Mechanics of Quarks and Hadrons: Proceedings of an International Symposium Held at the University of Bielefeld, F.R.G., August 24–31, 1980* (North-Holland, 1981), https://books.google.pl/books?id=OY8uAAAAIAAJ.
16. L. D. McLerran, *Rev. Mod. Phys.* **58**, 1021 (1986).
17. B. V. Jacak and B. Muller, *Science* **337**, 310 (2012).
18. K. Fukushima and C. Sasaki, *Prog. Part. Nucl. Phys.* **72**, 99 (2013), arXiv:1301.6377 [hep-ph].
19. D. J. Gross and F. Wilczek, *Phys. Rev. Lett.* **30**, 1343 (1973).
20. H. D. Politzer, *Phys. Rev. Lett.* **30**, 1346 (1973).
21. Y. B. Zel'dovich and Y. P. Raizer, *Physics of Shock Waves and High-Temperature Hydrodynamic Phenomena* (Courier Corporation, 1967).
22. J. R. Oppenheimer and G. M. Volkoff, *Phys. Rev.* **55**, 374 (1939).
23. G. Gamow, *Phys. Rev.* **70**, 572 (1946).
24. Y. B. Zel'dovich, *Zh. Eksp. Teor. Fiz.* **41**, 1609 (1961).
25. F. Karsch and E. Laermann, *Phys. Rev. D* **50**, 6954 (1994), arXiv:hep-lat/9406008.
26. F. Karsch, *Lect. Notes Phys.* **583**, 209 (2002), arXiv:hep-lat/0106019.
27. L. D. McLerran and B. Svetitsky, *Phys. Rev. D* **24**, 450 (1981).
28. P. de Forcrand, *PoS* **LAT2009**, 10 (2009), arXiv:1005.0539 [hep-lat].
29. Y. Nambu and G. Jona-Lasinio, *Phys. Rev.* **122**, 345 (1961).
30. Y. Nambu and G. Jona-Lasinio, *Phys. Rev.* **124**, 246 (1961).
31. S. P. Klevansky, *Rev. Mod. Phys.* **64**, 649 (1992).
32. M. Buballa, *Phys. Rep.* **407**, 205 (2005), arXiv:hep-ph/0402234.
33. C. Ratti, S. Roessner, M. A. Thaler and W. Weise, *Eur. Phys. J. C* **49**, 213 (2007), arXiv:hep-ph/0609218.
34. J. C. Collins and M. J. Perry, *Phys. Rev. Lett.* **34**, 1353 (1975).
35. R. Hagedorn, *Nuovo Cimento Suppl.* **3**, 147 (1965).
36. N. Cabibbo and G. Parisi, *Phys. Lett. B* **59**, 67 (1975).
37. E. V. Shuryak, *Phys. Lett. B* **78**, 150 (1978).
38. E. V. Shuryak, *Phys. Rep.* **61**, 71 (1980).
39. G. F. Chapline, M. H. Johnson, E. Teller and M. S. Weiss, *Phys. Rev. D* **8**, 4302 (1973).
40. W. Scheid, H. Muller and W. Greiner, *Phys. Rev. Lett.* **32**, 741 (1974).
41. M. I. Sobel, H. A. Bethe, P. J. Siemens and J. P. Bondorf, *Nucl. Phys. A* **251**, 502 (1975).
42. S. Nagamiya and M. Gyulassy, *Adv. Nucl. Phys.* **13**, 201 (1984).
43. R. Stock, *Phys. Rep.* **135**, 259 (1986).
44. H. Stoecker and W. Greiner, *Phys. Rep.* **137**, 277 (1986).
45. R. C. Hwa (ed.), *Quark-Gluon Plasma* (World Scientific, Singapore, 1990).
46. C. Y. Wong, *Introduction to High-Energy Heavy-Ion Collisions* (World Scientific, 1995).
47. B. Muller, *Rep. Prog. Phys.* **58**, 611 (1995), arXiv:nucl-th/9410005.
48. K. Yagi, T. Hatsuda and Y. Miake, *Quark-Gluon Plasma: From Big Bang to Little Bang, Cambridge Monographs on Particle Physics, Nuclear Physics and Cosmology*, Vol. 23 (Cambridge University Press, 2005), pp. 1–466.
49. R. Vogt, *Ultrarelativistic Heavy-Ion Collisions* (Elsevier, Amsterdam, 2007).
50. W. Florkowski, *Phenomenology of Ultra-Relativistic Heavy-Ion Collisions* (World Scientific, Singapore, 2010), https://cds.cern.ch/record/1321594.
51. W. Busza, K. Rajagopal and W. van der Schee, *Annu. Rev. Nucl. Part. Sci.* **68**, 339 (2018), arXiv:1802.04801 [hep-ph].

52. R. Stock, 7 Relativistic nucleus-nucleus collisions and the QCD matter phase diagram, in *Theory and Experiments*, Landolt-Börnstein-Group I Elementary Particles, Nuclei and Atoms, Vol. 21A (Springer-Verlag, Berlin, 2008), https://doi.org/10.1007/978-3-540-74203-6_7, https://materials.springer.com/lb/docs/sm_lbs_978-3-540-74203-6_7.

53. Z.-T. Liang, M. A. Lisa and X.-N. Wang, *Nucl. Phys. News* **30**, 10 (2020), arXiv:1912.07822 [nucl-th].

54. STAR Collab. (J. Adams *et al.*), *Nucl. Phys. A* **757**, 102 (2005), arXiv:nucl-ex/0501009.

55. W. Broniowski and W. Florkowski, *Phys. Rev. Lett.* **87**, 272302 (2001), arXiv:nucl-th/0106050.

56. L. D. Landau, *Izv. Akad. Nauk Ser. Fiz.* **17**, 51 (1953).

57. W. Broniowski, M. Chojnacki, W. Florkowski and A. Kisiel, *Phys. Rev. Lett.* **101**, 22301 (2008), arXiv:0801.4361 [nucl-th].

58. W. Florkowski, *Acta Phys. Pol. B* **45**, 2329 (2014), arXiv:1410.7904 [nucl-th].

59. WA80 Collab. (R. Albrecht *et al.*), *Phys. Rev. Lett.* **76**, 3506 (1996).

60. WA98 Collab. (M. M. Aggarwal *et al.*), *Phys. Rev. Lett.* **85**, 3595 (2000), arXiv:nucl-ex/0006008.

61. T. Matsui and H. Satz, *Phys. Lett. B* **178**, 416 (1986).

62. N. Brambilla *et al.*, *Eur. Phys. J. C* **71**, 1534 (2011), arXiv:1010.5827 [hep-ph].

63. A. Andronic *et al.*, *Eur. Phys. J. C* **76**, 107 (2016), arXiv:1506.03981 [nucl-ex].

64. J. Rafelski and B. Muller, *Phys. Rev. Lett.* **48**, 1066 (1982) [Erratum-ibid. **56**, 2334 (1986)].

65. P. Koch, B. Muller and J. Rafelski, *Phys. Rep.* **142**, 167 (1986).

66. M. Marczenko, D. Blaschke, K. Redlich and C. Sasaki, *Phys. Rev. D* **98**, 103021 (2018), arXiv:1805.06886 [nucl-th].

67. WA97 Collab., *Eur. Phys. J. C* **14**, 633 (2000).

68. M. Gyulassy and M. Plumer, *Phys. Lett. B* **243**, 432 (1990).

69. M. Gyulassy, I. Vitev, X.-N. Wang and B.-W. Zhang, arXiv:nucl-th/0302077.

70. ATLAS Collab. (G. Aad *et al.*), *Phys. Rev. Lett.* **105**, 252303 (2010), arXiv:1011.6182 [hep-ex].

71. CMS Collab. (S. Chatrchyan *et al.*), *Phys. Rev. C* **84**, 24906 (2011), arXiv:1102.1957 [nucl-ex].

72. U. W. Heinz and M. Jacob, arXiv:nucl-th/0002042.

73. PHENIX Collab. (K. Adcox *et al.*), *Nucl. Phys. A* **757**, 184 (2005), arXiv:nucl-ex/0410003.

74. M. Gyulassy and L. McLerran, *Nucl. Phys. A* **750**, 30 (2005), arXiv:nucl-th/0405013.

75. E. Shuryak, *Prog. Part. Nucl. Phys.* **62**, 48 (2009), arXiv:0807.3033 [hep-ph].

76. S. Giorgini, L. P. Pitaevskii and S. Stringari, *Rev. Mod. Phys.* **80**, 1215 (2008), arXiv:0706.3360 [cond-mat.other].

77. Fermi-LAT Collab. (R. P. Johnson), *PoS* **ICHEP2010**, 434 (2010).

78. C. Cao, E. Elliott, J. Joseph, H. Wu, J. Petricka, T. Schäfer and J. E. Thomas, *Science* **331**, 58 (2011), arXiv:1007.2625 [cond-mat.quant-gas].

79. G. Policastro, D. T. Son and A. O. Starinets, *Phys. Rev. Lett.* **87**, 81601 (2001), arXiv:hep-th/0104066.

80. P. Kovtun, D. T. Son and A. O. Starinets, *J. High Energy Phys.* **2003**, 64 (2003), arXiv:hep-th/0309213.

81. J. M. Maldacena, *Adv. Theor. Math. Phys.* **2**, 231 (1998), arXiv:hep-th/9711200 [hep-th].

82. A. Vilenkin, *Phys. Rev. D* **22**, 3080 (1980).

83. K. Fukushima, D. E. Kharzeev and H. J. Warringa, *Phys. Rev. D* **78**, 74033 (2008), arXiv:0808.3382 [hep-ph].

84. F. Becattini, F. Piccinini and J. Rizzo, *Phys. Rev. C* **77**, 24906 (2008), arXiv:0711.1253 [nucl-th].

85. STAR Collab. (L. Adamczyk *et al.*), *Nature* **548**, 62 (2017), arXiv:1701.06657 [nucl-ex].

86. W. Florkowski, R. Ryblewski and A. Kumar, *Prog. Part. Nucl. Phys.* **108**, 103709 (2019), arXiv:1811.04409 [nucl-th].

87. Z.-T. Liang and X.-N. Wang, *Phys. Rev. Lett.* **94**, 102301 (2005) [Erratum-ibid. **96**, 39901 (2006)], arXiv:nucl-th/0410079 [nucl-th].

88. A. Einstein and W. de Haas, *Deutsch. Phys. Ges., Verh.* **17**, 152 (1915).

89. S. J. Barnett, *Rev. Mod. Phys.* **7**, 129 (1935).

90. STAR Collab. (J. Adam *et al.*), *Phys. Rev. Lett.* **126**, 162301 (2021), arXiv:2012.13601 [nucl-ex].

91. STAR Collab. (J. Adam *et al.*), *Phys. Rev. C* **98**, 14910 (2018), arXiv:1805.04400 [nucl-ex].

92. F. Becattini, I. Karpenko, M. Lisa, I. Upsal and S. Voloshin, *Phys. Rev. C* **95**, 54902 (2017), arXiv:1610.02506 [nucl-th].

93. J.-H. Gao, Z.-T. Liang, Q. Wang and X.-N. Wang, *Lect. Notes Phys.* **987**, 195 (2021), arXiv:2009.04803 [nucl-th].

94. F. Becattini and M. A. Lisa, *Annu. Rev. Nucl. Part. Sci.* **70**, 395 (2020), arXiv:2003.03640 [nucl-ex].

95. E. Speranza and N. Weickgenannt, *Eur. Phys. J. A* **57**, 155 (2021), arXiv:2007.00138 [nucl-th].

96. A. Jaiswal *et al.*, *Int. J. Mod. Phys. E* **30**, 2130001 (2021), arXiv:2007.14959 [hep-ph].

97. F. Becattini, J. Liao and M. Lisa, *Lect. Notes Phys.* **987**, 1 (2021), arXiv:2102.00933 [nucl-th].

98. J. L. Francesco Becattini and M. Lisa (eds.), *Strongly Interacting Matter under Rotation, Lecture Notes in Physics* (Springer, 2021).

99. S. Bhadury, J. Bhatt, A. Jaiswal and A. Kumar, *Eur. Phys. J. Spec. Top.* **230**, 655 (2021), arXiv:2101.11964 [hep-ph].

100. F. Becattini, arXiv:2204.01144 [nucl-th].

101. J.-H. Gao, S.-W. Chen, W.-T. Deng, Z.-T. Liang, Q. Wang and X.-N. Wang, *Phys. Rev. C* **77**, 44902 (2008), arXiv:0710.2943 [nucl-th].

102. Q. Wang, *Nucl. Phys. A* **967**, 225 (2017), arXiv:1704.04022 [nucl-th].

103. D. E. Kharzeev and D. T. Son, *Phys. Rev. Lett.* **106**, 62301 (2011), arXiv:1010.0038 [hep-ph].

104. D. E. Kharzeev, J. Liao, S. A. Voloshin and G. Wang, *Prog. Part. Nucl. Phys.* **88**, 1 (2016), arXiv:1511.04050 [hep-ph].

105. W.-T. Deng and X.-G. Huang, *Phys. Rev. C* **93**, 64907 (2016), arXiv:1603.06117 [nucl-th].

106. Y. Hidaka, S. Pu and D.-L. Yang, *Phys. Rev. D* **95**, 91901 (2017), arXiv:1612.04630 [hep-th].

107. CMS Collab. (V. Khachatryan *et al.*), *Phys. Rev. Lett.* **118**, 122301 (2017), arXiv:1610.00263 [nucl-ex].

108. CMS Collab. (A. M. Sirunyan *et al.*), *Phys. Rev. C* **97**, 44912 (2018), arXiv:1708.01602 [nucl-ex].

109. Y.-C. Liu, L.-L. Gao, K. Mameda and X.-G. Huang, *Phys. Rev. D* **99**, 85014 (2019), arXiv:1812.10127 [hep-th].

110. J. Zhao and F. Wang, *Prog. Part. Nucl. Phys.* **107**, 200 (2019), arXiv:1906.11413 [nucl-ex].

111. STAR Collab. (J. Adam *et al.*), arXiv:2006.04251 [nucl-ex].

112. STAR Collab. (M. Abdallah *et al.*), *Phys. Rev. C* **105**, 14901 (2022), arXiv:2109.00131 [nucl-ex].

113. M. Buzzegoli, arXiv:2011.09974 [hep-th].
114. D. E. Kharzeev, L. D. McLerran and H. J. Warringa, *Nucl. Phys. A* **803**, 227 (2008), arXiv:0711.0950 [hep-ph].
115. STAR Collab. (B. I. Abelev *et al.*), *Phys. Rev. Lett.* **103**, 251601 (2009), arXiv:0909.1739 [nucl-ex].
116. D. E. Kharzeev and H.-U. Yee, *Phys. Rev. D* **83**, 85007 (2011), arXiv:1012.6026 [hep-th].
117. D. E. Kharzeev, *Prog. Part. Nucl. Phys.* **75**, 133 (2014), arXiv:1312.3348 [hep-ph].
118. P. Bozek, *Phys. Rev. C* **97**, 34907 (2018), arXiv:1711.02563 [nucl-th].
119. X. An *et al.*, *Nucl. Phys. A* **1017**, 122343 (2022), arXiv:2108.13867 [nucl-th].
120. STAR Collab. (M. Abdallah *et al.*), *Phys. Rev. Lett.* **128**, 92301 (2022), arXiv:2106.09243 [nucl-ex].
121. L. Yin, D. Hou and H.-C. Ren, *J. High Energy Phys.* **2021**, 117 (2021), arXiv:2102.04851 [hep-th].
122. D. E. Kharzeev, arXiv:2204.10903 [hep-ph].
123. F. Becattini and I. Karpenko, *Phys. Rev. Lett.* **120**, 12302 (2018), arXiv:1707.07984 [nucl-th].
124. S. A. Voloshin, *EPJ Web Conf.* **17**, 10700 (2018), arXiv:1710.08934 [nucl-ex].
125. STAR Collab. (J. Adam *et al.*), *Phys. Rev. Lett.* **123**, 132301 (2019), arXiv:1905.11917 [nucl-ex].
126. X.-L. Xia, H. Li, Z.-B. Tang and Q. Wang, *Phys. Rev. C* **98**, 24905 (2018), arXiv:1803.00867 [nucl-th].
127. I. Karpenko and F. Becattini, *Eur. Phys. J. C* **77**, 213 (2017), arXiv:1610.04717 [nucl-th].
128. Y. Sun and C. M. Ko, *Phys. Rev. C* **99**, 11903 (2019), arXiv:1810.10359 [nucl-th].
129. S. Bader and S. Parkin, *Annu. Rev. Condens. Matter Phys.* **1**, 71 (2010).
130. R. Takahashi, M. Matsuo, M. Ono, K. Harii, H. Chudo, S. Okayasu, J. Ieda, S. Takahashi, S. Maekawa and E. Saitoh, *Nat. Phys.* **12**, 52 (2016).
131. A. Hirohata, K. Yamada, Y. Nakatani, I.-L. Prejbeanu, B. Diény, P. Pirro and B. Hillebrands, *J. Magn. Magn. Mater.* **509**, 166711 (2020), https://hal.archives-ouvertes.fr/hal-03192774.
132. STAR Collab. (M. S. Abdallah *et al.*), *Phys. Rev. C* **104**, L061901 (2021), arXiv:2108.00044 [nucl-ex].
133. ALICE Collab. (S. Acharya *et al.*), *Phys. Rev. C* **101**, 44611 (2020), arXiv:1909.01281 [nucl-ex].
134. ALICE Collab. (S. Acharya *et al.*), arXiv:2107.11183 [nucl-ex].
135. Y. B. Ivanov, V. Toneev and A. Soldatov, *Phys. Rev. C* **100**, 14908 (2019), arXiv:1903.05455 [nucl-th].
136. X.-G. Deng, X.-G. Huang, Y.-G. Ma and S. Zhang, *Phys. Rev. C* **101**, 64908 (2020), arXiv:2001.01371 [nucl-th].
137. Y. Guo, J. Liao, E. Wang, H. Xing and H. Zhang, *Phys. Rev. C* **104**, L041902 (2021), arXiv:2105.13481 [nucl-th].
138. A. Ayala, I. Domínguez, I. Maldonado and M. E. Tejeda-Yeomans, *Phys. Rev. C* **105**, 34907 (2022), arXiv:2106.14379 [hep-ph].
139. X.-G. Deng, X.-G. Huang and Y.-G. Ma, arXiv:2109.09956 [nucl-th].
140. D. Teaney, J. Lauret and E. V. Shuryak, arXiv:nucl-th/0110037 [nucl-th].
141. P. Bozek and I. Wyskiel, *Phys. Rev. C* **79**, 44916 (2009), arXiv:0902.4121 [nucl-th].
142. P. Romatschke, *Int. J. Mod. Phys. E* **19**, 1 (2010), arXiv:0902.3663 [hep-ph].
143. C. Gale, S. Jeon and B. Schenke, *Int. J. Mod. Phys. A* **28**, 1340011 (2013), arXiv:1301.5893 [nucl-th].
144. A. Jaiswal and V. Roy, *Adv. High Energy Phys.* **2016**, 9623034 (2016), arXiv:1605.08694 [nucl-th].

145. P. Romatschke and U. Romatschke, *Relativistic Fluid Dynamics In and Out of Equilibrium, Cambridge Monographs on Mathematical Physics* (Cambridge University Press, 2019), arXiv:1712.05815 [nucl-th].

146. W. Florkowski, M. P. Heller and M. Spalinski, *Rep. Prog. Phys.* **81**, 46001 (2018), arXiv:1707.02282 [hep-ph].

147. S. Schlichting and D. Teaney, *Annu. Rev. Nucl. Part. Sci.* **69**, 447 (2019), arXiv:1908.02113 [nucl-th].

148. B. Schenke, *Rep. Prog. Phys.* **84**, 82301 (2021), arXiv:2102.11189 [nucl-th].

149. L. Rezzolla and O. Zanotti, *Relativistic Hydrodynamics* (Oxford University Press, Oxford, 2013).

150. C. Eckart, *Phys. Rev.* **58**, 267 (1940).

151. L. Landau and E. Lifshitz, *Fluid Mechanics, Course of Theoretical Physics*, Vol. 6 (Elsevier Science, 2013).

152. W. Hiscock and L. Lindblom, *Ann. Phys.* **151**, 466 (1983).

153. W. A. Hiscock and L. Lindblom, *Phys. Rev. D* **31**, 725 (1985).

154. D. A. Teaney, R. C. Hwa and X.-N. Wang, Viscous hydrodynamics and the quark gluon plasma, in *Quark-Gluon Plasma 4* (World Scientific, 2010), pp. 207–266, arXiv:0905.2433 [nucl-th].

155. G. S. Denicol, H. Niemi, E. Molnar and D. H. Rischke, *Phys. Rev. D* **85**, 114047 (2012) [Erratum-ibid. **91**, 39902 (2015)], arXiv:1202.4551 [nucl-th].

156. S. De Groot, W. Van Leeuwen and C. Van Weert, *Relativistic Kinetic Theory: Principles and Applications* (North Holland, 1980).

157. C. Cercignani and G. M. Kremer, *The Relativistic Boltzmann Equation: Theory and Applications* (Springer, 2002).

158. A. Jaiswal, Formulation of relativistic dissipative fluid dynamics and its applications in heavy-ion collisions, Ph.D. thesis, Tata Institute of Fundamental Research (2014).

159. F. Becattini, V. Chandra, L. Del Zanna and E. Grossi, *Ann. Phys.* **338**, 32 (2013), arXiv:1303.3431 [nucl-th].

160. F. Becattini, L. Csernai and D. J. Wang, *Phys. Rev. C* **88**, 34905 (2013) [Erratum-ibid. **93**, 69901 (2016)], arXiv:1304.4427 [nucl-th].

161. L.-G. Pang, H. Petersen, Q. Wang and X.-N. Wang, *Phys. Rev. Lett.* **117**, 192301 (2016), arXiv:1605.04024 [hep-ph].

162. Y. Xie, D. Wang and L. P. Csernai, *Phys. Rev. C* **95**, 31901 (2017), arXiv:1703.03770 [nucl-th].

163. Y. Sun and C. M. Ko, *Phys. Rev. C* **96**, 24906 (2017), arXiv:1706.09467 [nucl-th].

164. H. Li, L.-G. Pang, Q. Wang and X.-L. Xia, *Phys. Rev. C* **96**, 54908 (2017), arXiv:1704.01507 [nucl-th].

165. D.-X. Wei, W.-T. Deng and X.-G. Huang, *Phys. Rev. C* **99**, 14905 (2019), arXiv:1810.00151 [nucl-th].

166. J.-H. Gao and Z.-T. Liang, *Phys. Rev. D* **100**, 56021 (2019), arXiv:1902.06510 [hep-ph].

167. Y. B. Ivanov, V. Toneev and A. Soldatov, *Phys. At. Nucl.* **83**, 179 (2020), arXiv:1910.01332 [nucl-th].

168. J. I. Kapusta, E. Rrapaj and S. Rudaz, *Phys. Rev. C* **101**, 31901 (2020), arXiv:1910.12759 [nucl-th].

169. C. Yi, S. Pu and D.-L. Yang, *Phys. Rev. C* **104**, 64901 (2021), arXiv:2106.00238 [hep-ph].

170. S. Ryu, V. Jupic and C. Shen, *Phys. Rev. C* **104**, 54908 (2021), arXiv:2106.08125 [nucl-th].

171. X.-Y. Wu, C. Yi, G.-Y. Qin and S. Pu, *Phys. Rev. C* **105**, 64909 (2022), arXiv:2204.02218 [hep-ph].

R. Singh

172. W. Florkowski, A. Kumar, R. Ryblewski and A. Mazeliauskas, *Phys. Rev. C* **100**, 54907 (2019), arXiv:1904.00002 [nucl-th].

173. F. Becattini, M. Buzzegoli and A. Palermo, *Phys. Lett. B* **820**, 136519 (2021), arXiv:2103.10917 [nucl-th].

174. F. Becattini, M. Buzzegoli, G. Inghirami, I. Karpenko and A. Palermo, *Phys. Rev. Lett.* **127**, 272302 (2021), arXiv:2103.14621 [nucl-th].

175. B. Fu, S. Y. F. Liu, L. Pang, H. Song and Y. Yin, *Phys. Rev. Lett.* **127**, 142301 (2021), arXiv:2103.10403 [hep-ph].

176. W. Florkowski, A. Kumar, A. Mazeliauskas and R. Ryblewski, arXiv:2112.02799 [hep-ph].

177. X.-L. Sheng, Q. Wang and D. H. Rischke, arXiv:2202.10160 [nucl-th].

178. S. Lin and Z. Wang, arXiv:2206.12573 [hep-ph].

179. C. Itzykson and J. B. Zuber, *Quantum Field Theory, International Series in Pure and Applied Physics* (McGraw-Hill, New York, 1980), https://doi.org/10.1063/1.2916419.

180. S. Weinberg, *The Quantum Theory of Fields: Volume 1, Foundations* (Cambridge University Press, 2005).

181. F. J. Belinfante, *Physica* **7**, 449 (1940).

182. F. W. Hehl, *Rep. Math. Phys.* **9**, 55 (1976).

183. L. Tinti and W. Florkowski, arXiv:2007.04029 [nucl-th].

184. F. Belinfante, *Physica* **6**, 887 (1939).

185. L. Rosenfeld, *Mem. Acad. R. Belg.* **18**, 1 (1940).

186. J. Hilgevoord and S. Wouthuysen, *Nucl. Phys.* **40**, 1 (1963).

187. J. Hilgevoord and E. De Kerf, *Physica* **31**, 1002 (1965).

188. M. Buzzegoli, arXiv:2109.12084 [nucl-th].

189. N. Weickgenannt, D. Wagner and E. Speranza, *Phys. Rev. D* **105**, 116026 (2022), arXiv:2204.01797 [nucl-th].

190. E. Leader and C. Lorcé, *Phys. Rep.* **541**, 163 (2014), arXiv:1309.4235 [hep-ph].

191. A. D. Gallegos, U. Gürsoy and A. Yarom, *SciPost Phys.* **11**, 41 (2021), arXiv:2101.04759 [hep-th].

192. N. Weickgenannt, E. Speranza, X.-l. Sheng, Q. Wang and D. H. Rischke, *Phys. Rev. Lett.* **127**, 52301 (2021), arXiv:2005.01506 [hep-ph].

193. N. Weickgenannt, E. Speranza, X.-l. Sheng, Q. Wang and D. H. Rischke, *Phys. Rev. D* **104**, 16022 (2021), arXiv:2103.04896 [nucl-th].

194. D. Montenegro and G. Torrieri, *Phys. Rev. D* **100**, 56011 (2019), arXiv:1807.02796 [hep-th].

195. D. Montenegro and G. Torrieri, *Phys. Rev. D* **102**, 36007 (2020), arXiv:2004.10195 [hep-th].

196. W. M. Serenone, J. G. P. Barbon, D. D. Chinellato, M. A. Lisa, C. Shen, J. Takahashi and G. Torrieri, *Phys. Lett. B* **820**, 136500 (2021), arXiv:2102.11919 [hep-ph].

197. G. Torrieri and D. Montenegro, arXiv:2207.00537 [hep-th].

198. K. Hattori, M. Hongo, X.-G. Huang, M. Matsuo and H. Taya, *Phys. Lett. B* **795**, 100 (2019), arXiv:1901.06615 [hep-th].

199. K. Fukushima and S. Pu, *Phys. Lett. B* **817**, 136346 (2021), arXiv:2010.01608 [hep-th].

200. S. Li, M. A. Stephanov and H.-U. Yee, *Phys. Rev. Lett.* **127**, 82302 (2021), arXiv:2011.12318 [hep-th].

201. D. She, A. Huang, D. Hou and J. Liao, arXiv:2105.04060 [nucl-th].

202. A. Daher, A. Das, W. Florkowski and R. Ryblewski, arXiv:2202.12609 [nucl-th].

203. Z. Cao, K. Hattori, M. Hongo, X.-G. Huang and H. Taya, arXiv:2205.08051 [hep-th].

204. J. Hu, *Phys. Rev. D* **103**, 116015 (2021), arXiv:2101.08440 [hep-ph].

205. Y. Hidaka and D.-L. Yang, *Phys. Rev. D* **98**, 16012 (2018), arXiv:1801.08253 [hep-th].

206. D.-L. Yang, K. Hattori and Y. Hidaka, *J. High Energy Phys.* **2020**, 70 (2020), arXiv:2002.02612 [hep-ph].

207. Z. Wang, X. Guo and P. Zhuang, *Eur. Phys. J. C* **81**, 799 (2021), arXiv:2009.10930 [hep-th].

208. X.-L. Sheng, N. Weickgenannt, E. Speranza, D. H. Rischke and Q. Wang, *Phys. Rev. D* **104**, 16029 (2021), arXiv:2103.10636 [nucl-th].

209. N. Weickgenannt, D. Wagner, E. Speranza and D. Rischke, arXiv:2203.04766 [nucl-th].

210. J. Hu, *Phys. Rev. D* **105**, 76009 (2022), arXiv:2111.03571 [hep-ph].

211. M. A. Stephanov and Y. Yin, *Phys. Rev. Lett.* **109**, 162001 (2012), arXiv:1207.0747 [hep-th].

212. J.-Y. Chen, D. T. Son, M. A. Stephanov, H.-U. Yee and Y. Yin, *Phys. Rev. Lett.* **113**, 182302 (2014), arXiv:1404.5963 [hep-th].

213. E. V. Gorbar, D. O. Rybalka and I. A. Shovkovy, *Phys. Rev. D* **95**, 96010 (2017), arXiv:1702.07791 [hep-th].

214. S. Shi, C. Gale and S. Jeon, *Phys. Rev. C* **103**, 44906 (2021), arXiv:2008.08618 [nucl-th].

215. M. P. Heller, A. Serantes, M. Spaliński, V. Svensson and B. Withers, *SciPost Phys.* **10**, 123 (2021), arXiv:2012.15393 [hep-th].

216. A. D. Gallegos and U. Gürsoy, *J. High Energy Phys.* **2020**, 151 (2020), arXiv:2004.05148 [hep-th].

217. M. Garbiso and M. Kaminski, *J. High Energy Phys.* **2020**, 112 (2020), arXiv:2007.04345 [hep-th].

218. M. Hongo, X.-G. Huang, M. Kaminski, M. Stephanov and H.-U. Yee, *J. High Energy Phys.* **2021**, 150 (2021), arXiv:2107.14231 [hep-th].

219. A. D. Gallegos, U. Gursoy and A. Yarom, arXiv:2203.05044 [hep-th].

220. D. T. Son and P. Surowka, *Phys. Rev. Lett.* **103**, 191601 (2009), arXiv:0906.5044 [hep-th].

221. D. Montenegro, L. Tinti and G. Torrieri, *Phys. Rev. D* **96**, 56012 (2017), arXiv:1701.08263 [hep-th].

222. D. Montenegro, L. Tinti and G. Torrieri, *Phys. Rev. D* **96**, 76016 (2017), arXiv:1703.03079 [hep-th].

223. X.-G. Huang, P. Huovinen and X.-N. Wang, *Phys. Rev. C* **84**, 54910 (2011), arXiv:1108.5649 [nucl-th].

224. X. Li, Z.-F. Jiang, S. Cao and J. Deng, arXiv:2205.02409 [nucl-th].

225. W. Florkowski, B. Friman, A. Jaiswal and E. Speranza, *Phys. Rev. C* **97**, 41901 (2018), arXiv:1705.00587 [nucl-th].

226. W. Florkowski, B. Friman, A. Jaiswal, R. Ryblewski and E. Speranza, *Phys. Rev. D* **97**, 116017 (2018), arXiv:1712.07676 [nucl-th].

227. W. Florkowski, B. Friman, A. Jaiswal and E. Speranza, *Acta Phys. Pol. B Proc. Suppl.* **10**, 1139 (2017), arXiv:1708.04035 [hep-ph].

228. W. Florkowski, B. Friman, A. Jaiswal, R. Ryblewski and E. Speranza, *Acta Phys. Pol. B Proc. Suppl.* **11**, 507 (2018), arXiv:1810.01709 [nucl-th].

229. W. Florkowski, A. Kumar and R. Ryblewski, *Phys. Rev. C* **98**, 44906 (2018), arXiv:1806.02616 [hep-ph].

230. W. Florkowski, A. Kumar and R. Ryblewski, *Acta Phys. Pol. B* **51**, 945 (2020), arXiv:1907.09835 [nucl-th].

231. R. Hakim, *Introduction to Relativistic Statistical Mechanics: Classical and Quantum* (World Scientific, Singapore, 2011), https:cds.cern.ch/record/1379544.

232. H.-H. Peng, J.-J. Zhang, X.-L. Sheng and Q. Wang, *Chin. Phys. Lett.* **38**, 116701 (2021), arXiv:2107.00448 [hep-th].

233. H. T. Elze, M. Gyulassy and D. Vasak, *Nucl. Phys. B* **276**, 706 (1986).

234. D. Vasak, M. Gyulassy and H. T. Elze, *Ann. Phys.* **173**, 462 (1987).

235. H.-T. Elze and U. W. Heinz, *Phys. Rep.* **183**, 81 (1989) [Erratum-ibid. **183**, 117 (1989)], https://inspirehep.net/literature/278581.

236. P. Zhuang and U. W. Heinz, *Ann. Phys.* **245**, 311 (1996), arXiv:nucl-th/9502034 [nucl-th].

237. W. Florkowski, J. Hufner, S. P. Klevansky and L. Neise, *Ann. Phys.* **245**, 445 (1996), arXiv:hep-ph/9505407 [hep-ph].

238. W. Florkowski, A. Kumar, R. Ryblewski and R. Singh, *Phys. Rev. C* **99**, 44910 (2019), arXiv:1901.09655 [hep-ph].

239. R. Singh, G. Sophys and R. Ryblewski, *Phys. Rev. D* **103**, 74024 (2021), arXiv:2011.14907 [hep-ph].

240. R. Singh, M. Shokri and R. Ryblewski, *Phys. Rev. D* **103**, 94034 (2021), arXiv:2103.02592 [hep-ph].

241. W. Florkowski, R. Ryblewski, R. Singh and G. Sophys, *Phys. Rev. D* **105**, 54007 (2022), arXiv:2112.01856 [hep-ph].

242. R. Singh, M. Shokri and S. M. A. T. Mehr, *Nucl. Phys. A* **1035**, 122656 (2023), arXiv:2202.11504 [hep-ph].

243. R. Singh, *Acta Phys. Pol. B Proc. Suppl.* **15**, 35 (2022).

244. S. Bhadury, W. Florkowski, A. Jaiswal and R. Ryblewski, *Phys. Rev. C* **102**, 64910 (2020), arXiv:2006.14252 [hep-ph].

245. S. Bhadury, W. Florkowski, A. Jaiswal, A. Kumar and R. Ryblewski, *Phys. Lett. B* **814**, 136096 (2021), arXiv:2002.03937 [hep-ph].

246. S. Bhadury, W. Florkowski, A. Jaiswal, A. Kumar and R. Ryblewski, *Phys. Rev. D* **103**, 14030 (2021), arXiv:2008.10976 [nucl-th].

247. S. Bhadury, W. Florkowski, A. Jaiswal, A. Kumar and R. Ryblewski, arXiv:2204.01357 [nucl-th].

248. V. E. Ambrus, R. Ryblewski and R. Singh, *Phys. Rev. D* **106**, 14018 (2022), arXiv:2202.03952 [hep-ph].

249. A. Monnai, Relativistic dissipative hydrodynamic description of the quark-gluon plasma, Ph.D. thesis, University of Tokyo (2014).

250. V. E. Ambrus, *Phys. Rev. C* **97**, 24914 (2018), arXiv:1706.05310 [physics.flu-dyn].

251. J. Hu, *Phys. Rev. D* **105**, 96021 (2022), arXiv:2204.12946 [hep-ph].

252. J. Hu, arXiv:2202.07373 [hep-ph].

253. J. Hu and Z. Xu, arXiv:2205.15755 [hep-ph].

254. R. Singh, *Acta Phys. Pol. B Proc. Suppl.* **13**, 931 (2020), arXiv:2001.05592 [hep-ph].

255. R. Ryblewski and R. Singh, *Acta Phys. Pol. B* **51**, 1537 (2020), arXiv:2004.02544 [nucl-th].

256. R. Singh, *PoS* **LHCP2020**, 236 (2021), arXiv:2009.05130 [nucl-th].

257. R. Singh, *Acta Phys. Pol. B Proc. Suppl.* **14**, 461 (2021), arXiv:2009.07067 [nucl-th].

258. A. Das, W. Florkowski, R. Ryblewski and R. Singh, *Acta Phys. Pol. B* **52**, 1395 (2021), arXiv:2012.05662 [hep-ph].

259. A. Das, W. Florkowski, R. Ryblewski and R. Singh, *Phys. Rev. D* **103**, L091502 (2021), arXiv:2103.01013 [nucl-th].

260. R. Singh, *Acta Phys. Pol. B* **52**, 1081 (2021), arXiv:2104.01009 [hep-ph].

261. A. Das, W. Florkowski, R. Ryblewski and R. Singh, *Acta Phys. Pol. B* **53**, 5 (2022), arXiv:2105.02125 [nucl-th].

262. R. Singh, *PoS* **LHCP2021**, 161 (2021), arXiv:2109.11068 [quant-ph].

263. R. Singh, *J. Phys., Conf. Ser.* **2105**, 12006 (2021), arXiv:2109.08201 [nucl-th].

264. R. Singh, *Rev. Mex. Fis. Suppl.* **3**, 308115 (2022), arXiv:2110.02727 [nucl-th].

265. A. Das, W. Florkowski, A. Kumar, R. Ryblewski and R. Singh, arXiv:2203.15562 [hep-th].

266. C. Cercignani, *J. Fluid Mech.* **81**, 793–794 (1977).

267. G. S. Denicol and D. H. Rischke, *Microscopic Foundations of Relativistic Fluid Dynamics* (Springer, Cham, 2021).

268. E. Wigner, *Phys. Rev.* **40**, 749 (1932).

269. X.-l. Sheng, D. H. Rischke, D. Vasak and Q. Wang, *Eur. Phys. J. A* **54**, 21 (2018), arXiv:1707.01388 [hep-ph].

270. D. Kobayashi, T. Yoshikawa, M. Matsuo, R. Iguchi, S. Maekawa, E. Saitoh and Y. Nozaki, *Phys. Rev. Lett.* **119**, 77202 (2017).

271. M. Matsuo, Y. Ohnuma and S. Maekawa, *Phys. Rev. B* **96**, 20401 (2017).

272. J.-H. Gao, S. Pu and Q. Wang, *Phys. Rev. D* **96**, 16002 (2017), arXiv:1704.00244 [nucl-th].

273. D.-L. Yang, *Phys. Rev. D* **98**, 76019 (2018), arXiv:1807.02395 [nucl-th].

274. J.-H. Gao, J.-Y. Pang and Q. Wang, *Phys. Rev. D* **100**, 16008 (2019), arXiv:1810.02028 [nucl-th].

275. A. Huang, S. Shi, Y. Jiang, J. Liao and P. Zhuang, *Phys. Rev. D* **98**, 36010 (2018), arXiv:1801.03640 [hep-th].

276. H. T. Elze, M. Gyulassy and D. Vasak, *Phys. Lett. B* **177**, 402 (1986).

277. J.-H. Gao, Z.-T. Liang, S. Pu, Q. Wang and X.-N. Wang, *Phys. Rev. Lett.* **109**, 232301 (2012), arXiv:1203.0725 [hep-ph].

278. R.-H. Fang, L.-G. Pang, Q. Wang and X.-N. Wang, *Phys. Rev. C* **94**, 24904 (2016), arXiv:1604.04036 [nucl-th].

279. R.-H. Fang, J.-Y. Pang, Q. Wang and X.-N. Wang, *Phys. Rev. D* **95**, 14032 (2017), arXiv:1611.04670 [nucl-th].

280. N. Weickgenannt, X.-L. Sheng, E. Speranza, Q. Wang and D. H. Rischke, *Phys. Rev. D* **100**, 56018 (2019), arXiv:1902.06513 [hep-ph].

281. X.-G. Huang, P. Mitkin, A. V. Sadofyev and E. Speranza, *J. High Energy Phys.* **2020**, 117 (2020), arXiv:2006.03591 [hep-th].

282. J.-H. Gao, Z.-T. Liang and Q. Wang, *Int. J. Mod. Phys. A* **36**, 2130001 (2021), arXiv:2011.02629 [hep-ph].

283. Y. Hidaka, S. Pu, Q. Wang and D.-L. Yang, arXiv:2201.07644 [hep-ph].

284. J. Zamanian, M. Marklund and G. Brodin, *New J. Phys.* **12**, 43019 (2010), arXiv:0910.5165 [cond-mat.quant-gas].

285. R. Ekman, F. A. Asenjo and J. Zamanian, *Phys. Rev. E* **96**, 23207 (2017), arXiv:1702.00722 [physics.plasm-ph].

286. R. Ekman, H. Al-Naseri, J. Zamanian and G. Brodin, *Phys. Rev. E* **100**, 23201 (2019), arXiv:1904.08727 [physics.plasm-ph].

287. F. Jüttner, *Ann. Phys.* **339**, 856 (1911).

288. C. Cercignani and G. M. Kremer, *The Relativistic Boltzmann Equation: Theory and Applications* (Birkhäuser, Basel, 2002).

289. F. W. Olver, D. W. Lozier, R. F. Boisvert and C. W. Clark, *NIST Handbook of Mathematical Functions Hardback and CD-ROM* (Cambridge University Press, 2010).

290. M. Mathisson, *Acta Phys. Pol.* **6**, 163 (1937).

291. J. D. Jackson, *Classical Electrodynamics* (John Wiley & Sons Inc., 1998), p. 808.

292. W. Florkowski and R. Ryblewski, *Phys. Rev. C* **85**, 44902 (2012), arXiv:1111.5997 [nucl-th].

293. M. Martinez, R. Ryblewski and M. Strickland, *Phys. Rev. C* **85**, 64913 (2012), arXiv:1204.1473 [nucl-th].

294. M. Alqahtani, M. Nopoush and M. Strickland, *Phys. Rev. C* **92**, 54910 (2015), arXiv:1509.02913 [hep-ph].

295. P. A. M. Dirac, *Proc. R. Soc. Lond. A* **112**, 661 (1926).

296. A. Zannoni, arXiv:cond-mat/9912229 [cond-mat.stat-mech].

297. S. Floerchinger and M. Martinez, *Phys. Rev. C* **92**, 64906 (2015), arXiv:1507.05569 [nucl-th].
298. J. K. Lubański, *Physica* **9**, 310 (1942).
299. L. H. Ryder, *Quantum Field Theory* (Cambridge University Press, 1996).
300. F. Cooper and G. Frye, *Phys. Rev. D* **10**, 186 (1974).
301. R. Ryblewski, Monte-Carlo statistical hadronization in relativistic heavy-ion collisions, in *53rd Winter School of Theoretical Physics: Understanding the Origin of Matter from QCD*, Karpacz, Poland, 2017, arXiv:1712.05213 [nucl-th].
302. E. Leader, *Spin in Particle Physics* (Cambridge University Press, 2001).
303. BRAHMS Collab. (I. G. Bearden *et al.*), *Phys. Rev. Lett.* **94**, 162301 (2005), arXiv:nucl-ex/0403050.
304. R. P. Feynman, *Phys. Rev. Lett.* **23**, 1415 (1969).
305. J. Bjorken, *Phys. Rev. D* **27**, 140 (1983).
306. Particle Data Group Collab. (P. A. Zyla *et al.*), *Prog. Theor. Exp. Phys.* **2020**, 83C01 (2020).
307. S. A. Voloshin, arXiv:nucl-th/0410089 [nucl-th].
308. W. Florkowski and R. Ryblewski, arXiv:2102.02890 [hep-ph].
309. M. Abramowitz and I. A. Stegun (eds.), *Handbook of Mathematical Functions with Formulas, Graphs and Mathematical Tables* (Dover Publications, New York, 1965).
310. STAR Collab. (Z. Xu), *Nucl. Phys. A* **1005**, 121894 (2021).
311. X.-G. Huang, *Rep. Prog. Phys.* **79**, 76302 (2016), arXiv:1509.04073 [nucl-th].
312. L. Oliva, P. Moreau, V. Voronyuk and E. Bratkovskaya, *Phys. Rev. C* **101**, 14917 (2020), arXiv:1909.06770 [nucl-th].
313. V. Roy, S. Pu, L. Rezzolla and D. Rischke, *Phys. Lett. B* **750**, 45 (2015), arXiv:1506.06620 [nucl-th].
314. S. Pu, V. Roy, L. Rezzolla and D. H. Rischke, *Phys. Rev. D* **93**, 74022 (2016), arXiv:1602.04953 [nucl-th].
315. G. Inghirami, L. Del Zanna, A. Beraudo, M. H. Moghaddam, F. Becattini and M. Bleicher, *Eur. Phys. J. C* **76**, 659 (2016), arXiv:1609.03042 [hep-ph].
316. A. Dash, V. Roy and B. Mohanty, *J. Phys. G* **46**, 15103 (2019), arXiv:1705.05657 [nucl-th].
317. M. Shokri and N. Sadooghi, *Phys. Rev. D* **96**, 116008 (2017), arXiv:1705.00536 [nucl-th].
318. W. Florkowski, A. Kumar and R. Ryblewski, *Eur. Phys. J. A* **54**, 184 (2018), arXiv:1803.06695 [nucl-th].
319. M. Shokri and N. Sadooghi, *J. High Energy Phys.* **2018**, 181 (2018), arXiv:1807.09487 [nucl-th].
320. M. Shokri, *J. High Energy Phys.* **2020**, 11 (2020), arXiv:1911.06196 [hep-th].
321. G. Inghirami, M. Mace, Y. Hirono, L. Del Zanna, D. E. Kharzeev and M. Bleicher, *Eur. Phys. J. C* **80**, 293 (2020), arXiv:1908.07605 [hep-ph].
322. U. Gangopadhyaya and V. Roy, arXiv:2206.12197 [nucl-th].
323. J. Hernandez and P. Kovtun, *J. High Energy Phys.* **2017**, 1 (2017), arXiv:1703.08757 [hep-th].
324. J. D. Bekenstein and E. Oron, *Phys. Rev. D* **18**, 1809 (1978), https://doi.org/10.1103/PhysRevD.18.1809.
325. L. Parker and D. Toms, *Quantum Field Theory in Curved Spacetime: Quantized Fields and Gravity, Cambridge Monographs on Mathematical Physics* (Cambridge University Press, 2009).
326. P. Kovtun, *J. High Energy Phys.* **2016**, 28 (2016), arXiv:1606.01226 [hep-th].
327. I. A. Karpenko, P. Huovinen, H. Petersen and M. Bleicher, *Phys. Rev. C* **91**, 64901 (2015), arXiv:1502.01978 [nucl-th].

328. V. B. Jovanovic, S. R. Ignjatovic, D. Borka and P. Jovanovic, *Phys. Rev. D* **82**, 117501 (2010), arXiv:1011.1749 [hep-ph].

329. S. S. Gubser, S. S. Pufu and A. Yarom, *Phys. Rev. D* **78**, 66014 (2008), arXiv:0805.1551 [hep-th].

330. G. Aarts and A. Nikolaev, *Eur. Phys. J. A* **57**, 118 (2021), arXiv:2008.12326 [hep-lat].

331. S. S. Gubser, *Phys. Rev. D* **82**, 85027 (2010), arXiv:1006.0006 [hep-th].

332. S. S. Gubser and A. Yarom, *Nucl. Phys. B* **846**, 469 (2011), arXiv:1012.1314 [hep-th].

333. S. S. Gubser, S. S. Pufu and A. Yarom, *J. High Energy Phys.* **2009**, 50 (2009), arXiv:0902.4062 [hep-th].

334. M. Csanád, M. I. Nagy and S. Lökös, *Eur. Phys. J. A* **48**, 173 (2012), arXiv:1205.5965 [nucl-th].

335. H. Marrochio, J. Noronha, G. S. Denicol, M. Luzum, S. Jeon and C. Gale, *Phys. Rev. C* **91**, 14903 (2015), arXiv:1307.6130 [nucl-th].

336. S. S. Gubser and W. van der Schee, *J. High Energy Phys.* **2015**, 28 (2015), arXiv:1410.7408 [hep-th].

337. G. S. Denicol, U. W. Heinz, M. Martinez, J. Noronha and M. Strickland, *Phys. Rev. Lett.* **113**, 202301 (2014), arXiv:1408.5646 [hep-ph].

338. M. Nopoush, R. Ryblewski and M. Strickland, *Phys. Rev. D* **91**, 45007 (2015), arXiv:1410.6790 [nucl-th].

339. M. Strickland, M. Nopoush and R. Ryblewski, *Nucl. Phys. A* **956**, 268 (2016), arXiv:1512.07334 [nucl-th].

340. A. Dash and V. Roy, *Phys. Lett. B* **806**, 135481 (2020), arXiv:2001.10756 [nucl-th].

341. D.-L. Wang, X.-Q. Xie, S. Fang and S. Pu, *Phys. Rev. D* **105**, 114050 (2022), arXiv:2112.15535 [hep-ph].

342. R. Baier, P. Romatschke, D. T. Son, A. O. Starinets and M. A. Stephanov, *J. High Energy Phys.* **2008**, 100 (2008), arXiv:0712.2451 [hep-th].

343. S. Bhattacharyya, S. Lahiri, R. Loganayagam and S. Minwalla, *J. High Energy Phys.* **2008**, 54 (2008), arXiv:0708.1770 [hep-th].

344. R. Loganayagam, *J. High Energy Phys.* **2008**, 87 (2008), arXiv:0801.3701 [hep-th].

345. H. Kastrup, *Ann. Phys.* **17**, 631 (2008), arXiv:0808.2730 [physics.hist-ph].

346. L. Fabbri, *Ann. Fond. Louis Broglie* **38**, 155 (2013), arXiv:1101.2334 [gr-qc].

347. E. B. Christoffel, *J. Reine Angew. Math.* **B70**, 46 (1869), http://gdz.sub.uni-goettingen.de/dms/load/img/?PPN=GDZPPN002153882&IDDOC=266356.

348. M. M. G. Ricci and T. Levi-Civita, *Math. Ann.* **54**, 125 (1901), https://link.springer.com/article/10.1007/BF01454201.

349. S. Carroll, *Spacetime and Geometry: An Introduction to General Relativity* (Benjamin Cummings, 2003).

350. R. M. Wald, *General Relativity* (Chicago University Press, Chicago, IL, 1984), https://cds.cern.ch/record/106274.

351. V. Faraoni, E. Gunzig and P. Nardone, *Fundam. Cosm. Phys.* **20**, 121 (1999), arXiv:gr-qc/9811047.

352. J. Callan, G. Curtis, S. R. Coleman and R. Jackiw, *Ann. Phys.* **59**, 42 (1970).

353. J. Wess, *Springer Tracts Mod. Phys.* **60**, 1 (1971).

354. P. Di Francesco, P. Mathieu and D. Senechal, *Conformal Field Theory*, Graduate Texts in Contemporary Physics (Springer-Verlag, New York, 1997).

355. M. Forger and H. Romer, *Ann. Phys.* **309**, 306 (2004), arXiv:hep-th/0307199.